\documentclass[a4paper,fleqn,usenatbib]{mnras}

\usepackage{newtxtext,newtxmath}

\usepackage[T1]{fontenc}
\usepackage{ae,aecompl}

\usepackage{graphicx}	
\usepackage{amsmath}	
\usepackage{amssymb}	
\usepackage{txfonts}
\usepackage{amsfonts}
\usepackage{pdflscape}
\usepackage{hyperref}
\usepackage{bm}
\usepackage{pdfpages}

\title[Calibrating the metallicity of M dwarfs in wide physical binaries -- I]
{Calibrating the metallicity of M dwarfs in wide physical binaries with F-, G-, and K- primaries -- I: High-resolution spectroscopy with HERMES: stellar parameters, abundances, and kinematics\thanks{Based on observations obtained with the HERMES spectrograph mounted on the 1.2~m Mercator Telescope at the Spanish Observatorio del Roque de los Muchachos of the Instituto de Astrof{\'i}sica de Canarias}}

\author[D.~Montes, et al.]{D. Montes$^{1}$\thanks{E-mail: \url{dmontes@ucm.es}},
R.~Gonz{\'a}lez-Peinado$^{1}$,
H.~M. Tabernero$^{2,1}$,
J.~A. Caballero$^{3}$,
E.~Marfil$^{1}$,
\newauthor
F.~J. Alonso-Floriano$^{4,1}$,
M.~Cort\'es-Contreras$^{3}$,
J.~I. Gonz{\'a}lez Hern{\'a}ndez$^{5,6}$,
A.~Klutsch$^{7,1}$,
\newauthor
and C.~Moreno-J{\'o}dar$^{8,1}$
\\
$^{1}$Departamento de F\'{\i}sica de la Tierra y Astrof\'{\i}sica \& UPARCOS-UCM (Unidad de F\'{\i}sica de Part\'{\i}culas y del Cosmos de la UCM), Facultad de\\ Ciencias F\'{\i}sicas, Universidad Complutense de Madrid, E-28040 Madrid, Spain\\
$^{2}$Departamento de F\'{\i}sica, Ingenier\'{\i}a de Sistemas y Teor\'{\i}a de la Se\~nal, Universidad de Alicante, Apdo. 99 E-03080, Alicante, Spain\\
$^{3}$Centro de Astrobiolog\'ia (INTA--CSIC), ESAC campus, Camino Bajo del Castillo s/n, E-28691 Villanueva de la Ca\~nada, Madrid, Spain\\
$^{4}$Leiden Observatory, Leiden University, 2300 RA Leiden, Netherlands\\
$^{5}$Instituto de Astrof\'isica de Canarias (IAC), Calle V{\'i}a L\'actea s/n, E-38200 La Laguna, Tenerife, Spain\\ 
$^{6}$Departamento Astrof{\'i}sica, Universidad de La Laguna, E-38206 La Laguna, Tenerife, Spain\\
$^{7}$INAF - Osservatorio Astrofisico di Catania, via S. Sofia 78, 95123 Catania, Italy\\
$^{8}$Escuela T{\'e}cnica Superior de Ingenier{\'i}a Aeron{\'a}utica y del Espacio, Plaza de Cardenal Cisneros 3, E-28040, Madrid, Spain\\
}

\date{Accepted 2018 May 09. Received 2018 May 09; in original form 2018 March 02}

\pubyear{2018}

\begin{document}
\label{firstpage}
\pagerange{\pageref{firstpage}--\pageref{lastpage}}
\maketitle

\begin{abstract}
We investigated almost 500 stars distributed among 193 binary or multiple systems made of late-F, G-, or early-K primaries and late-K or M dwarf companion candidates.
For all of them, we compiled or measured coordinates, $J$-band magnitudes, spectral types, distances, and proper motions.
With these data, we established a sample of 192 physically bound systems.
In parallel, we carried out observations with HERMES/Mercator and obtained high-resolution spectra for the 192 primaries and five secondaries.
We used these spectra and the automatic {\sc StePar} code for deriving precise stellar atmospheric parameters: $T_{\rm eff}$, $\log{g}$, $\xi$, and chemical abundances for 13 atomic species, including [Fe/H].
After computing Galactocentric space velocities for all the primary stars, we performed a kinematic analysis and classified them in different Galactic populations and stellar kinematic groups of very different ages, which match our own metallicity determinations and isochronal age estimations.
In particular, we identified three systems in the halo and 33 systems in the young Local Association, Ursa Major and Castor moving groups, and IC~2391 and Hyades Superclusters.
We finally studied the exoplanet-metallicity relation in our 193 primaries and made a list 13 M-dwarf companions with very high metallicity that can be the targets of new dedicated exoplanet surveys.
All in all, our dataset will be of great help for future works on the accurate determination of metallicity of M dwarfs.

%
%
%
%
%
%
\end{abstract}




\section{Introduction}
\label{introduction}


Cool, low-mass dwarfs of M spectral type are, by far, the most numerous stellar constituents of the Milky Way.
Having main-sequence lifetimes that exceed the current age of the Universe \citep{Baraffe1998,Henry2006}, M dwarfs stand as excellent objects in order to probe the structure and evolution of the Milky Way's thin and thick discs. 
Because of their ubiquity, M dwarfs may also be the largest population of planet-hosting stars.
As a result, a large fraction of low-mass planets are expected to orbit an M-type star within its habitable zone, which is considerably closer than for solar-like ones.

 \begin{figure*}
	\includegraphics[width=\columnwidth]{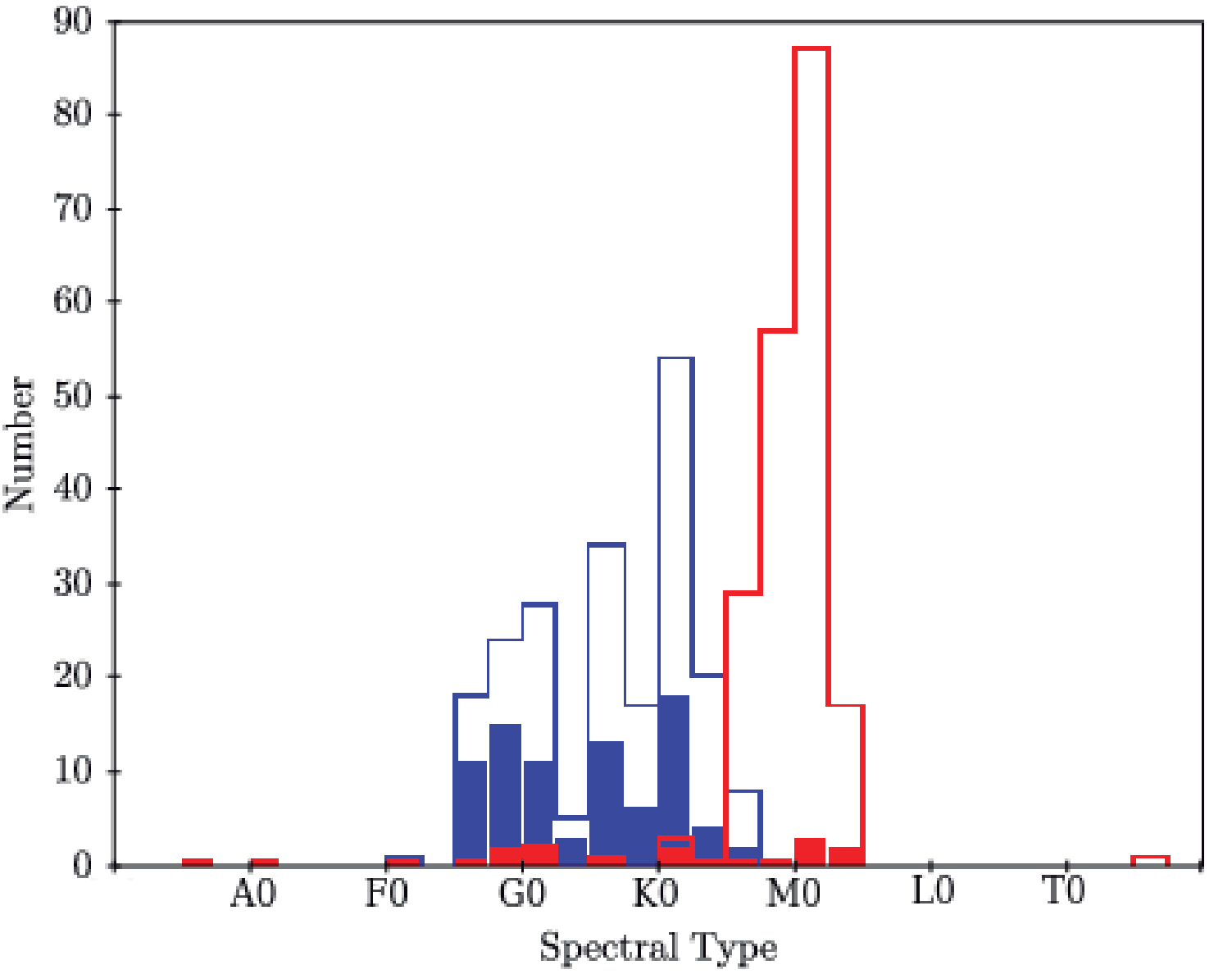}
	\includegraphics[width=\columnwidth]{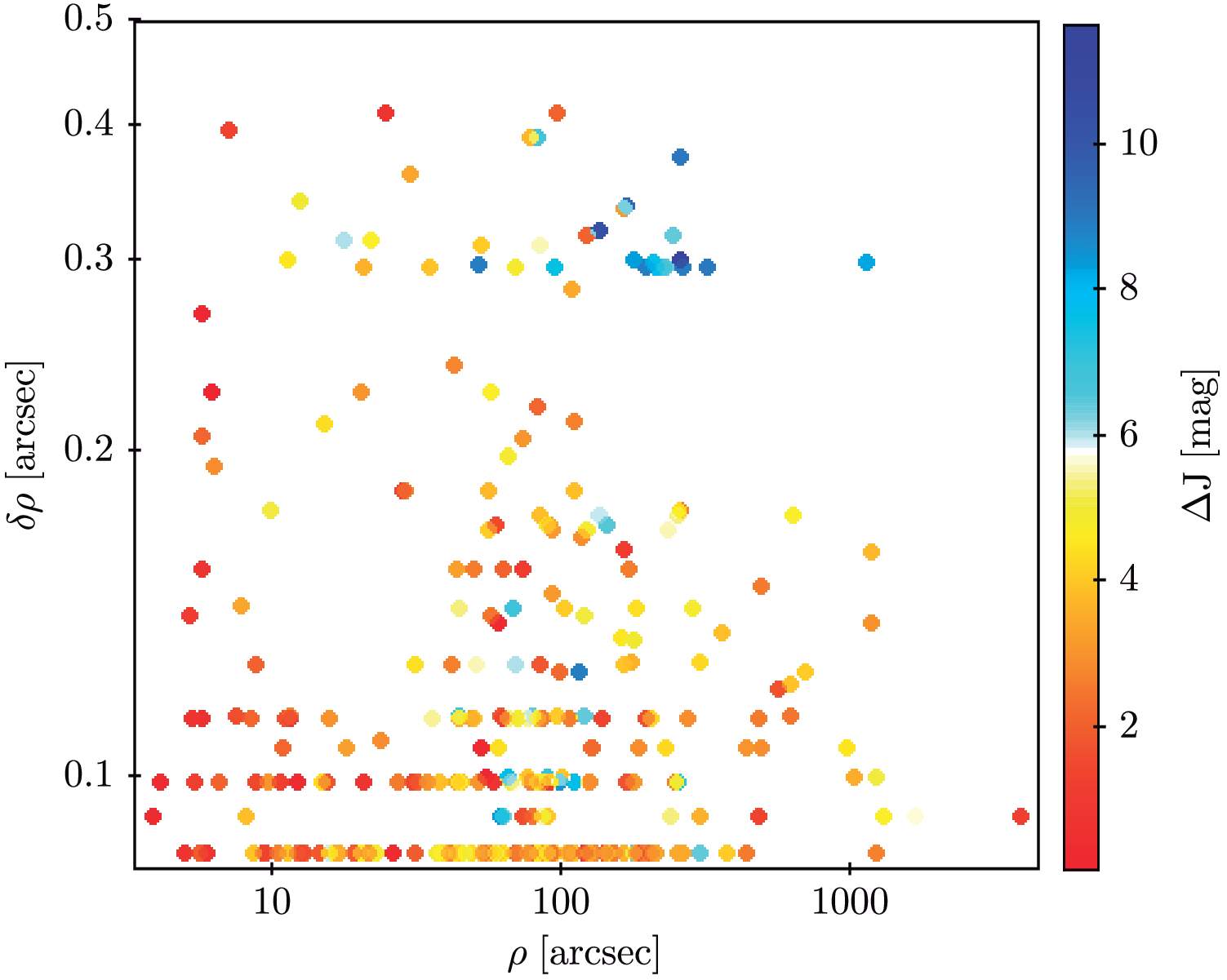}
    \caption{{\em Left panel:} distribution of spectral types for the stars of our sample. 
    Blue and red bars represent primaries and secondary candidates, respectively, while open and filled bars represent physical and optical components, respectively.
    Note the tail of optical secondaries in the background with spectral types much earlier than primaries.
    The late T dwarf is GJ~570\,D \citep{Burgasser2000}. 
    {\em Right panel:} angular separation between stars in pairs and their uncertainties, colour-coded with the difference in $J$ magnitude.}
    \label{fig:rhoerho_spt}
\end{figure*}

More importantly, the detectability of any such planet via the transit and radial-velocity techniques is enhanced by the lower masses and smaller radii of M dwarfs \citep{Clanton2014,Reiners2018}.
Therefore, M dwarfs have become key targets for planet hunting \citep[e.g. MEarth --][]{Charbonneau2009, Berta-Thompson2015, Dittmann2017}.
This is also illsutrated by new spectrographs optimised for exoplanet searches around M dwarfs \citep[e.g. CARMENES --][]{Alonso-Floriano2015, Quirrenbach2016, Reiners2018}.

The observational efficiency of exoplanet searches around M dwarfs could be vastly increased with prior knowledge of stellar metallicity.
In this sense, previous studies have already pointed out that planets are more likely to be found orbiting metal-rich, solar-like stars 
(\citealt{Santos2001,Santos2004, FischerValenti2005} -- but see below).
However, the metallicity of low-mass dwarfs has been an elusive fundamental property due to the complexity of modeling their atmospheres.
Fortunately, the advent of new observational techniques, as well as independent theoretical improvements in atmospheric models, now seem to link the metallicity of M dwarfs to both their photospheric and spectroscopic features 
\citep{Bonfils205, Bean2006b,  WoolfWallerstein2006, JohnsonApps2009, HauschildtBaron2010, Rojas-Ayala2010, Rojas-Ayala2012, Onehag2012, Neves2014, Maldonado2015, Passegger2018}.
Not only do these metallicity studies have deep implications in the realm of stellar astrophysics, but they also play a crucial role in the analysis of the Galactic evolution \citep{West2011,WoolfWest2012}.
 
There were preliminary indications that the M dwarfs with known planets have sub-solar metallicities 
\citep{Bonfils205, Bean2006b}, 
in contrast to their earlier counterparts.
Actually, while giant planets preferentially form around metal-rich stars, Neptunes and super-Earths are not necessarily more abundant in metal-rich stars but they are abundant at solar metallicity \citep{Sousa2008,Adibekyan2012,Buchhave2012}. 
However, more recent results showed instead that planet-hosting M dwarfs appear to be metal-rich \citep{JohnsonApps2009, Rojas-Ayala2010, Terrien2012}.
We refer the reader to \cite{Hobson2018} for a recent review on the planet-metallicty relation in M dwarfs.

A few studies have estimated M-dwarf metallicities using wide multiple systems that consist of at least an M dwarf and a higher-mass star, typically of late F-, G-, or early-K spectral type. 
Since binaries are assumed to be born in a common parental cloud and be coeval, the composition of the FGK star, which can be accurately derived from a careful comparison with theoretical models and current tools, can be extrapolated to its companion M dwarf. 
However,  in some cases small differences in composition between components (often at a level of $\approx$0.05 dex) may arise if they are comoving but not coeval, there originally was chemical heterogeneity within the birth cloud, or some of the components underwent accretion of planetary material after birth 
\citep[see][and references therein]{Desidera2004, Teske2015, Brewer2016, Andrews2018, Oh2018ApJ...854..138O}.
Some of these studies have used optical and infrared spectroscopy to tie spectroscopic features to a metallicity scale 
\citep{Valenti1998,WoolfWallerstein2005,WoolfWallerstein2006,Bean2006a,Bean2006b,Woolf2009,Rojas-Ayala2010, Rojas-Ayala2012,Terrien2012,Mann2013,Mann2014,Mann2015,GaidosMann2014,Newton2014,Souto2017}. 
Other studies have used photometric calibrations.
For example, \cite{Bonfils205} and \cite{JohnsonApps2009} used M dwarfs in wide binaries to derive a relation between metallicity, absolute $K$-band magnitude, and the $V-K$ colour index
\citep[higher metallicity M dwarfs are slightly brighter at a given colour -- see also][]{Casagrande2008,SchlaufmanLaughlin2010,Johnson2012,Neves2012}.

To date, different authors with different methods have analysed only slightly over one hundred wide FGK+M benchmark systems, which results in a lack of homogeneity in the literature.
A larger and homogeneous sample of wide visual binaries and multiple systems covering a large range in metallicity and spectral type is needed to reduce the scatter of the current calibrations and to get a good calibration relationship that would be valid throughout the parameter space.
Here we start a series of papers devoted to improve the spectroscopic calibration of the M-dwarf metallicity.
In this first article, we present our sample with a total of nearly 500 stars, study the common proper motion of the multiple systems, and derive stellar atmospheric parameters of the FGK ``primaries'' ($T_{\rm eff}$, $\log{g}$, $\xi$, and chemical abundances for 13 atomic species). 


\section{Analysis}
\label{analysis}

\begin{figure*}
	\includegraphics[width=\columnwidth]{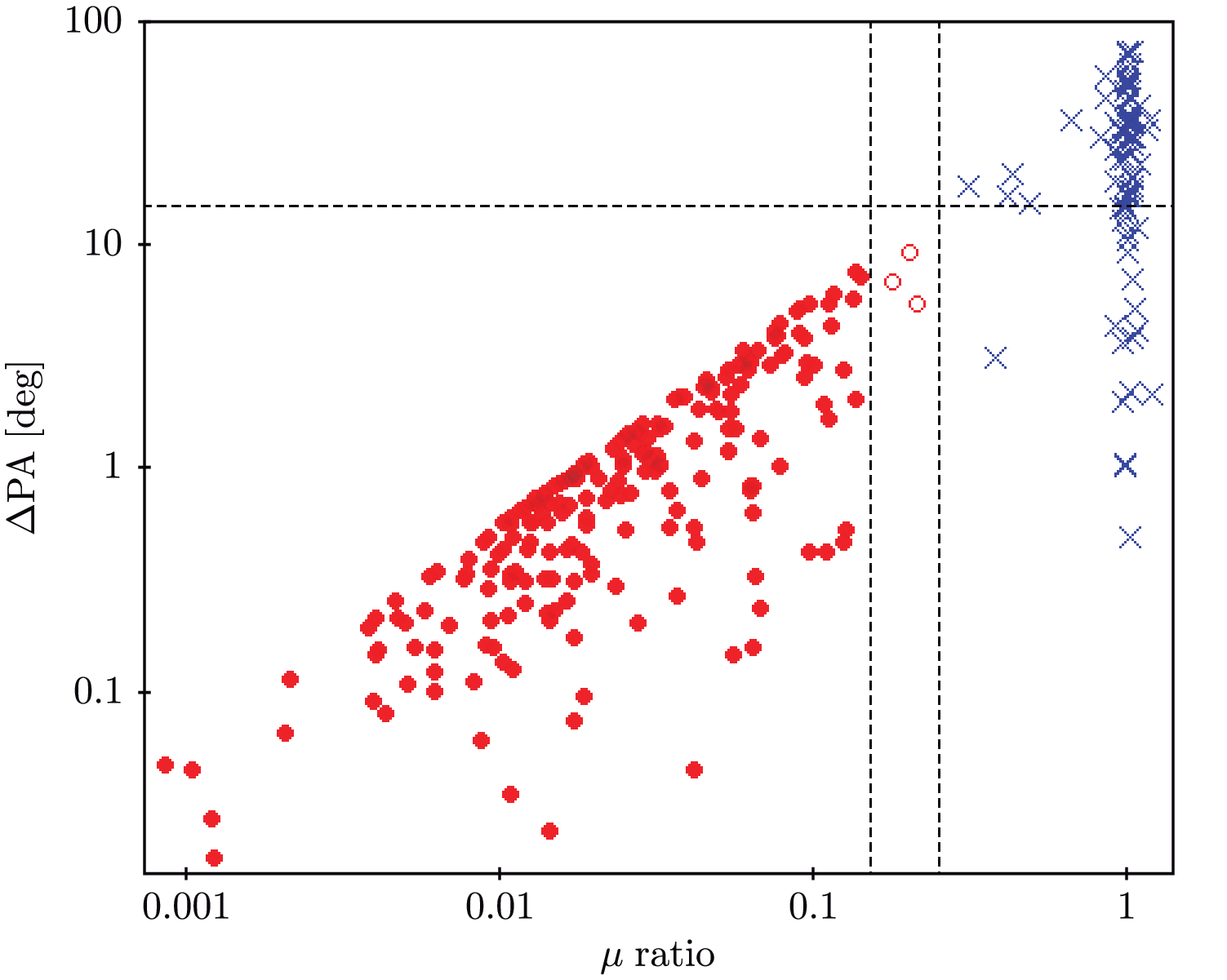}
	\includegraphics[width=\columnwidth]{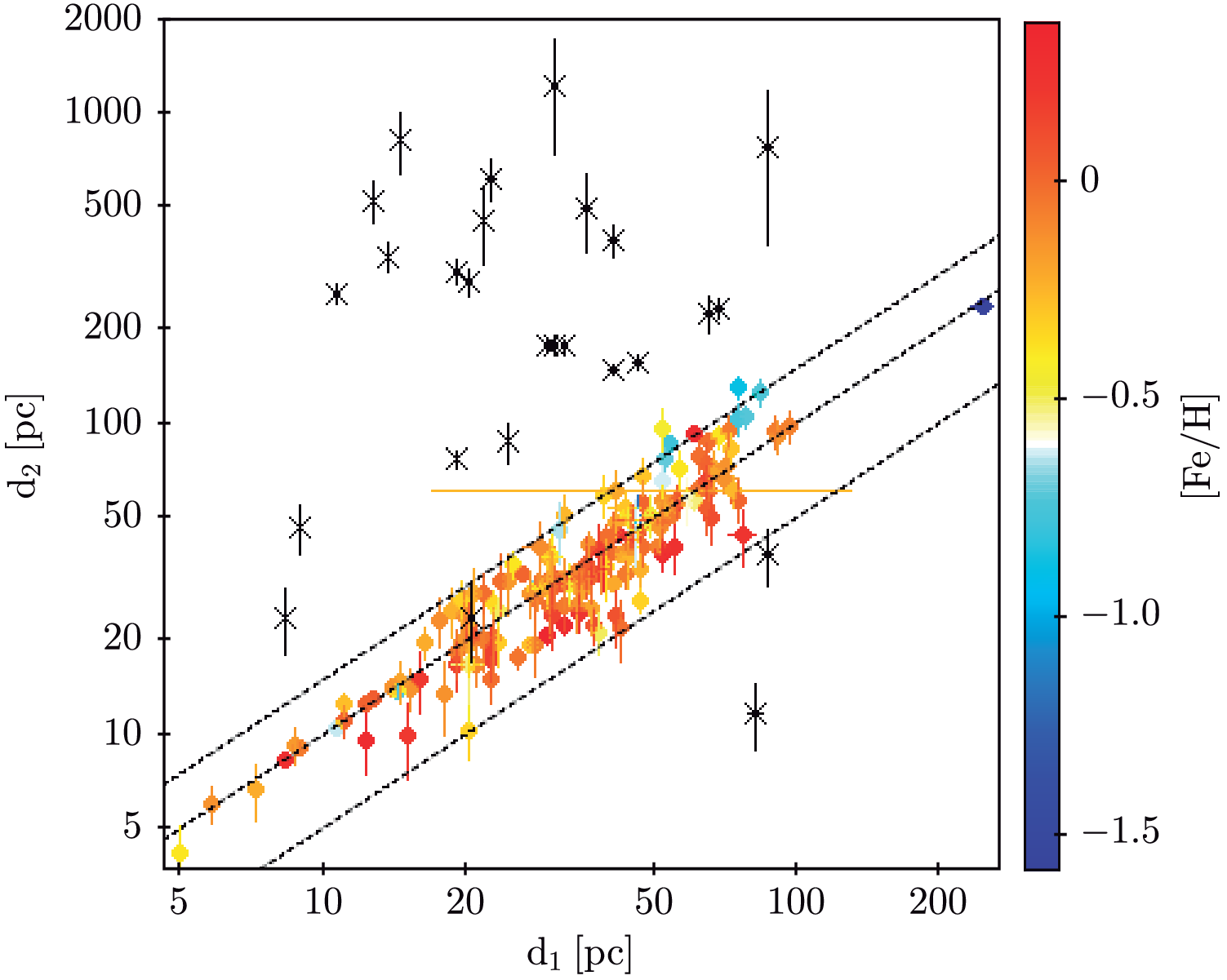}
    \caption{{\em Left panel:} $\Delta {\rm PA}$ vs. $\mu~{\rm ratio}$ diagram.
    Physical (red filled circles), doubtful physical (red open circles) and optical (blue crosses) attending to our criteria. 
    Dashed vertical and horizontal lines mark the 0.15 and 0.25 $\mu$ ratio and 15 deg $\Delta$PA.
    {\em Right panel:} heliocentric distances for primary (1) and companion (2) stars colour-coded with metallicity. Dashed lines indicate 1.5:1, 1:1, and 0.5:1 $d$ relationships, respectively. Black crosses represent optical pairs. 
    Low-metallicity stars tend to lie in the upper part of the 1:1 distance relation.}
    \label{fig:pm_distances}
\end{figure*}

First of all, we collected from the literature {193} binary or multiple system candidates formed by late-F, G-, or early K-type primaries and late-K or M-type secondaries observable from Calar Alto, in Southern Spain ($\delta$\,$\textgreater$\,--23 deg). 
The main sources used to gather our initial sample were searches for common proper motion companions \citep{GlieseJahreiss1991,Poveda1994,Poveda2009,Simons1996,Tokovinin1997,GouldChaname2004,Zapatero2004,Caballero2007,Caballero2009,LepineBongiorno2007,Raghavan2010}, as well as previous metallicity calibrations of M dwarfs based on photometric and/or spectroscopic data (see Section~\ref{introduction}).
The sample consists on 489 stars distributed in 193 binary or multiple candidate systems, from which 193 are late-F, G-, or early K-type primaries, and 296 are companion candidates.

Table \ref{tab:investigated_stars} lists the surveyed systems studied in this paper.
For each of the {296} pairs of primaries and companion candidates, we tabulate its number and discoverer code as provided by the Washington Double Star Catalog \citep[WDS --][]{Mason2001}.
To avoid including too many spurious sources in the analysis, we tabulate all components with designation A to D regardless of their WDS notes (such as ``Proper motion or other technique indicates that this pair is non-physical''), and all the physical pair candidates regardless of their designation (e.g. {GJ~570\,D} is component G in WDS, HD~211472\,B is component T in WDS).
We were not able to identify faint optical companions found in deep adaptive optics surveys (e.g. \citealt{Lafreniere2007,Ehrenheich2010,Janson2013,Ammler2016}) and the LDS 585 ``D'' companion of the system WDS~17050--0504 (according to WDS, LDS 585 ``D'' is a dubious double\footnote{``A dubious double (or Bogus binary) may represent a positional typo in the original publication [...], an optical double dissappearing due to radically different proper motions, a plate flaw, or simply a pair not at a magnitude, separation, etc., sufficiently similar to those noted when the first measure was added'' \citep{Mason2001}.}).
{Three} pairs have no WDS entry, and are marked with ``...'' in the `Discoverer code' field.

In Table \ref{tab:investigated_stars} we also provide angular separation $\rho$ and position angle $\theta$ measured by us with the Virtual Observatory tool TOPCAT \citep{Taylor2005} from Two-Micron All Sky Survey \citep[2MASS --][]{Skrutskie2006} data, Simbad's star name, equatorial coordinates, $J$-band magnitude, and spectral type from the literature.
Fig.~\ref{fig:rhoerho_spt} shows the distribution of spectral types of primaries and companions.
Most spectral types of primaries range from F4\,V to K5\,V, and of physical companions from K7\,V to M7\,V, while angular separations range from 4 to about 4000\,arcsec, with uncertainties lower than 0.4\,arcsec.

The close companion candidates in four systems with very bright primaries, namely WDS~04359+1631 (Aldebaran B), WDS~16147+3352 ($\sigma$~CrB ``C''), WDS~19553+0624 ($\beta$~Aql B), and WDS~20462+3358 ($\epsilon$~Cyg B and C), were not tabulated by 2MASS in spite of being visible in their images.
Besides, the 88-arcsec wide companion candidate BD--13~5608B in system WDS~20124-1237 was not tabulated by 2MASS due to a nearby speckle from the primary ($\xi$~Cap).
In these {five} cases, we computed $\rho$ and $\theta$ with the raw 2MASS $H$-band images and Aladin Sky Atlas \citep{Bonnarel2000}.

In Table~\ref{tab:pms} we list heliocentric distances and proper motions of all the investigated stars, which we used for discarding optical (non-physical) pairs.
First, we compiled parallactic distances in the following order from the Tycho-{\em Gaia} Astrometric Solution \citep[TGAS --][]{GaiaColaboration2016}, the new \citep[HIP2 --][]{vanLeeuwen2007} and old \citep[HIP1 --][]{Perryman1997} {\em Hipparcos} reductions, \cite{vanAltena1995}, and \citeauthor{Prieur2014} (\citeyear{Prieur2014}; only for $\mu^{02}$~Her\,BC).
All {193} primaries have parallactic distances, while only {52} companions do. 
Of the remaining {244} companion candidates, we derived our own spectro-photometric distances for {165} late K, and early and intermediate M dwarfs resolved by 2MASS, using the spectral type--$M_J$ relation of \cite{Miriam2017}. 
As discussed in Section~\ref{metallicity}, this relationship is applicable only to main-sequence late-type dwarfs of solar metallicity, and the tabulated spectro-photometric distances of low-metallicity dwarfs must be handled with care.
For {two} secondaries with $J$ magnitude and reliable spectral type ($\eta$~Cas\,B in WDS~00491+5749, and BD+48~3952B in WDS~23104+4901) we did not derive any distance because their 2MASS quality flags indicate a poor photometry.
Besides, there are two physical companions, a white dwarf and a brown dwarf, with both spectral type and near-infrared magnitudes without a distance derived by us, namely $o^{02}$~Eri\,B (DA2.3) in WDS~J04153-0739, and GJ~570\,D (T8) in WDS~14575-2125.
Atogether, there are only {74} companion candidates without any heliocentric distance determination. 

Next, we compiled proper motions for the {193} primaries and {293} (all but three) companions from the following catalogues and works: 
TGAS,
Hot Stuff for One Year \citep[HSOY --][]{Altmann2017},
HIP2,
UCAC5 \citep{Zacharias2017},
Tycho-2 \citep{Hogg2000},
PPMXL \citep{Roeser2010},
UCAC4 \citep{Zacharias2012},
\cite{Caballero2009},
\citet[][for GJ~570\,D]{Faherty2009},
and \citet[][for Aldebaran~B]{Ivanov2008}, in this order.
For {37} stars ({36} secondaries and the primary 39~Leo\,A in WDS~10172+2306) with probably wrong proper motions or  no proper motions whatsoever, we improved or measured their values for the first time.
To do so, we used the method used by \cite{Caballero2009} and the astrometric epochs from 
DENIS \citep{Epchtein1997},
USNO-A2 \citep{Monet1998}, 
2MASS, GSC2.3 \citep{Lasker2008}, 
AllWISE \citep{Cutri2014}, 
CMC15 \citep{MuinosEvans2014},  
{\em Gaia} DR1 \citep{GaiaColaboration2016}, and, in the most difficult cases, the SuperCOSMOS digitalization of the Digital Sky Survey photographic plates \citep{Hambly2001}. 
The time baseline varied between 4.5 and 119.3 years, with a median of seven astrometric epochs per star.
As for the distances, we did not assign proper motions of primaries to companions.
We were not able to compile or measure by ourselves any proper motions of the secondaries in the systems  WDS 00491+5749 ($\eta$ Cas AB; first measured in 1779), WDS~11378+4150 (BD+42 2230 AC; first detected in 1998) and WDS J21546--0318 (HD 208177 AB; first observed in 1829).

With the distances and proper motions in Table~\ref{tab:pms}, we set a uniform criterion to distinguish between physical (bound) and optical (unbound) systems (Fig.~\ref{fig:pm_distances}, left panel).
First, we computed two parameters for each pair of stars: the $\mu~{\rm ratio}$, defined as:
\begin{equation}
\left(\mu\,\rm ratio\right)^2\,$=$\,\dfrac{\left(\mu_\alpha \cos{\delta}\,_{1}-\mu_\alpha \cos{\delta}\,_{2}\right)^{2}\, \text{+} \,\left(\mu_{\delta\,1}-\mu_{\delta\,2}\right)^{2}}{\left(\mu_\alpha \cos{\delta}\,_{1}\right)^{2} \text{+} \,\left(\mu_{\delta\, 1}\right)^{2}},
\end{equation}

\noindent and the proper motion position angle difference:
\begin{equation}
\Delta {PA}\,$=$\,\left|{PA_{1} - PA_{2}}\right|,
\end{equation}

\noindent where $PA_i$ is the angle between $\mu_\alpha \cos{\delta}_i$ and $\mu_{\delta,i}$, being $i$\,=\,1 for the primary star and $i$\,=\,2 for the companion candidate.

We discarded {84} pairs of stars that have:
($i$) $\mu~{\rm ratio} >$ 0.15, and/or
($ii$) proper motion position angle difference $\Delta {\rm PA}>$ 15\,deg (compare with the selection criteria in e.g. \citealt{LepineBongiorno2007}, \citealt{Dhital2010}, and \citealt{AF2015}).  
Besides, we investigated in detail the three pairs with $\Delta {\rm PA}<$ 15\,deg and 0.15 $< \mu~{\rm ratio} <$ 0.25.
Two of them, namely WDS~15282-0921~AC and WDS~23026+2948 AC, are very wide pairs ($\rho$ > 1000 arcsec) that are affected by high proper motion projection effect (as between $\alpha$~Cen AB and Proxima).
The third pair, WDS~23536+1207~AB (VYS 11), is a close binary of $\rho$ = 5.7 arcsec already investigated by \cite{TokovininKiyaeva2015}. 
We also classified these three systems as physical despite they did not pass our $\mu$ ratio criterion.
We must wait for $Gaia$ DR2 to confirm them.
Overall, we have {209} physical pairs distributed in {192} systems.
We only discarded the source WDS~10585-1046 (LDS 4041).

As a double check, we compared the compiled and derived heliocentric distances of primaries and companions  (Fig.~\ref{fig:pm_distances}, right panel).
For systems with parallactic distances only, they vary less than 15\,\%, while for systems with spectro-photometric distances, they vary less than 50\,\%, except for three pairs with low metallicities (Section~\ref{metallicity}).
To assure that we did not reject any physical pair because of abnormal metallicity, we did not discard any pair based on different heliocentric distances.
 New parallax-based distances, such as the ones provided by {\em Gaia} DR2 \citep{Gaia-DR2_col_Brown2018}, are invaluable since they are independent of metallicity and stellar parameter analyses.

\begin{figure}
	\includegraphics[width=\columnwidth]{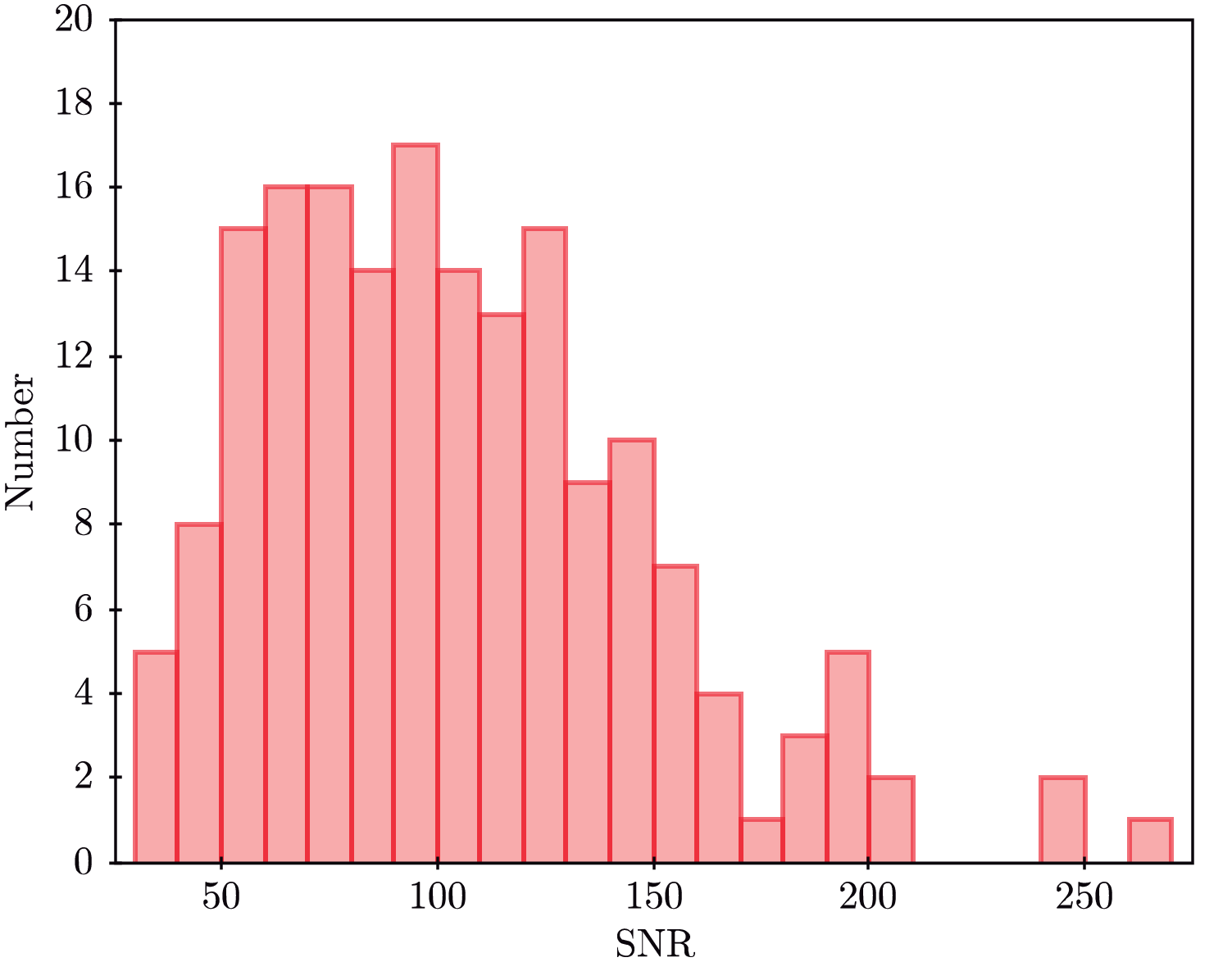}
    \caption{Histogram of signal-to-noise ratios measured in our HERMES spectra.}
    \label{fig:snrhist}
\end{figure}


 
\section{Spectroscopy and kinematics}

\subsection{Observations and reduction}
\label{obsred}

\begin{figure*}
	\includegraphics{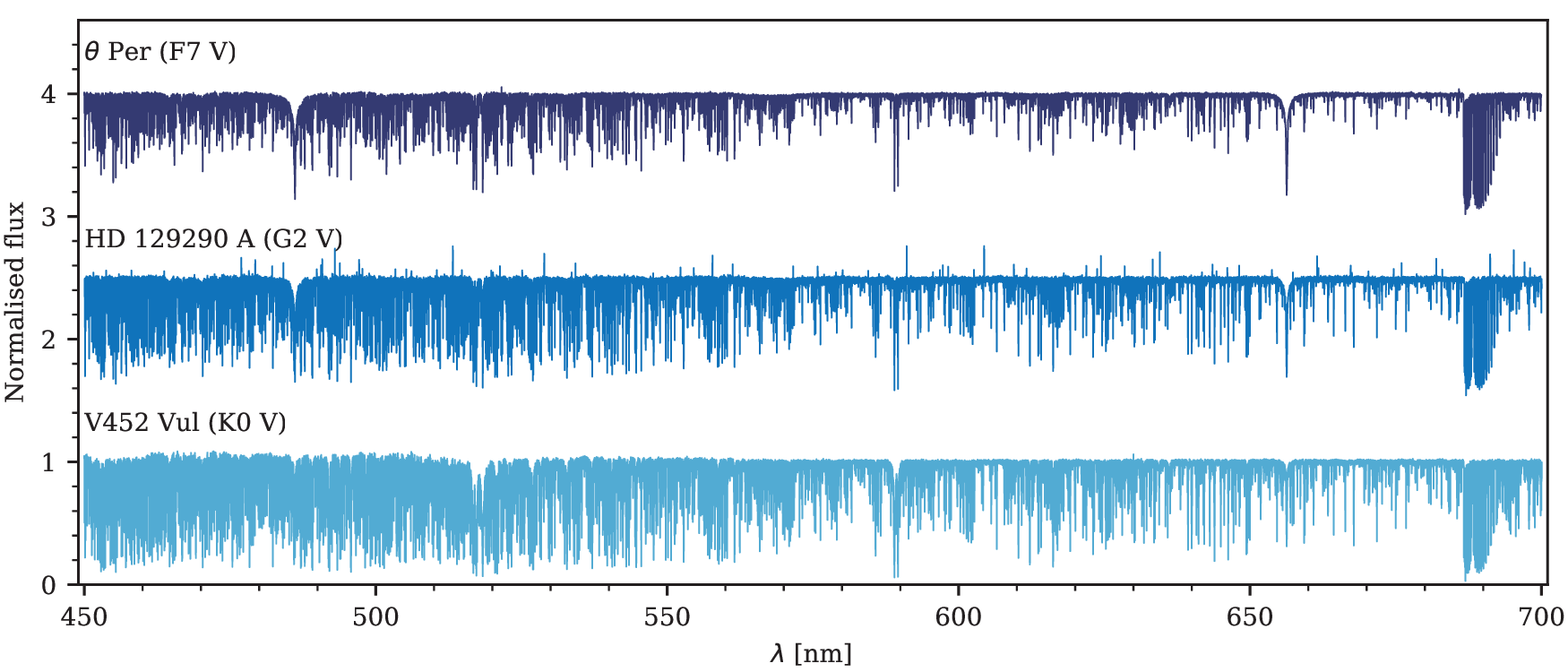}
	\includegraphics{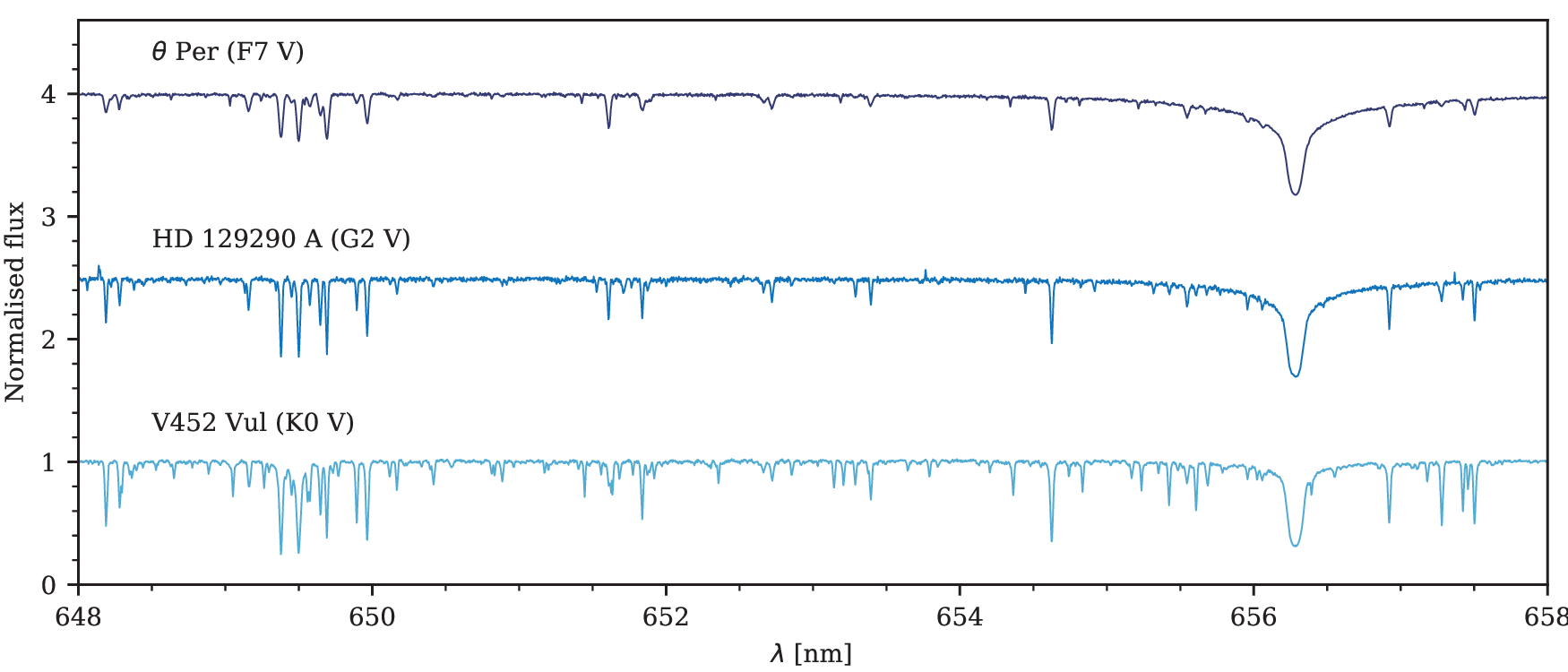}
    \caption{High-resolution spectra of three representative primaries from our sample (\textit{from top to botom}): $\theta$ Per, HD 129290 A, and V452 Vul (HD~189733).
    \textit{Top}: Full investigated wavelenght range.
     \textit{Bottom}: zoomed range, 10~nm wide, near H$\alpha$ $\lambda$~656.3~nm.}
    \label{fig:spectra}
\end{figure*}

FGK-type stars of the multiple systems described above are relatively bright, $J <$ 9.0\,mag ($V <$ 11.0\,mag), which allowed us to obtain high signal-to-noise ratio, high-resolution, optical spectra with reasonable exposure times ($t_{\rm exp} \le$ 20\,min), and to  derive reliable stellar parameters and abundances.
We took high-resolution {echelle} spectra of {192 primaries and 5 secondaries} with HERMES \citep[High Efficiency and Resolution Mercator Echelle Spectrograph -- ][]{Raskin2011} at the 1.2\,m Mercator Telescope at the Observatorio del Roque de los Muchachos (La Palma, Spain) between January 2010 and December 2017.
We used the high resolution mode, which provides with a spectral resolution of 86,000 in the approximate wavelength range from $\lambda$\,380\,nm to $\lambda$\,875\,nm.
Most of the spectra have a signal-to-noise ratio (SNR) between 60 and 140 in the $V$ band, as shown in the third column of Table~\ref{tab:parameters} and Fig.~\ref{fig:snrhist}. 
Additionally, we took several spectra of the asteroid Vesta with the same spectrograph configuration.
All the obtained spectra were reduced with the automatic pipeline for HERMES \citep{Raskin2011}.
Next, we used several standard tasks within the IRAF environment for normalising the spectra, using a low-order polynomial fit to the observed continuum, and for applying the corresponding Doppler correction.
To do so, we computed the observed radial velocity ($V_{r}$), which is the sum of the spectrum relative velocity (measured with the IRAF function {\tt fxcor}) and barycentric correction (obtained from the FITS header).
When several exposures were available for the same star, we combined all the individual spectra and obtained a unique spectrum with higher SNR.
For our analysis we used only the wavelength range from 450~nm to 700~nm (Fig.~\ref{fig:spectra}).
The {197} stars observed with HERMES are marked with ``H'' in the last column of Table \ref{tab:investigated_stars}.

We also observed many M-dwarf companions with the low-resolution optical spectrograph CAFOS at the 2.2\,m Calar Alto telescope.
They are marked with ``C'' \citep{Alonso-Floriano2015} and ``C*'' (unpublished) in the last column of Table~\ref{tab:investigated_stars}. 
We are using these spectra for calibrating spectral indices and abundance determinations with features analysed at high spectral resolution, and will appear in forthcoming publications.
In particular, seven of our M-dwarf companions (namely BX~Cet, $o^{02}$~Eri~C, HD~233153, BD--02~2198, $\rho^{01}$~Cnc~B, $\theta$~Boo~B, and HD~154363~B) have also been observed with the CARMENES spectrograph with very high SNR and spectral resolution \citep{Quirrenbach2016,Reiners2018}.

\subsection{Stellar parameters}
\label{stellarparameters}

\begin{figure}
	\includegraphics[width=\columnwidth]{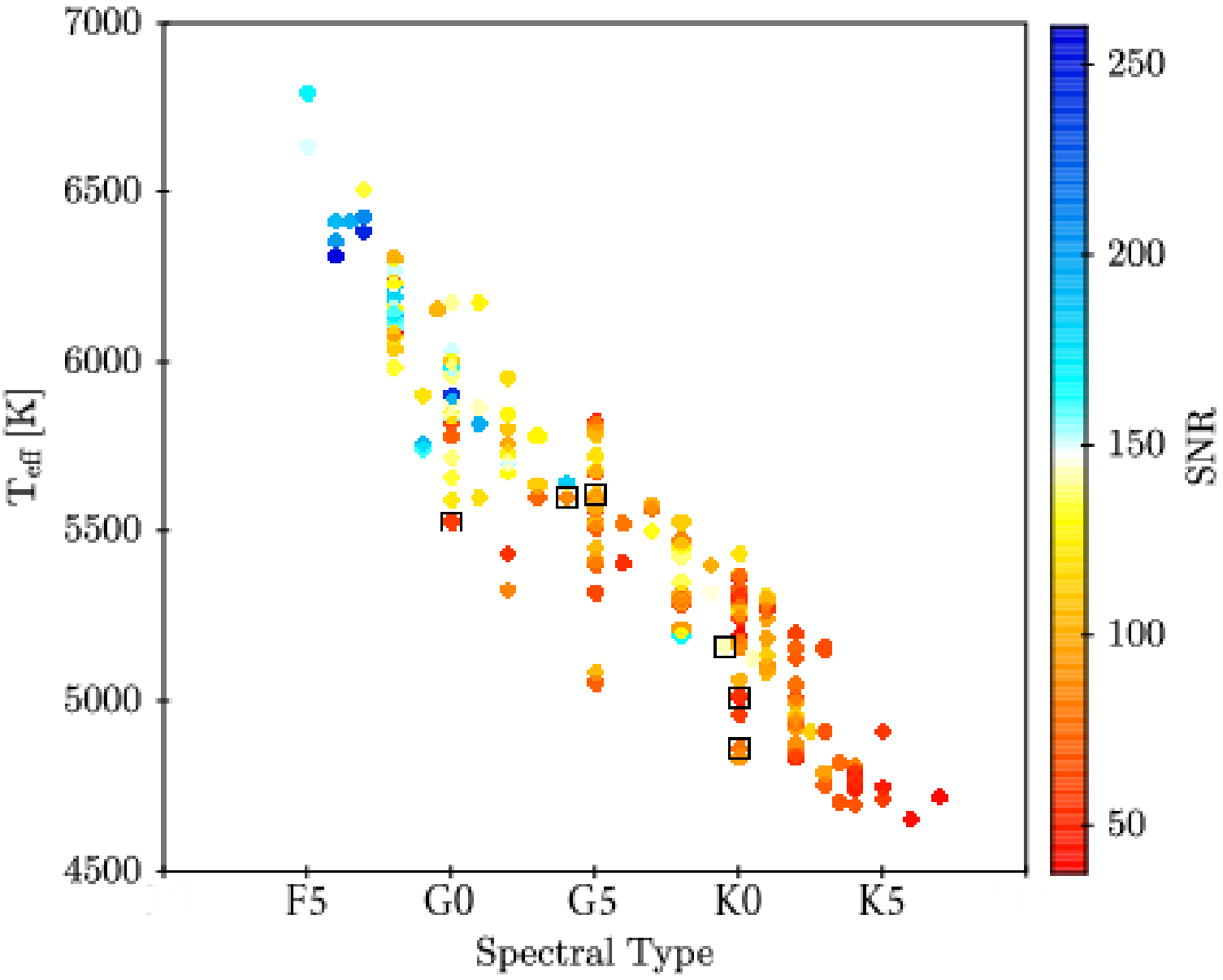}
	\includegraphics[width=\columnwidth]{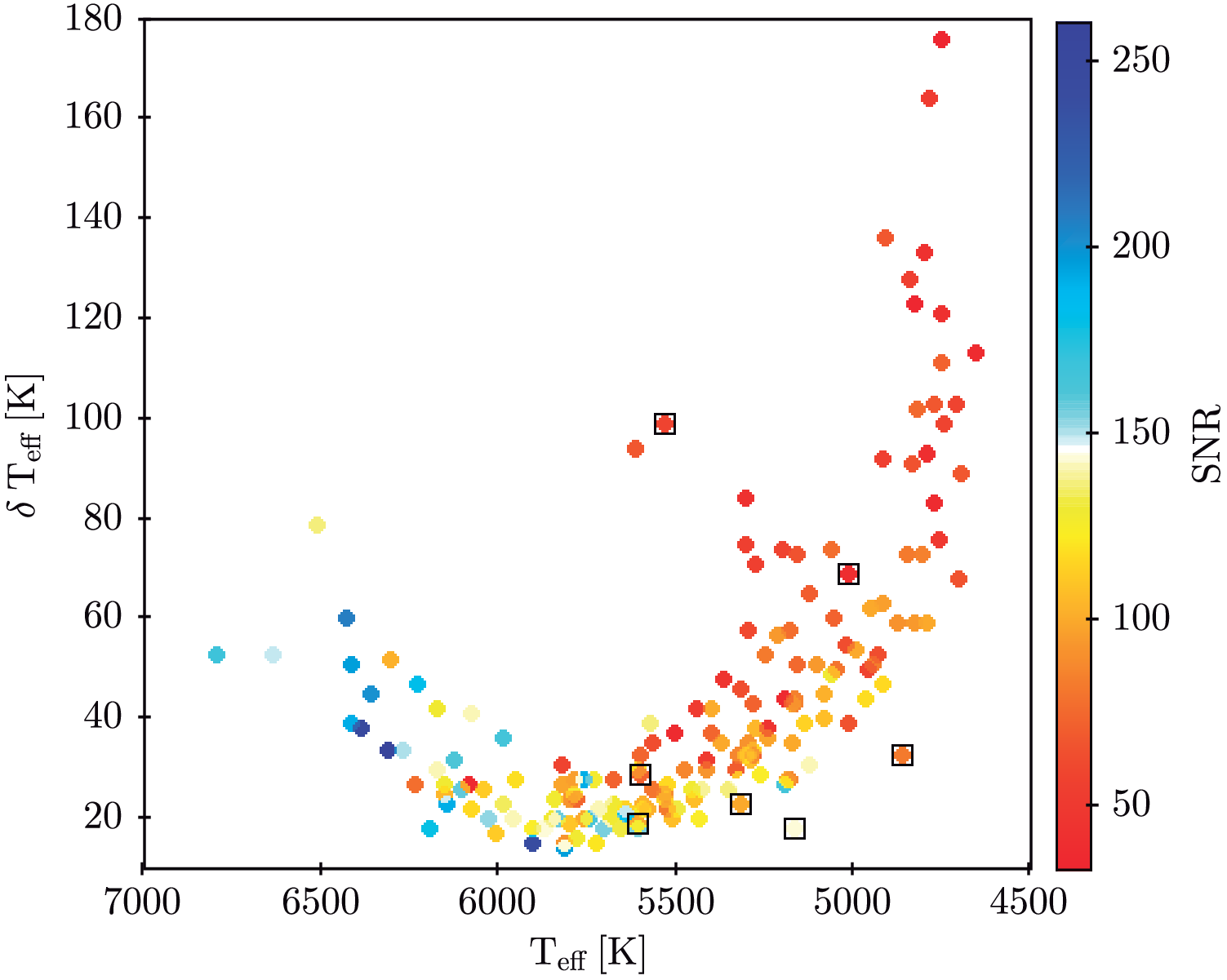}
	\includegraphics[width=\columnwidth]{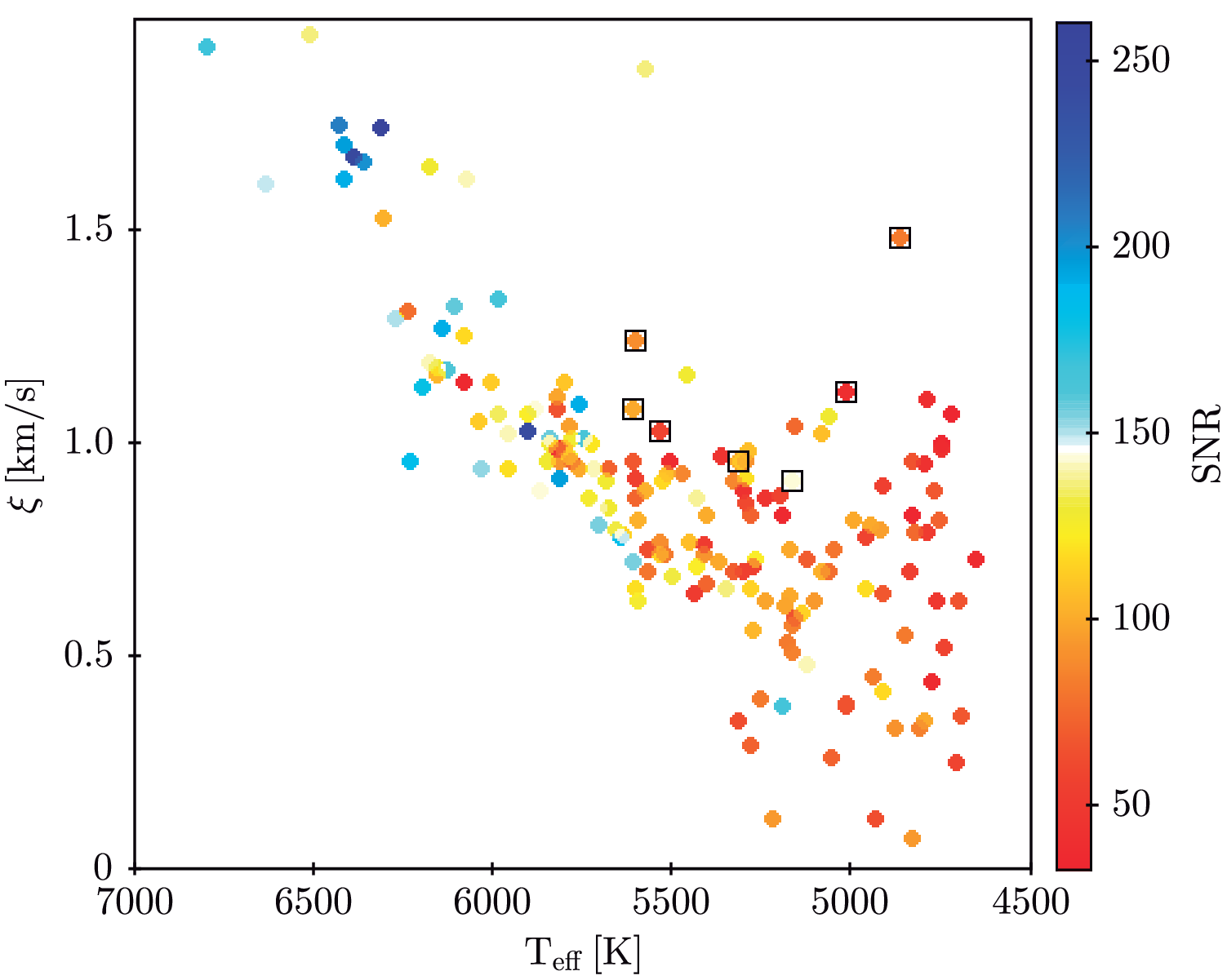}
    \caption{\textit{Upper panel:} Effective temperature as a function of spectral type.
    \textit{Middle panel:} Error in effective temperature as a function of the effective temperature. \textit{Down panel:} Microturbulence velocity ($\xi$) against effective temperature. All the symbols are colour-coded with SNR. Black-ensquared stars represent low-gravity stars (Section~\ref{lowgrav}).}
    \label{fig:teff}
\end{figure}

\begin{table}
\caption{Solar parameters and element abundances.}
\label{tab:sun_abu}
\centering
\begin{tabular}{lc} 
\hline
\hline
 Parameter &     \\
\hline
$T_{\rm eff}$  [K] &  5777$\pm$18  \\ 
$\log{g}$   &  4.41$\pm$0.05   \\ 
$\xi$ [km\,s$^{-1}$] & 0.91$\pm$0.03\\
\hline
Element & $\log{\epsilon}$ (X) \\ 
\hline
Fe & 7.48$\pm$0.01 \\
Na & 6.44$\pm$0.03 \\
Mg & 7.69$\pm$0.05 \\
Al  & 6.51$\pm$0.01 \\
 Si &  7.59$\pm$0.07 \\
 Ca &  6.44$\pm$0.05 \\
Sc  & 3.15$\pm$0.03  \\ 
Ti & 5.02$\pm$0.05 \\
V & 4.03$\pm$0.03 \\
Cr &  5.70$\pm$0.05 \\
Mn & 5.51$\pm$0.03  \\
Co &  4.95$\pm$0.03 \\
Ni & 6.30$\pm$0.07 \\
\hline
 \end{tabular}
\end{table}

Stellar atmospheric parameters (effective temperature $T_{\rm eff}$, surface gravity $\log{g}$, microturbulence velocity $\xi$, and iron abundance [Fe/H], Section \ref{metallicity}) were computed using the automatic {\scshape StePar} code \citep*{Tabernero2012}, which relies on the equivalent width ($EW$) method. 
We employed the 2014 version of the MOOG code \citep{Sneden1973} and a grid of Kurucz ATLAS9 plane-parallel model atmospheres \citep{Kurutz1993}. 
As the damping prescription, we used the Uns\"{o}ld approximation multiplied by a factor recommended by the Blackwell group (``option 2'' within MOOG). 
We employed the line list of $\sim$300 solar-calibrated $\ion{Fe}{i}$ and $\ion{Fe}{ii}$ lines from \citet{Sousa2008}. 
We measured their $EW$s using ARESv2 \citep{Sousa2015}. 
ARES input parameters were set to those recommended in its manual\footnote{\tt \url{https://github.com/sousasag/ARES}, \url{http://www.astro.up.pt/~sousasag/ares/}}. 
The {\scshape StePar} code iterates within the parameter space until the slopes of $\chi$ vs. $\log{\epsilon}$(Fe {\sc i}) and
$\log{(EW / \lambda)}$ vs. $\log{\epsilon}$(Fe {\sc i}) are zero (i.e. the iron atoms are in excitation equilibrium). 
In addition, it imposes the ionisation equilibrium, such that $\log{\epsilon}$(Fe {\sc i})=$\log{\epsilon}$(Fe {\sc ii}). 
We also imposed that the [Fe/H] average of the MOOG output is equal to the iron abundance of the atmospheric model.

Table \ref{tab:parameters} shows the stellar atmospheric parameters of {198} F-, G-, and K- stars in our sample {(193 primaries and 5 secondaries)}.
The {\sc StePar} code is based on an $EW$ method that is meant to work for a limited range of $T_{\rm eff}$.
We were not able to determine with {\scshape StePar} the stellar atmospheric parameters of {21} stars:
\begin{itemize}

\item {\em Hot}. 
Stars with spectral types earlier than F6 ($T_{\rm eff} \approx$ 6700\,K) do not have enough iron lines for our analysis.
The triple system 9~Aur\,Aa,Ab,B comprises three stars of spectral types F2\,V and early M, and the effective temperature reported in the literature is $\sim$7000\,K \citep{AllendePrieto1999,LeBorgne2003}. 
The star HD 27887\,A, with an F5\,V spectral type and $T_{\rm eff} \approx$ 6500\,K \citep{AllendePrieto1999,Katz2011} is at the boundary of our grid, and {\scshape StePar} did not converge either.

\item {\em Cool}. 
Stars with spectral types later than K4 ($T_{\rm eff} \approx$ 4500\,K), on the contrary to hot stars, have too many overlapping iron lines.
We were not able to derive parameters for Aldebaran (K5\,III, 3900\,K; \citealt{Soubiran1998,Prugniel2011}) and SZ~Crt (K7\,V, 4200\,K; \citealt{Wright2011,Luck2017}).

\item {\em Fast}. 
At high rotational velocities, iron lines become so broad that overlap, too.
Six stars rotate too fast for {\scshape StePar}, i.e. have $v \sin{i} \gtrsim$ 10 km\,s$^{-1}$.
Published $v \sin{i}$ values for the six of them range from 16.2\,km\,s$^{-1}$ for V368~Cep \citep{Mishenina2012} to 84.8\,km\,s$^{-1}$ for $\eta$~UMi~A \citep{Schroeder2009}.

\item {\em SB2}. 
We discarded double-line spectroscopic binaries with blended or partially blended lines.
We found double peaks in spectral lines of {eight} primary stars; as discussed in Section~\ref{sbs}, {four} are reported here for the first time.
We were able to derive stellar atmospheric parameters for HD~200077\,Aa1, the primary of a known SB2 (see Section~\ref{sbs}).
There is a ninth SB2 in our sample, namely $\sigma$ CrB Aa,Ab \citep{Bakos1984}.

\item {\em No obs}. 
We could not observe only one primary, the SB2  $\sigma$~CrB\,Aa,Ab.

\end{itemize}

To sum up, we derived reliable spectroscopic stellar parameters for {175} primaries and {5} companions. 
Only $\sigma$~CrB\,Aa,Ab (the ``{193rd}'' primary star) lacks our homogeneous spectroscopy.
Stellar parameters derived with {\scshape StePar} are given in Table \ref{tab:parameters}, together with [Fe/H] from the literature, when available.
Fig.~\ref{fig:teff} shows the effective temperature and other parameters derived by us. 
 In addition, as can be seen in the top part of Table~\ref{tab:sun_abu}, we have successfully derived the atmospheric parameters of the Sun 
($T_{\rm eff}$, $\log{g}$, $\xi$) by means of a solar spectrum (Vesta) taken with the HERMES spectrograph.

\subsection{Abundances}
\label{abundances}

In order to calculate the individual chemical abundances of the {180} stars, we assumed the stellar parameters derived with {\scshape StePar}. 
We obtained abundances for 13 different chemical species: Fe, the $\alpha$-elements (Mg, Si, Ca, and Ti), the Fe-peak elements (Cr, Mn, Co, and Ni), and the odd-Z elements (Na, Al, Sc, and V). 
We calculated chemical abundances using the $EW$ method, Kurucz ATLAS9 plane-parallel model atmospheres \citep{Kurutz1993}, and the MOOG code \citep{Sneden1973}, as in \cite{Tabernero2012,Tabernero2017}. 
The $EW$s were determined using the ARES code \citep{Sousa2015}, following the approach described in Section~\ref{stellarparameters}. 
We also re-measured manually the EWs with the task {\tt splot} within the IRAF environment when any individual abundance determination of particular lines was separated from the general trend.
We computed final abundances in a differential manner (i.e., in a line-by-line basis) with respect to our solar spectrum (Vesta) observed with HERMES. 
See the resulting solar element abundances ($\log{\epsilon}$ (X))  in the bottom part of Table~\ref{tab:sun_abu}.
Table~\ref{tab:abundances} reports these differential abundances ([X/H]) thus derived for our star sample.

\subsection{Kinematics}

Stellar kinematic groups (SKGs), superclusters (SCs), and moving groups (MGs) are kinematic coherent groups of stars that may share a common origin and, therefore, age and chemical composition \citep{BoesgaardFriel1990,Eggen1994,DeSilva2007,Famaey2008,Antoja2009}.
Among them, the youngest SKGs are: the Hyades SC ($\sim$600\,Myr), 
Ursa Major MG (Sirius SC -- $\sim$400\,Myr), 
Castor MG ($\sim$300\,Myr), 
Local Association (Pleiades MG -- 20 to 150\,Myr), 
and IC~2391 SC (35--55\,Myr), 
We refer the reader to \cite{Montes2001}, \cite{LS2006}, \cite{Klutsch2014}, \cite{Riedel2017} and references therein for more details.

Other very young SKGs, such as the $\epsilon$~Chamaeleontis, TW~Hydrae, $\beta$~Pictoris, Tucana-Horologium, AB~Doradus, Columba, Carina, and Hercules-Lyra moving groups, have kinematics close to the Local Association, as well as Argus' to IC 2391, and Octans and Octans-Near's to Castor \citep{ZS2004,Torres2008,Montes2010,Montes2015,Bell2015}.
Even new associations are identified, such as the All Sky Young Association (ASYA -- \citealt{Torres2016}).

With the coordinates in  Table~\ref{tab:investigated_stars}, parallactic distances and proper motions in Table~\ref{tab:pms}, and radial velocities measured in Section~\ref{obsred}, we computed Galactocentric space velocities as in \cite{Montes2001} with the procedure established by \cite{JohnsonSoderblom1987}.
For the single- and double-lined spectroscopic binaries (Section~\ref{sbs}) and the unobserved star $\sigma$~CrB~Aa,Ab, we adopted their systemic radial-velocity values $\gamma$ from the literature.
Table~\ref{tab:kinematics} lists the used radial velocities $V_r$ along with the computed space velocities $U$, $V$, and $W$ of our 198 F-, G-, and K- stars.

\section{Results and discussion}

We investigated {489} stars distributed in {193} systems, formed by {193} primary F-, G-, and K- stars and {296} common proper-motion companions and candidates (Table~\ref{tab:investigated_stars}). 
For these systems, we studied their proper motions and distances, as explained in Section~\ref{analysis}.
We got a final sample of {192} physical systems, of which {135} are double and {57} are multiple ({43} triple, {9} quadruple, and {5} quintuple).
In Table~\ref{tab:pms} we marked the 84 discarded stars, along with other useful remarks for the remaining stars.

\subsection{Spectroscopic binaries}
\label{sbs}

\begin{table}
\caption{Primary spectroscopic binaries.}
\label{tab:sb}
\begin{tabular}{llll} 
\hline
\hline
WDS & Name & Type & Reference$^{a}$\\ 
\hline
00452+0015 & HD 4271 Aa,Ab & SB1 & Gri01\\


00491+5749 & Archid Aa,Ab & SB1 & A\&L76 \\

02291+2252 & BD+22 353Aa,Ab & SB1 & Hal12 \\

02482+2704 & BC Ari Aa,Ab & SB1 & Lat02 \\

03206+0902 & HD 20727 Aa,Ab & SB1 & D\&M91 \\

03396+1823 & V1082 Tau Aa,Ab & SB2 & Lat92\\

03398+3328$^b$ & HD 278874 Aa,Ab & SB2 & This work \\

03566+5042 & 43 Per Aa,Ab & SB2 & Wal73 \\

05067+5136 & 9 Aur Aa,Ab & SB1 & Abt65 \\

05289+1233 & HD 35956 Aa,Ab & SB1 & Kat13 \\

06173+0506$^c$ & HD 43587 & SB1 & Kat13 \\

09245+0621$^b$ & HD 81212 AB & SB2 & This work \\

09393+1319 & HD 83509 Aa,Ab & SB2 & Gri03 \\

15282-0921$^c$ & HD 137763 & SB1 & D\&M92 \\

16147+3352$^d$ & $\sigma$ CrB Aa,Ab & SB2 & Bak84 \\

16329+0315$^c$ & HD 149162 & SB1 & Lat02 \\

16348-0412 & HD 149414 Aa,Ab & SB1 & Lat02 \\

20169+5017 & HD 193216 Aa,Ab & SB1 & Gri02 \\

20462+3358 & $\epsilon$ Cyg Aa,Ab & SB1 & Gra15 \\

20599+4016$^{c}$ & HD 200077 Aa1,Aa2,Ab & SB2 & Gol02 \\

23026+2948$^b$ & BD+29 4841Aa,Ab & SB2 & This work \\

23581+2420$^b$ & HD 224459 Aa,Ab & SB2 & This work \\
 \hline
 \multicolumn{4}{l}{$^a$ Reference -- Abt65: \cite{Abt1965}; A\&L76: \cite{AbtLevy1976};}\\
  \multicolumn{4}{l}{Bak84: \cite{Bakos1984}; D\&M91: \cite{DuquennoyMayor1991};}\\
 \multicolumn{4}{l}{D\&M92: \cite{DuquennoyMayor1992}; Gol02: \cite{Goldberg2002};}\\
 \multicolumn{4}{l}{Gra15: \cite{Gray2015}; Gri01: \cite{Griffin2001}; Gri02: \cite{Griffin2002};}\\
 \multicolumn{4}{l}{Gri03: \cite{Griffin2003}; Hal12: \cite{Halbwachs2012};}\\
 \multicolumn{4}{l}{Kat13: \cite{Katoh2013}; Lat92 \cite{Latham1992};}\\
  \multicolumn{4}{l}{Lat02: \cite{Latham2002}; Wal73: \cite{Wallerstein1973}.}\\
 \multicolumn{4}{l}{$^b$ New SB2, discovered in this work.}\\
\multicolumn{4}{l}{$^c$ Resolved close multiple system described in text.}\\
\multicolumn{4}{l}{$^d$ Not observed by us.}\\
 \end{tabular}
\end{table}

As discussed in Section~\ref{stellarparameters}, we were not able to determine stellar parameters for seven double-peak spectroscopic binaries (SB2s).
They are listed in Table~\ref{tab:sb}, together with the other SB2s $\sigma$~CrB~Aa,Ab (not observed) and HD~200077\,Aa,Ab (with stellar parameters).
We report for the first time four SB2s, namely HD~278874\,Aa,Ab, HD~81212\,AB, BD+29~4841\,Aa,Ab, and HD~224459\,Aa,Ab.
Only for the later, we have two HERMES spectra separated by one day but we could not see any significant difference between them, 
so the orbital period must be  $P_{\rm orb} \gg$ 1\,d.
Interestingly, HD~81212\,AB was found to be an astrometric binary by F.~G.~W. Struve in 1831.
The pair is separated by $\rho$ = 1.1--1.9\,arcsec and, thus, unresolved by us.
The very small magnitude difference between A and B, $\Delta m \approx$ 0.12\,mag, indicates a mass ratio close to unity. 
We estimate an orbital period of 200--300\,yr for the astrometric pair, and a radial-velocity difference of about 6\,km\,s$^{-1}$, 
which is consistent with what we observe in the double-line spectroscopic binary.
Therefore, the spectroscopic binary can actually be the astrometric binary.
This fact could also explain the apparently wrong parallax tabulated by TGAS.
The other three new SB2 stars are not known close astrometric binaries.

\begin{figure*}
	\includegraphics[width=\columnwidth]{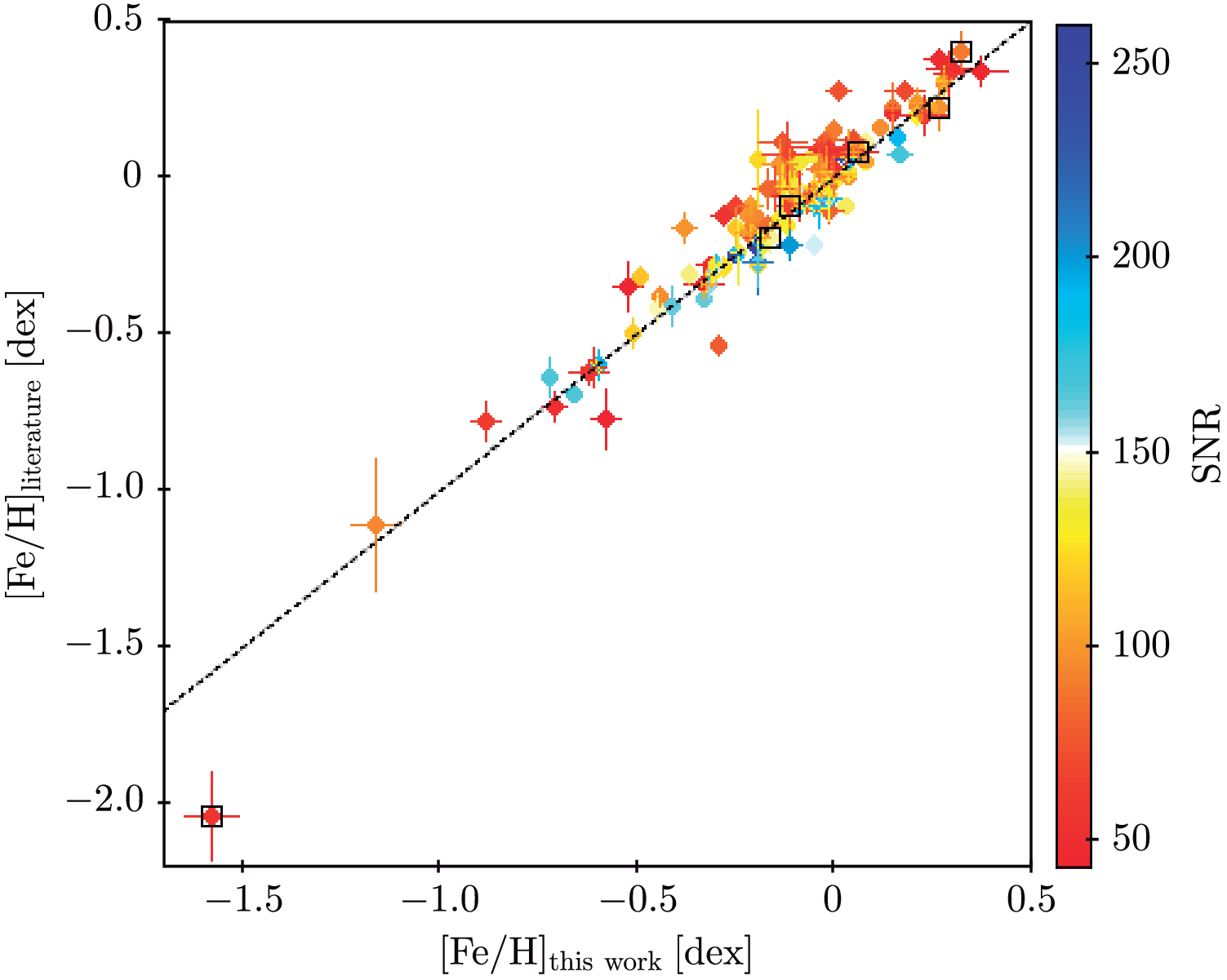}
	\includegraphics[width=\columnwidth]{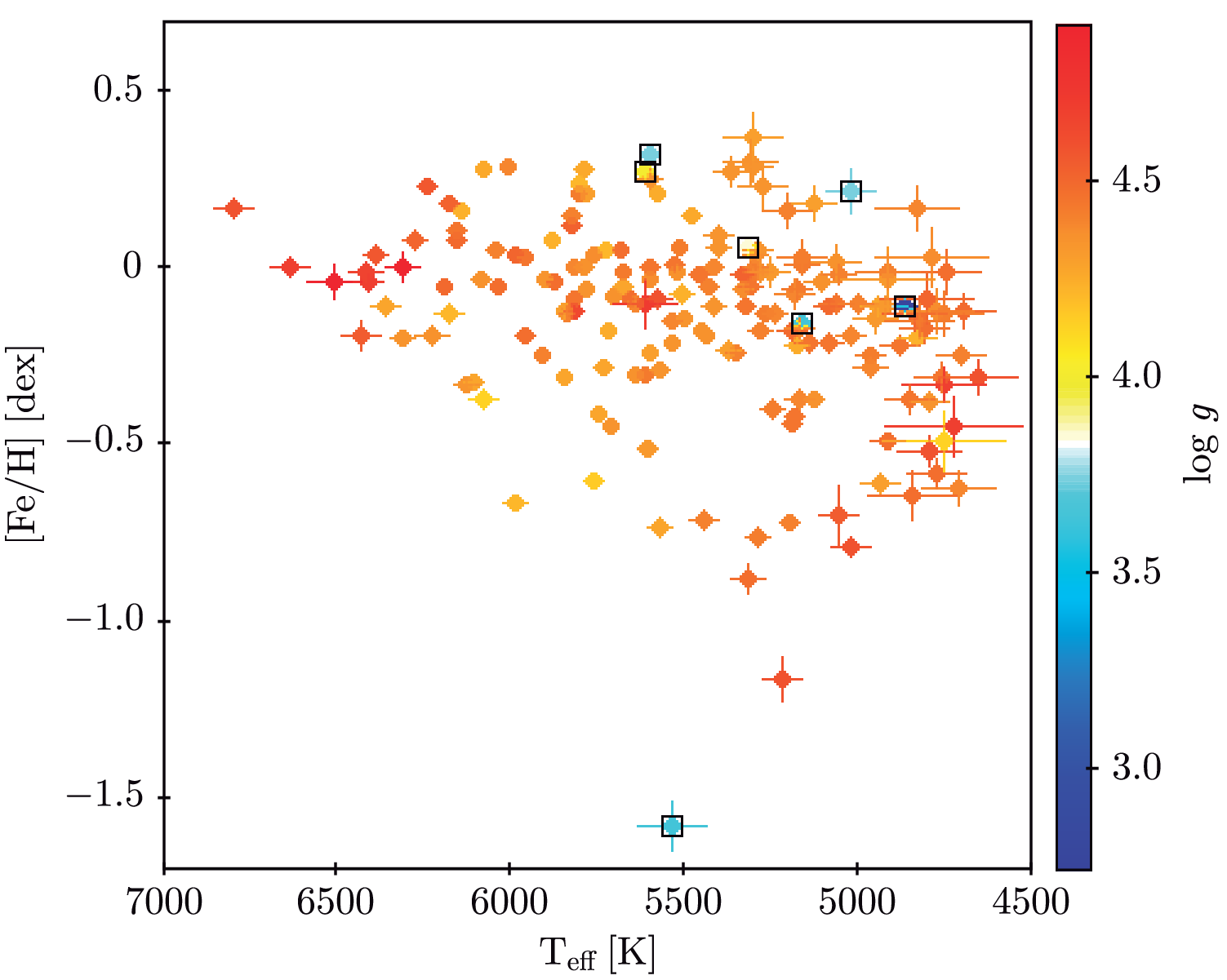}
    \caption{\textit{Left panel:} iron abundance published in the literature against that obtained in this work, colour-coded with SNR. Black dashed line represents 1:1 relation. \textit{Right panel:} iron abundance as a function of effective temperature for our primary stars, colour-coded with $\log{g}$. Black-ensquared stars represent low-gravity stars. }
    \label{fig:metprimlit}
\end{figure*}

There are full orbital parameters ($P$, $e$, $\gamma$, $K_1$, $K_2$) available in the literature for the other five stars, including $\sigma$~CrB~Aa,Ab and HD~200077 \citep[S$_{{\rm B}^9}$ --][]{Pourbaix2004}.
The later is part of a quintuple system containing a close SB2 (F8\,V + G6--9:, $P$ = 112.5\,d) first resolved by \citealt{Horch2012} (LSC~1, $\rho \approx$ 0.022\,arcsec), a close companion resolved by {\em Hipparcos} (late K; COU 2431, $\rho$ = 2.2\,arcsec), and the wide cool companion G~210--44 (K7\,V + M0--1:), which is in turn another close binary \citep{Latham1988, Goldberg2002, Mazeh2003, Caballero2009}.
In our HERMES spectra of HD~200077, the Aa1 component (late F) dominates over Aa2 (late G) and Ab (late K, not visible), and its lines were well separated from those of the other components. 

Besides, there are 13 known single-line spectroscopic binaries (SB1) in our sample.
We did not discard them in our analysis because the determined stellar parameters correspond to the primary in the system and were not significantly affected by the companion.
Four of the SB1s were also resolved astrometrically:

\begin{itemize}
\item HD~43587 (CAT~1, $\rho \approx$ 0.90\,arcsec).
The orbital period of $P$ = 34.2\,yr determined by \cite{Katoh2013} from radial-velocity monitorisation matches reasonably well the adaptive optics observations by \cite{Catala2006}. The system deserves a new analysis given the low mass of the companion, several magnitudes fainter than the primary. 

\item HD~137763 (BAG~25, $\rho \approx$ 0.10\,arcsec).
The orbital period of $P$ = 2.44\,yr determined by \cite{DuquennoyMayor1992} also matches the measured projected physical separations measured astrometrically \citep{Jancart2005,Balega2006,Horch2015}, and, therefore, the dynamical masses of the two stars can be determined precisely.

\item HD~149162 (DSG~7, $\rho \approx$ 0.0148\,arcsec and $\rho \approx$ 0.284\,arcsec).
Again, the astrometric measurements of \cite{Horch2015} agree with the spectroscopic measurements of \cite{Latham2002}, who determined an orbital period of 0.620\,yr.
This is a hierarchical triple system, and the seven-month period of the SB1 corresponds to the closest pair.
The effect of the component at $\sim$0.3~arcsec is not discernible spectroscopically.
The wide common proper motion companion, at 4.2\,arcmin to the south east, is in turn a binary made of an M3.0\,V star and a white dwarf, which makes HD~149162 a quintuple system. 

\item $\epsilon$ Cyg (CHR 100, $\rho$ = 0.041\,arcsec).
This well-studied, binary giant star has been the subject of numerous radial-velocity surveys (e.g. \citealt{Griffin1994,Gray2015}) and has also been resolved with optical interferometry \citep{Hartkopf1994}.

\end{itemize}

\subsection{[Fe/H]}
\label{metallicity}

\begin{table*}
\caption{Estimated ages for the seven low-gravity stars ($\log{g}$ < 4.1) with stellar parameters in our sample.}
\label{tab:ages}
\begin{tabular}{llccccl} 
\hline
\hline
 WDS & Simbad & $\log{g}$ & [Fe/H] & Estimated & Published & Reference$^{a}$ \\
   &  &  &  &  age [Gyr]  & age [Gyr] &  \\
\hline
01572-1015 & HD 11964 A & 3.85$\pm$0.06 & 0.06$\pm$0.02 & $\sim$5--10 & 9.77$\pm$0.52 & Tsa13\\ 
05466+0110 & HD 38529 A & 3.75$\pm$0.07 & 0.32$\pm$0.02 & $\sim$2--5 & 3.77$\pm$0.36 & Ram12\\
08110+7955 & BD+80 245 & 3.63$\pm$0.20 & --1.58$\pm$0.07 & $\sim$13--14 & ... & ... \\
11523+0957 & HD 103112 & 3.75$\pm$0.19 & 0.22$\pm$0.06 & $\sim$10 & ... & ... \\
17465+2743 & $\mu^{01}$ Her A & 4.03$\pm$0.04 & 0.27$\pm$0.02 & $\sim$10 & 7.88$\pm$0.24 & Ram12\\
19553+0624 & $\beta$ Aql A & 3.64$\pm$0.06 & --0.16$\pm$0.01 & $\sim$2--5 & 4.08$\pm$3.95 & Ram13\\
20462+3358 & $\epsilon$ Cyg A & 2.74$\pm$0.11 & --0.11$\pm$0.03 & $\sim$1 & 0.90$\pm$0.20 & daS15\\
 \hline
 \multicolumn{7}{l}{$^a$Reference -- Ram12: \cite{Ramirez2012}; Ram13: \cite{Ramirez2013}; Tsa13: \cite{Tsantaki2013}; }\\
  \multicolumn{7}{l}{daS15: \cite{daSilva2015}.}\\
 \end{tabular}
\end{table*}

We derived stellar atmospheric parameters of {175} primaries and {five} secondaries (Section~\ref{stellarparameters} and Table~\ref{tab:parameters}), from which {50} are presented here for the first time.
One of the parameters is the iron abundance [Fe/H], which is the most used proxy for metallicity.
In the left panel of Fig.~\ref{fig:metprimlit}, we depict spectroscopic [Fe/H] collected from the literature against the ones derived by us.
For a fair comparison, we only collected spectroscopic [Fe/H] from the literature \citep[e.g.,][]{,ValentiFischer2005,Sousa2011,Ramirez2013,Santos2013}, and did not take into account the ones derived photometrically \citep[e.g.,][]{Bonfils205,JohnsonApps2009,SchlaufmanLaughlin2010}. 
We compiled and selected spectroscopic [Fe/H] with the PASTEL Catalogue \citep{Soubiran2016}, giving priority to the most recent works. 
According to the diagram, our values agree very well with the published ones, mainly in the range of --1.0 < [Fe/H] < 0.5  and no significant offset is detected.
The iron abundance determined by us does not display any trend as a function of $T_{\rm eff}$ or $\log{g}$, as shown in the right panel of Fig.~\ref{fig:metprimlit},

The least metallic star in our sample is the red giant branch star BD+80~245 (G0\,IV). 
We measured [Fe/H]  = --1.58$\pm$0.07, a value slightly higher than those provided by \citet[][[{Fe/H]}  = --2.05]{Fulbright2000}, \citet[][{[Fe/H]}  = --1.76]{Stephens2002}, and \citet[][{[Fe/H]}  = --2.04]{Roederer2014}. 
BD+80~245 was also studied by \cite{Ivans2003}, who classified it as a halo star based on its chemical composition (we also classified it as a halo star in Section~\ref{kinematics} based on kinematics), and explained a possible formation from material polluted by the earliest supernovae Ia events that occurred in the Milky Way. 
BD+80~245 is the only star that stays away from the general trend in the left panel of Fig.~\ref{fig:metprimlit}.

In general, [Fe/H] has an effect on the derivation of spectro-photometric distances (Section~\ref{analysis}).
As illustrated in Fig.~\ref{fig:pm_distances}, stars with [Fe/H] < --0.4 tend to lie in the upper part of the 1:1 relation between primary and ``secondary'' distances. 
This effect may be due to an intrinsic offset in the spectral type--$M_{J}$ relation used to derive spectro-photometric distances to our late-K and M dwarfs, as \cite{Miriam2017} assumed solar metallicity (all {192} physical primaries have parallactic distances but {160} companions have spectro-photometric distances).
The two low-metallicity systems that suffer more from this offset are WDS~03150+0101 (BD+00~549A, with [Fe/H] = --0.88,  and BD+00~549B), and WDS~22090-1754 (HD 210190, with [Fe/H] = --0.42, and LP 819-37, with $\zeta$ = 0.856, where $\zeta$ is a metallicity spectral index defined by \citealt{Lepine2007} and measured by \citealt{Alonso-Floriano2015}).
For these systems, the spectro-photometric distances for the secondary component are about twice as large as the parallactic distance of the  primary.

Besides, for WDS~16348-0412 (HD 149414 Aa,Ab, with [Fe/H] = --1.16, and GJ 629.2B, with $\zeta$ = 0.664) we did not derive a spectro-photometric distance for the secondary because it is a sub-dwarf candidate (sdM0:, \citealt{Alonso-Floriano2015})\footnote{Note the wrong spectral type of GJ 629.2B in Simbad.}.
Instead, we adopted the spectro-photometric distance of 48$^{\text{+}12}_{-9}$\,pc from the $M_J$-SpT relationship for subdwarfs in \cite{Zhang2013}, which agrees with the distance to its very low metallicity primary tabulated by TGAS of 46.3$\pm$0.9\,pc.
We concluded that the metallicity affects the derivation of our spectro-photometric distances, but heterogeneously and in extreme cases.




\subsection{Abundances}

Apart from iron, we measured chemical abundances of 12 different elements for the 180 F-, G-, and K- stars in our sample (Na, Mg, Al, Si, Ca, Sc, Ti, V, Cr, Mn, Co, and Ni -- see Section~\ref{abundances} and Table~\ref{tab:abundances}). 
Galactic trends are depicted in Figs.~\ref{fig:abunds-A1} and~\ref{fig:abunds-A2}  where we plot the abundance ratios of [X/Fe] versus [Fe/H] for each element X.
We compared them to the FGK stellar sample from \cite{Adibekyan2012}. 
Our sample covers a wide range of [Fe/H] and includes a few low-metallicity stars ([Fe/H] $<$ --1.0) that fall well below the range studied by \cite{Adibekyan2012} and, as expected, have enhanced content in $\alpha$ elements \citep{Bensby2014,Jofre2015}. 

Using our line-by-line differential analysis we reproduced the expected behaviour of the different chemical species, with manganese being a remarkable exception. 
Useful Mn lines are scarce and difficult to measure in our HERMES spectra either by hand (with IRAF {\tt splot}) or with a semiautomatic method (using the ARES code), and thus our results present an offset that reflects this fact. 
Interestingly, we also reproduced the scatter found by \cite{Adibekyan2012} for vanadium and scandium, which is a known issue for stars cooler than 5000\,K (see \citealt{Neves2009} and \citealt{Tabernero2012} for further details). 
Giants and subgiants tend to deviate from the general trends.
Although this effect appears to be entirely real \citep{Smiljanic2012,Tabernero2012}, it is not observed in these cases, and, therefore, may be an effect only on very low gravity stars ($\log{g} \leq $ 2.5).

\subsection{Giants and subgiants}
\label{lowgrav}

\begin{figure}
	\includegraphics[width=\columnwidth]{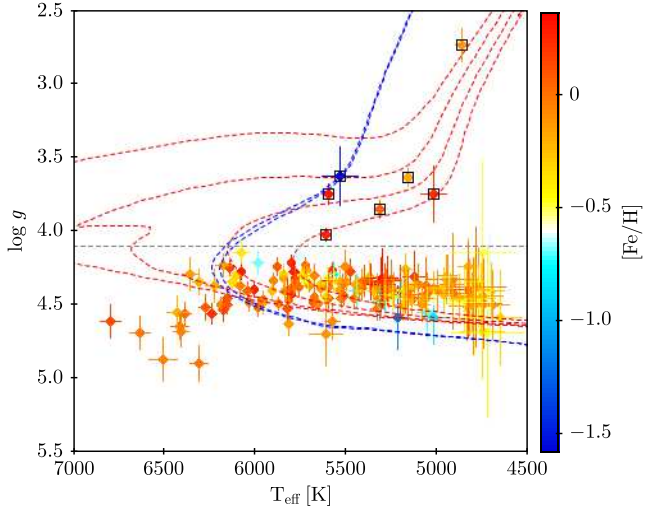}
    \caption{Surface gravity as a function of effective temperature for our primary stars, colour-coded with metallicity. 
    Red dashed lines correspond to isochrones of [Fe/H]\,=\,0.0 and ages of 1, 2, 5, and 10\,Gyr, from top to bottom. 
    Blue dashed lines correspond to isochrones of [Fe/H]\,=\,--1.5 and ages of 13 and 14\,Gyr. 
    All isochrones are from YaPSI \citep{Spada2017}.
    The grey, horizontal, dashed line correspond to $\log{g}$\,=\,4.1.}
    \label{fig:teffloggprim}
\end{figure}

Among our list of {192} physical primaries there are eight stars with surface gravities lower than $\log{g}$ = 4.1 (see Fig.~\ref{fig:teffloggprim}).
They are Aldebaran ($\log{g}$ = 1.66; \citealt{Prugniel2011}), for which we were not able to determine stellar parameters with {\sc StePar}, the giant star $\epsilon$~Cyg\,A ($\log{g}$ = 2.74), BD+80~245 ($\log{g}$ = 3.63), which is the red giant branch star with the lowest metallicity in our sample, and five subgiant stars with $\log{g}$ = 3.64--4.03.
Of them, HD~103112 had not been reported before to display any subgiant class or low-gravity feature in its spectra (but see \citealt{McDonald2017} and their photometric analysis). 
The remaining four subgiants are quite well investigated, either because of their brightness ($\beta$~Aql\,A and $\mu^{01}$~Her\,A) or presence of exoplanets (HD~11964\,A and HD~38529\,A; Section~\ref{planets}).

For the {seven} low-gravity stars with derived stellar parameters, we estimated their ages using the Yale-Potsdam Stellar Isochrones \citep[YaPSI --][]{Spada2017} with two different iron abundances ([Fe/H] = 0.0 and [Fe/H] = --1.5) and fixed solar helium abundance (Y = 0.28;  see again Fig.~\ref{fig:teffloggprim}). 
Estimated ages agree within uncertainties with published values in five cases (Table~\ref{tab:ages}).
We determined ages for the first time for the two remaining stars: the poorly-investigated subgiant HD~103112 and the very low-metallicity star BD+80~245.
For the later, we infer an age similar to that of the Universe (limited by the accuracy of the YaPSI models), which is consistent with the hypothesis of \cite{Ivans2003} of it being a halo star polluted by the earliest supernova explosions.

\subsection{Kinematics}
\label{kinematics}

\begin{figure}	
	\includegraphics[width=\columnwidth]{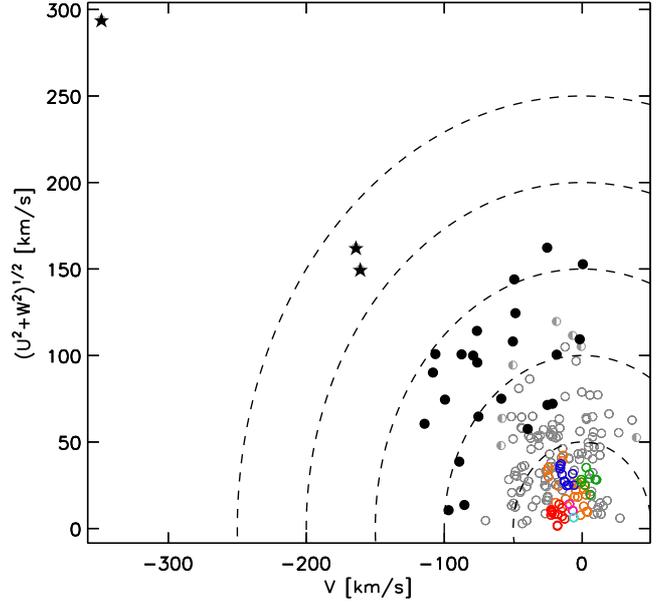}
    	\caption{Toomre diagram of our F-, G-, and K- stars. 
	Black stars: halo; black filled circles: thick disc; grey semi-filled circles: thick-to-thin transition disc; grey open circles: thin disc.
	Blue: Hyades SC; red: Local Association; green: Ursa Major MG; magenta: IC~2391 SC; cyan: Castor MG; orange: other young stars. 
	Dashed grey lines represent constant values of the total space velocity $v_{tot}$=($U^{2} \text{+} V^{2} \text{+} W^{2}$)$^{1/2}$ in steps of 50\,km\,s$^{-1}$.
	The Galactocentric space velocities $U$, $V$, and $W$ are referred to the local standard of rest.}
	\label{fig:Toomreyo}
\end{figure}

\begin{figure*}
	\includegraphics[width=2\columnwidth]{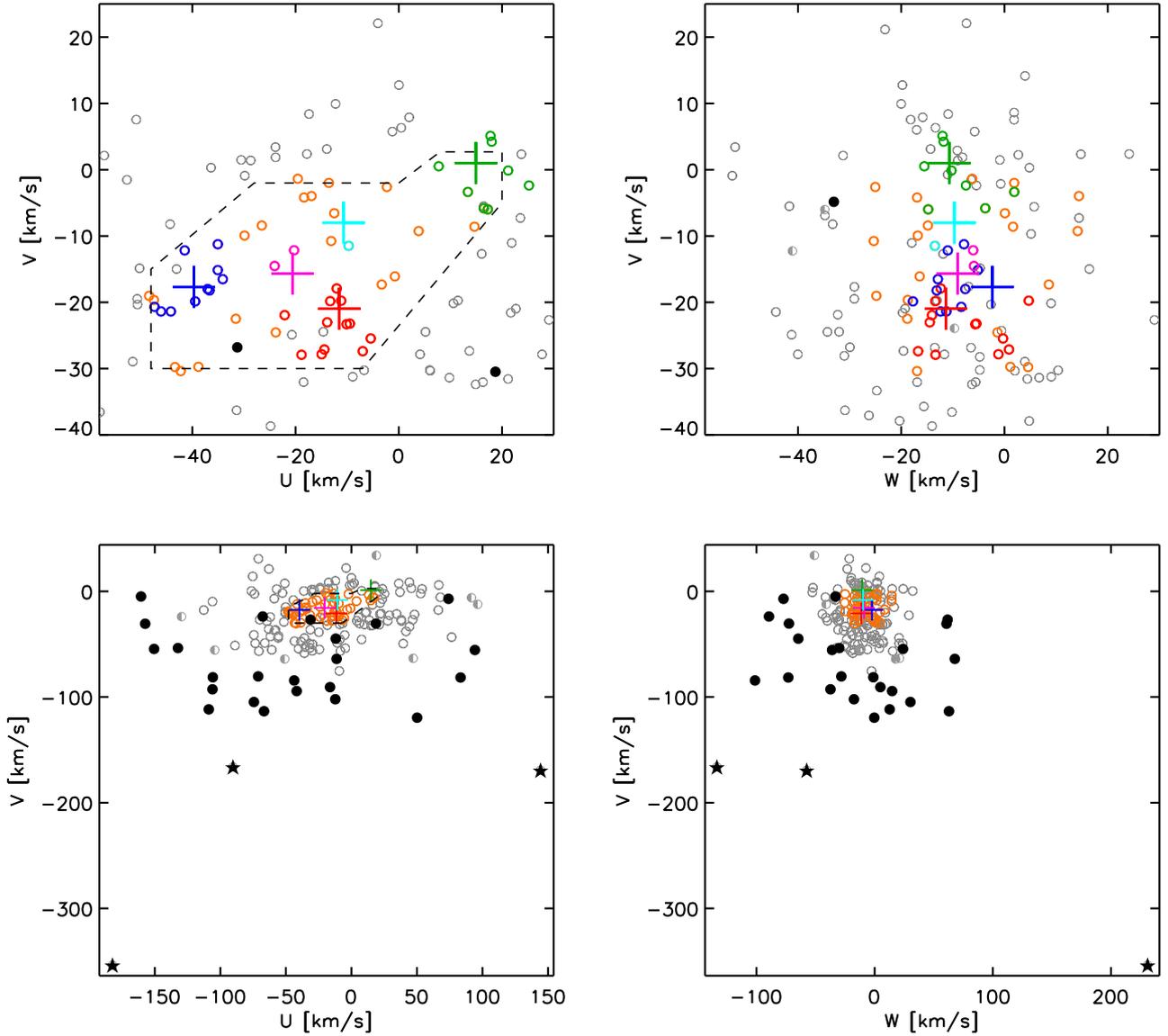}
	 \caption{Same as Fig~\ref{fig:Toomreyo}, but for the B\"ottlinger diagrams.
	 Crosses mark the centres of each young SKG.
	 Upper panels represent zoomed areas of lower panels.
	 In the top left panel, the dashed grey line confines the young disc population as defined by \citet{Eggen1984,Eggen1989}. }
	 \label{fig:Boettlinger}
\end{figure*}

As illustrated by the Toomre diagram (Fig.~\ref{fig:Toomreyo}), we classified each star in the different Galactic populations, halo (H), thick disc (TD), thick-to-thin transition disc (TD-D), and thin disc (D), as in \cite{Bensby2003,Bensby2005}.
For that, we assumed Gaussian distributions of space velocities $U$, $V$, and $W$.
We found 165 stars in the thin disc, 23 in the thick disc, 7 in the thick-to-thin transition disc, and 3 in the halo, as it is shown in Table~\ref{tab:kinematics} and Fig.~\ref{fig:Toomreyo}.
The three stars in the halo are:

\begin{itemize}
\item Ross~413. 
It is a halo star catalogued by \cite{Allen2014} in the context of MACHO studies.
The iron abundance derived by us, [Fe/H] = --0.58, is again slightly higher than the published value ([Fe/H] = --0.77;  \citealt{WoolfWallerstein2005}).

\item BD+80 245. 
It is the old, low-metallicity, subgiant star discussed above.
Our classification as an halo star agrees with the one published in \citeauthor{Ivans2003} (\citeyear{Ivans2003}; see Section~\ref{metallicity}).

\item HD~149414.
It is a well-studied halo star (e.g. \citealt{Sandage1969,Tomkin1999,Gratton2003,Allen2014}), and the star with the second lowest iron abundance in our sample ([Fe/H] = --1.16). 
Besides, it is also a single-lined spectroscopic binary (Table~\ref{sbs}).
\end{itemize}
 
In general, thick-disc (and also thick-to-thin-transition-disc) stars have subsolar metallicities.
However, there are some remarkable outliers, such as HD~102326 ([Fe/H] = +0.15$\pm$0.02), HD~103112 ([Fe/H] = +0.22$\pm$0.06), and HD~190360 ([Fe/H] = +0.21$\pm$0.02), which may represent the tail of the distribution towards high metallicities or, conversely, a kinematic classification at the boundary with the thin disc.
 
Next, for the stars in the thin disc, we separated between young disc stars and non-young disc stars (designated with the symbol '$\times$' in Table~\ref{tab:kinematics}) as defined by \cite{Eggen1984,Eggen1989} and depicted in the B\"{o}ttlinger diagram (Fig~\ref{fig:Boettlinger}).
Thick disc, thick-to-thin transition disc, and halo stars are also non-young disc stars. 
Besides, for each young disc star, we studied its membership in known SKGs, also as in \cite{Montes2001}.
In particular, we identified {33} star candidates in SKGs: {12} stars in the Local Association, {10} in the Hyades SC, {eight} in the Ursa Major MG, {two} in the IC 2391 SC, and {one} in the Castor MG, whereas 20 are young disc star with no apparent SKG membership (designated with 'YD' in Table~\ref{tab:kinematics}).
Also, we checked the membership of our 33 young Galactic disc stars with the LACEwING \citep{Riedel2017} and BANYAN $\Sigma$ \citep{Gagne2018}\footnote{\url{http://www.exoplanetes.umontreal.ca/banyan/banyansigma.php}} algorithms.
As summarised in Table~\ref{tab:kinematics}, of the 33 stars candidates in known SKGs, 17 had been already proposed to belong to some of them.

\begin{itemize}
\item {Local Association.} 
Seven of our 12 LA candidates were already classified as probable LA members by \cite{Montes2001}.
All of them except HD~98736 had been later assigned to SKGs linked to the LA, which supports our classification:
Hercules-Lyra (V538~Aur and V382~Ser, by \citealt{Fuhrmann2004}; DX~Leo and HH~Leo, by \citealt{Eisenbeiss2013}), AB~Doradus (V577~Per, by \citealt{Riedel2017}), and Columba (V368~Cep, with BANYAN $\Sigma$).
The eighth known LA star candidate is HD~82939, which proper-motion companion MCC~549 has been subject of debate: \cite{Schlieder2012} proposed it to be a member in the $\beta$~Pictoris moving group, also linked to the LA, but this statement was later denied by \cite{Malo2014} and \cite{Skholnik2017}.
There is no trail of lithium in our HERMES spectrum of HD~82939 (G5\,V), which supports the \cite{Malo2014} conclusion.

\item {Hyades super-cluster.} 
We recovered two stars previously considered as members of the Hyades SC using different methods: the multi-planet host $\rho^{01}$~Cnc\,A (aka Copernicus, 55~Cnc), with chemical tagging \citep{Tabernero2012}, and HD~51067~A, with LACEwING \citep{Riedel2017}.
Besides, HD~116963 could be a Carina-Near star according to BANYAN $\Sigma$.

\item {Ursa Major moving group.} 
Four stars were confirmed as UMa MG members by chemical tagging in \cite{Tabernero2017}: the pair WDS~J05445-2227 ($\gamma$ Lep and AK Lep), V869~Mon, and HD~167389.
Another two stars, HD~24961 and SZ~Crt, were also classified as UMa MG members by \cite{Montes2001}, \cite{King2003}, and \cite{LS2010}.
\item {IC 2391 super-cluster.} 
V447~Lac was classified as a doubtful member in Hercules-Lyra by \cite{Eisenbeiss2013}, but a more recent analysis by \cite{Riedel2017} located it in the Argus moving group, which is kinematically linked to the IC 2391 super-cluster.

\end{itemize}

We computed the mean iron abundance for each SKG and compared them with the ones published in the literature \citep{BoesgaardFriel1990,Randich2001,Paulson2003,Vauclair2008,Pompeia2011,Tabernero2012,deSilva2013,Tabernero2017}, and found a good agreement between them.
Although this is not an exhaustive chemical tagging analysis, it supports our kinematic classification.

\subsection{Planetary systems with late-type dwarfs}
\label{planets}

\begin{figure}
	\includegraphics[width=\columnwidth]{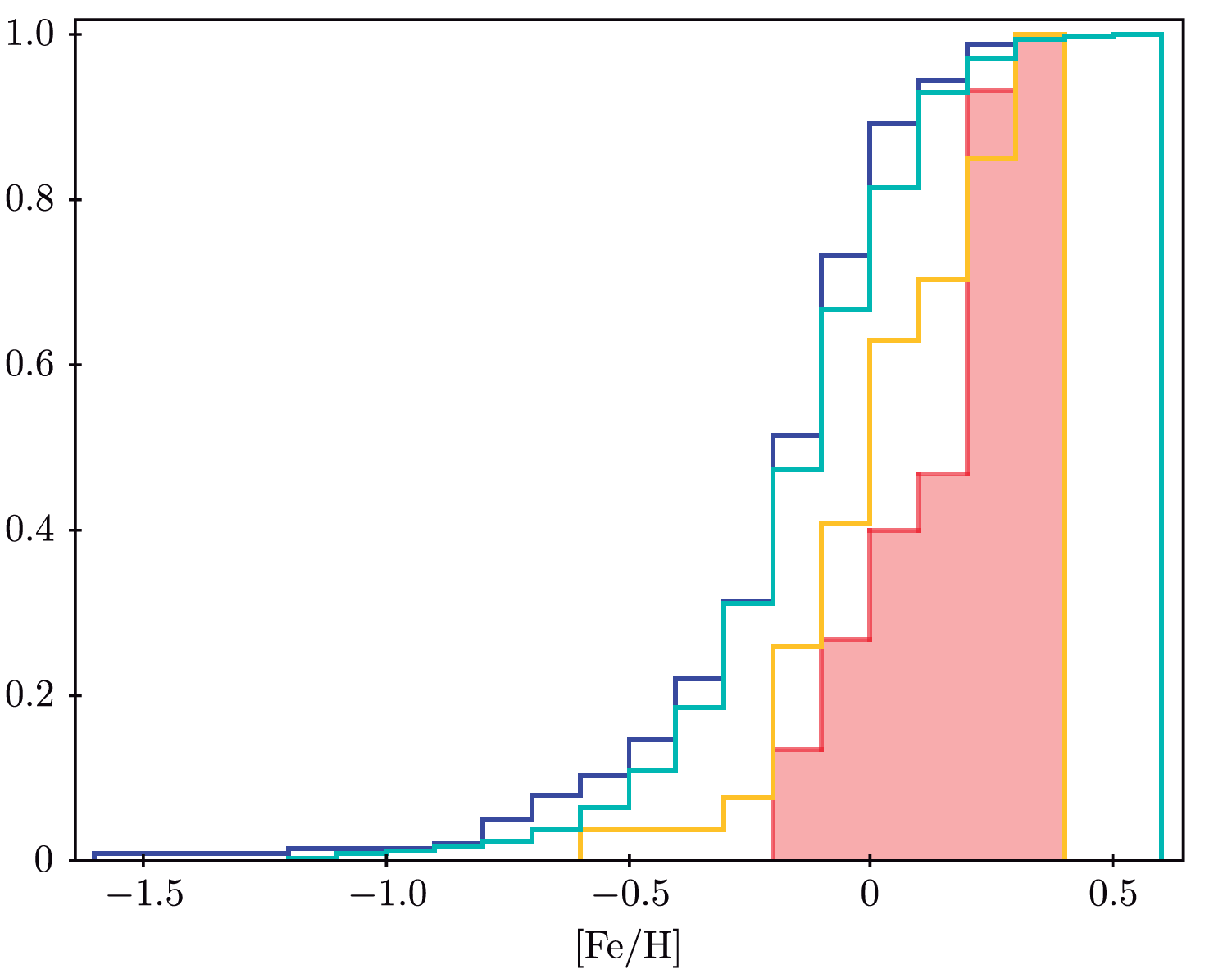}
	\caption{Normalised cumulative iron abundance histogram of the stars in our sample with (red, shaded) and without planets (blue), and of the \citet{Sousa2011} sample with (yellow) and without planets (cyan).}
    \label{fig:histfeplanets}
\end{figure}

\begin{table*}
\caption{F, G, K- stars with confirmed exoplanets in our sample.}
\label{tab:exoplanets}
\begin{tabular}{llclllc} 
\hline
\hline
WDS & Primary & [Fe/H] & Planet(s) & Reference$^a$ & Secondary & $s$ \\
 &  &  & &  & & [au]\\
\hline
01572-1015 & HD 11964 A & 0.06$\pm$0.02 & b, c & Wri09 & HD 11964 B & 974$\pm$23\\
03480+4032 & HD 23596 & 0.28$\pm$0.02 & b & Per03 & J03480588+4032226 & 3693$\pm$52\\
04359+1631 & Aldebaran & --0.27$\pm$0.05$^b$ & b & Hat15 & Aldebaran B & 573$\pm$11\\
05466+0100 & HD 38529 A  & 0.32$\pm$0.02 & b, c & Ben10 & HD 38529 B & 11148$\pm$175\\
06332+0528 & HD 46375 A & 0.23$\pm$0.06 & b & WF11 & HD 46375 B & 363$\pm$12\\
08526+2820 & $\rho^{01}$ Cnc A & 0.29$\pm$0.04 & b, c, d, e, f & Nel14 & $\rho^{01}$ Cnc B & 1044$\pm$10\\
09152+2323 & HD 79498 & 0.21$\pm$0.02 & b & Rob12 & BD+23 2063B & 2768$\pm$80\\
13018+6337 & HD 113337 A & 0.17$\pm$0.03 & b & Bor14 & LSPM J1301+6337 & 4419$\pm$48\\
18006+2943 & HD 164595 A & --0.08$\pm$0.01 & b & Cou15 & HD 164595 B & 2509$\pm$27\\
18292+1142 & HD 170469 & 0.28$\pm$0.02 & b & Fis07 & J18291369+1141271 & 2617$\pm$37\\
20007+2243 & V452 Vul & --0.10$\pm$0.03 & b & Sou10 & J20004297+2242342 & 224$\pm$2\\
20036+2954 & HD 190360 A & 0.21$\pm$0.02 & b, c & Vog05 & HD 190360 B & 2854$\pm$27\\
21324-2058 & HD 204941 & --0.19$\pm$0.03 & b & Dum11 & LP 873-74 & 1610$\pm$12\\
23419-0559 & HD 222582 A & 0.00$\pm$0.02 & b & But06 & HD 222582 B & 4637$\pm$59\\
 \hline
 \multicolumn{7}{l}{$^a$ Reference -- Ben10: \cite{Benedict2010}; Bor14: \cite{Borgniet2014}; But06:  \cite{Butler2006};}\\
  \multicolumn{7}{l}{Cou15: \cite{Courcol2015}; Dum11: \cite{Dumusque2011}; Fis07: \cite{Fischer2007}; Hat15: \cite{Hatzes2015};}\\
 \multicolumn{7}{l}{Nel14: \cite{Nelson2014}; Per03: \cite{Perrier2003}; Rob12: \cite{Robertson2012}; Sou10: \cite{Southworth2010};}\\
 \multicolumn{7}{l}{Vog05: \cite{Vogt2005}; WF11: \cite{WangFord2011}; Wri09: \cite{Wright2009}.}\\
  \multicolumn{7}{l}{$^b$: Iron abundance from \cite{Hatzes2015}.}\\

 \end{tabular}
\end{table*}

As \cite{Gonzalez1997}, \cite{Santos2001} and \cite{FischerValenti2005} reported for the first time, the probability of hosting a giant exoplanet tends to increase with the iron abundance (metallicity) of F-, G-, and K- dwarf stars. 
This relationship has been investigated and confirmed in detail afterwards by many other authors \citep{Gonzalez2006,Guillot2006,Pasquini2007,Ghezzi2010,Johnson2010,Sousa2011,Buchhave2018}.
However, this relationship has not been found to hold for M dwarfs \citep{Laughlin2004,JohnsonApps2009,Hobson2018}.

We searched for confirmed exoplanet discoveries around our 193 primaries using The Extrasolar Planet Encyclopaedia\footnote{\tt \url{http://exoplanet.eu/}}.
The identified planetary systems are listed in Table~\ref{tab:exoplanets}, along with our derived iron abundance and the projected physical separation $s$ between primary and companion. 
The separations between stars in a system is much larger than between star and planet (e.g. $\sim$7000 times in the case of V452 Vul, also known as HD~189733; \citealt{Martin2018}).

We found that 14 of our FGK primaries have, at least, one confirmed exoplanet.
Of them, ten have a derived iron abundance [Fe/H] $\geq$ 0.0, and of these ten, eight have [Fe/H] $\geq$ 0.15, including the five-planet host $\rho^{01}$~Cnc~A.
This new proof of the planet-metallicity relation is illustrated by Fig.~\ref{fig:histfeplanets}. 
As depicted, the detection probability of exoplanets in both \cite{Sousa2011} sample and ours tends to be higher when the iron abundance increases.

We performed a Kolmogorov-Smirnoff test to assess the difference between the stars with and without planets in our sample. At the 2$\sigma$ level, we can safely say that they are significantly different. 
We also repeated the test for the \cite{Sousa2011} sample and found the same result at the same confidence level. 
In addition, we compared our subsamples with \cite{Sousa2011}'s and found no difference between them (with and without planets) at the 2$\sigma$ level confidence, in spite of the iron abundance range of stars with exoplanets in the \cite{Sousa2011} sample (--0.50 $\leq$ [Fe/H] $\leq$ +0.40) being slightly wider than ours (--0.20 $\leq$ [Fe/H] $\leq$ +0.40).

In order to prove the planet-metallicity relation in M dwarfs, we have prepared Table~\ref{tab:target}. It tabulates the {13} most metallic ([Fe/H] $\geq$ 0.16), late-K- and M-dwarf companions of our sample brighter than $J$ = 10.5\,mag and without companion at $\rho$ < 5 arcsec (exactly as in the CARMENES radial-velocity survey -- \citealt{Caballero2016}).
They should be high-priority targets of exoplanet searches, as they have the same iron abundance as their primaries if they were born in the same molecular cloud.
Six late-type dwarfs in Table~\ref{tab:target} are already common proper-motion companions to FGK-type stars with known exoplanets (Table~\ref{tab:exoplanets}).
Interestingly,  of the 13 M dwarfs, only one is being monitored in radial velocity by CARMENES, which sample is unbiased by metallicity or activity \citep{Reiners2018}.

\begin{table}
\caption{Single late-K and M dwarf companions with [Fe/H] $>$ 0.16.}
\label{tab:target}
\begin{tabular}{lllc} 
\hline
\hline
WDS & Late-type & SpType & [Fe/H] \\
 & companion & & \\
\hline
02556+2652 & HD 18143 B & K7\,V & 0.18$\pm$0.05 \\ 
                        & HD 18143 C & M4.0\,V & 0.18$\pm$0.05 \\
03480+4032$^a$ & J03480588+4032226 & M1.5\,V & 0.28$\pm$0.02\\
04429+1843 & HD 285970 & K5\,V & 0.24$\pm$0.02 \\ 
05466+0100$^a$ & HD 38529 B & M2.5\,V & 0.32$\pm$0.02 \\
06332+0528$^a$ & HD 46375 B & M2.0\,V & 0.23$\pm$0.06 \\
07191+6644 & HD 55745 B & M0.0\,V & 0.23$\pm$0.02 \\
08526+2820$^{a,b}$ & $\rho^{01}$ Cnc B & M4.5\,V & 0.29$\pm$0.04 \\
09152+2323$^a$ & BD+23 2063B & M0.0\,V & 0.21$\pm$0.02 \\
10010+3155 & 20 LMi B & M6.0\,V & 0.21$\pm$0.01 \\ 
11218+1811 & HD 98736 & M0.0\,V & 0.30$\pm$0.06 \\
20036+2954$^a$ & HD 190360 B & M4.5\,V & 0.21$\pm$0.02 \\
23104+4901 & HD 218790 & K5\,V & 0.29$\pm$0.01\\ 
\hline
\multicolumn{4}{l}{$^a$: FGK-type primary with confirmed exoplanet. See Table~\ref{tab:exoplanets}.}\\
\multicolumn{4}{l}{$^b$: M dwarf being monitored by CARMENES.}\\
 \end{tabular}
\end{table}

\section{Conclusions}

This is the first item of a series of papers devoted to improve the metallicity calibration and to investigate the abundances of M dwarfs.
For that, we investigate wide binary and multiple benchmark systems containing solar-like primaries and M-dwarf companions. 
Here we present the sample and our first results on physical companionship, stellar parameters, abundances, and kinematics of the primaries.

Here we characterised a sample of 489 stars distributed in 193 binary and multiple candidate systems formed by a late-F-, G-, or early-K primaries and at least late-K- or M-dwarf companion candidate.
For each of them, we compiled or derived coordinates, spectral types, $J$-band magnitudes, proper motions, and heliocentric distances.
After a common proper-motion analysis, we ended up with a sample of 192 binary and multiple physical FGK+M systems.
With HERMES at the 1.2 m Mercator Telescope, we obtained high-resolution optical spectra of 197 stars and, after excluding spectroscopic binaries, fast rotators, and hot and cool stars, we derived stellar atmospheric parameters for 175 primaries and five companions with the the {\sc StePar} code.
We measured effective temperature $T_{\rm eff}$, surface gravity $\log{g}$, microturbulence velocity $\xi$, and photospheric chemical abundances for 13 atomic species, including iron.
For 50 stars we tabulated the first measure of [Fe/H].
We estimated ages for the seven stars with the lowest surface gravity using isochrones for different iron abundances.
We computed Galactocentric space velocities $U$, $V$, and $W$ for the 198 FGK stars, and compared them with the ones published in the literature.
We identified three systems in the Galactic halo, 23 systems in the thick disc, and 33 systems in young stellar kinematic groups (of which 16 are new candidates).
Finally, we studied the presence of exoplanets around our F-, G- and K-type primaries, and provided a list of late-type dwarf companions useful to test the planet-metallicity relation in M dwarfs under the assumption that companions have the same metallicity as primary stars.

Forthcoming papers of this series will focus on the calibration of spectral indices from optical to infrared low-resolution spectra and photometry of M-dwarf companions with the metallicity of their primaries.
We will also derive stellar atmospheric parameters and abundances of the M-dwarf companions with spectral synthesis on high-resolution spectra, and compare the results with the values presented here, which will be very useful for other groups worldwide.

\section*{Acknowledgements}

This research made use of the SIMBAD database and VizieR catalogue access tool, operated at Centre de Donn\'ees astronomiques de Strasbourg, France, the Spanish Virtual Observatory, the NASA's Astrophysics Data System, and the Washington Double Star Catalog maintained at the U.S. Naval Observatory.
Financial support was provided by the {Universidad Complutense de Madrid}, the Comunidad Aut\'onoma de Madrid, the Spanish Ministerio de Econom\'ia y Competitividad (MINECO) and the Fondo Europeo de Desarrollo Regional (FEDER/ERF) under grants 
AYA2016-79425-C3-1/2-P, 
AYA2015-68012-C2-2-P, 
AYA2014-56359-P, 
RYC-2013-14875, 
and FJCI-2014-23001, 
the Conserjer\'ia de Educaci\'on, Juventud y Deporte de la Comunidad de Madrid, and the Fondo Social Europeo y la Iniciativa de Empleo Juvenil (YEI) under grant
PEJD-2016/TIC-2347, 
and the Spanish Ministerio de Educaci\'on, Cultura y Deporte, programa de Formaci\'on de Profesorado Universitario under fellowship FPU15/01476. 
 Finally, we would like to thank the anonymous referee for helpful comments and corrections.





\bibliographystyle{mnras}
\bibliography{Mcomp_FGK_I} 








\appendix

\section{Additional graphs}
\label{sec:graphics}

\begin{figure*}
 	\includegraphics[width=\columnwidth]{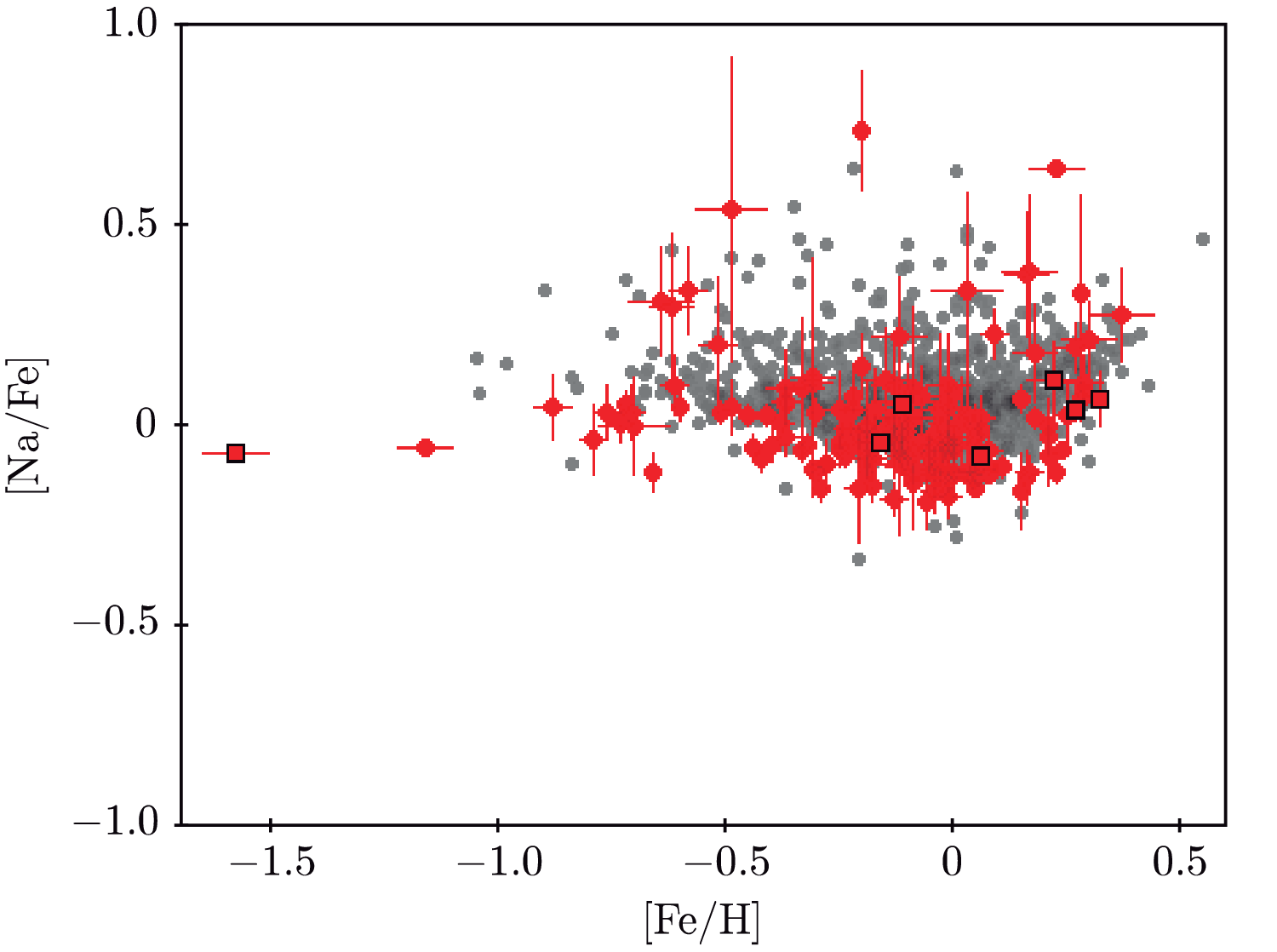}
	 \includegraphics[width=\columnwidth]{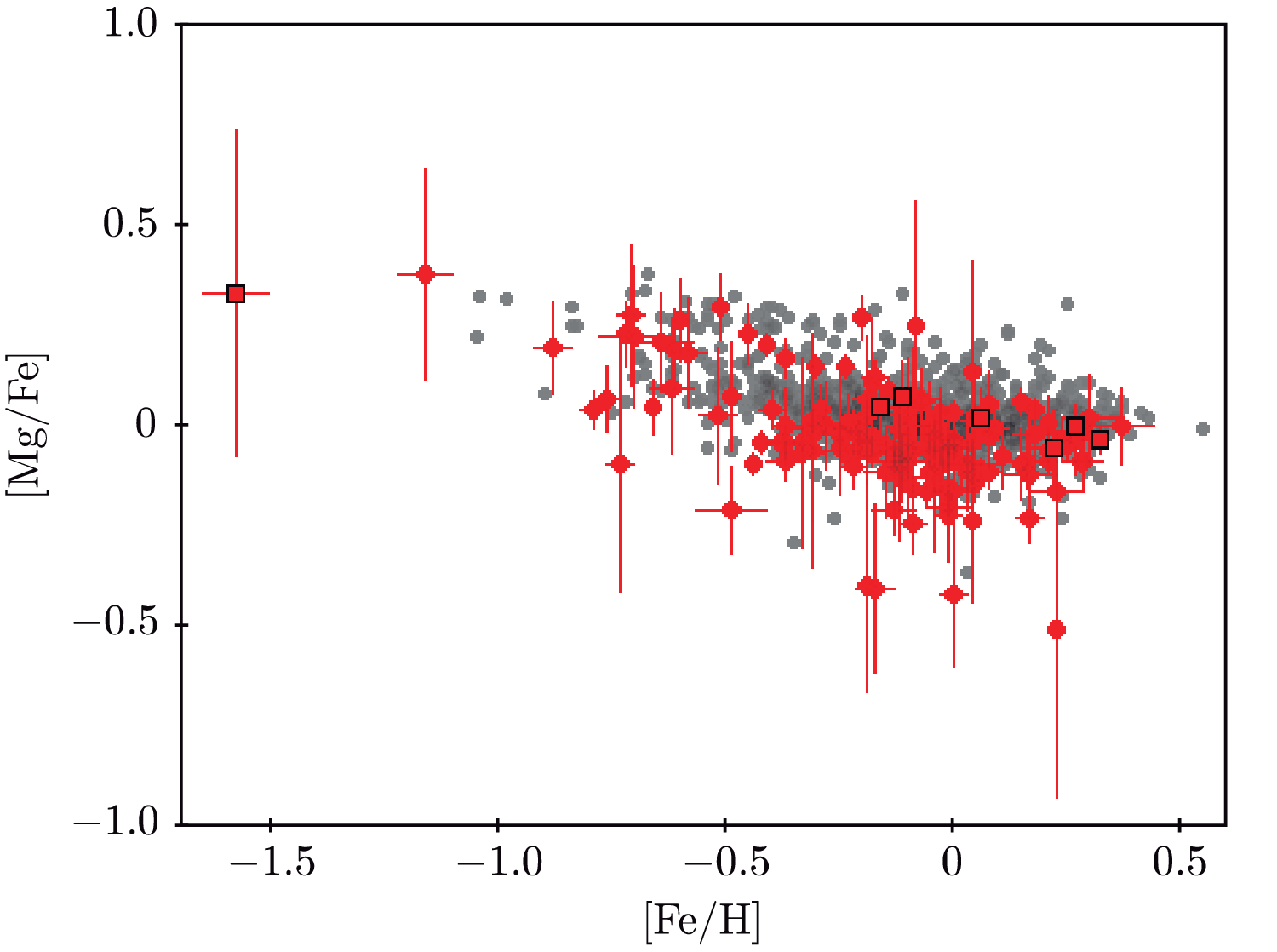}
	\includegraphics[width=\columnwidth]{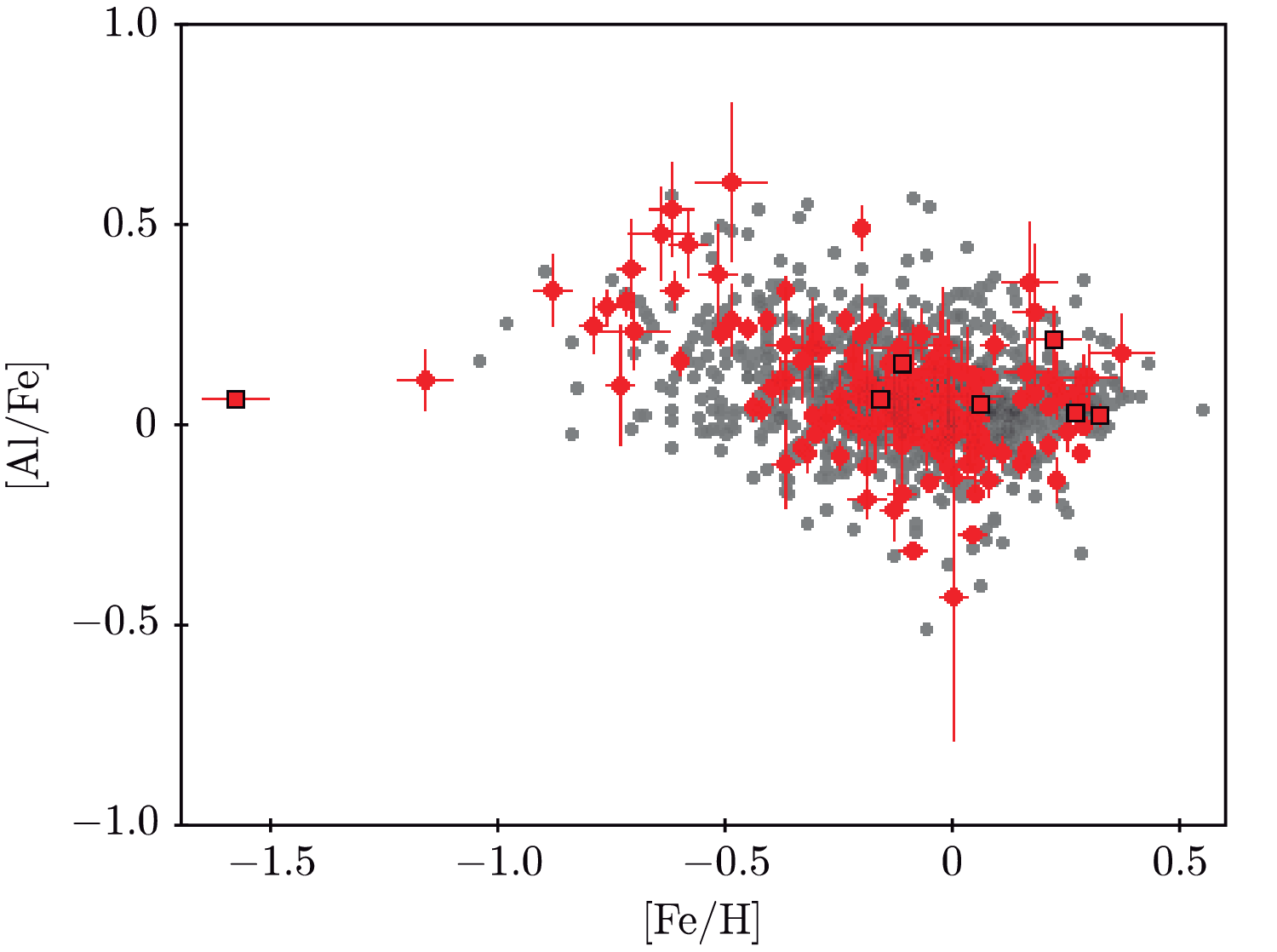}
	\includegraphics[width=\columnwidth]{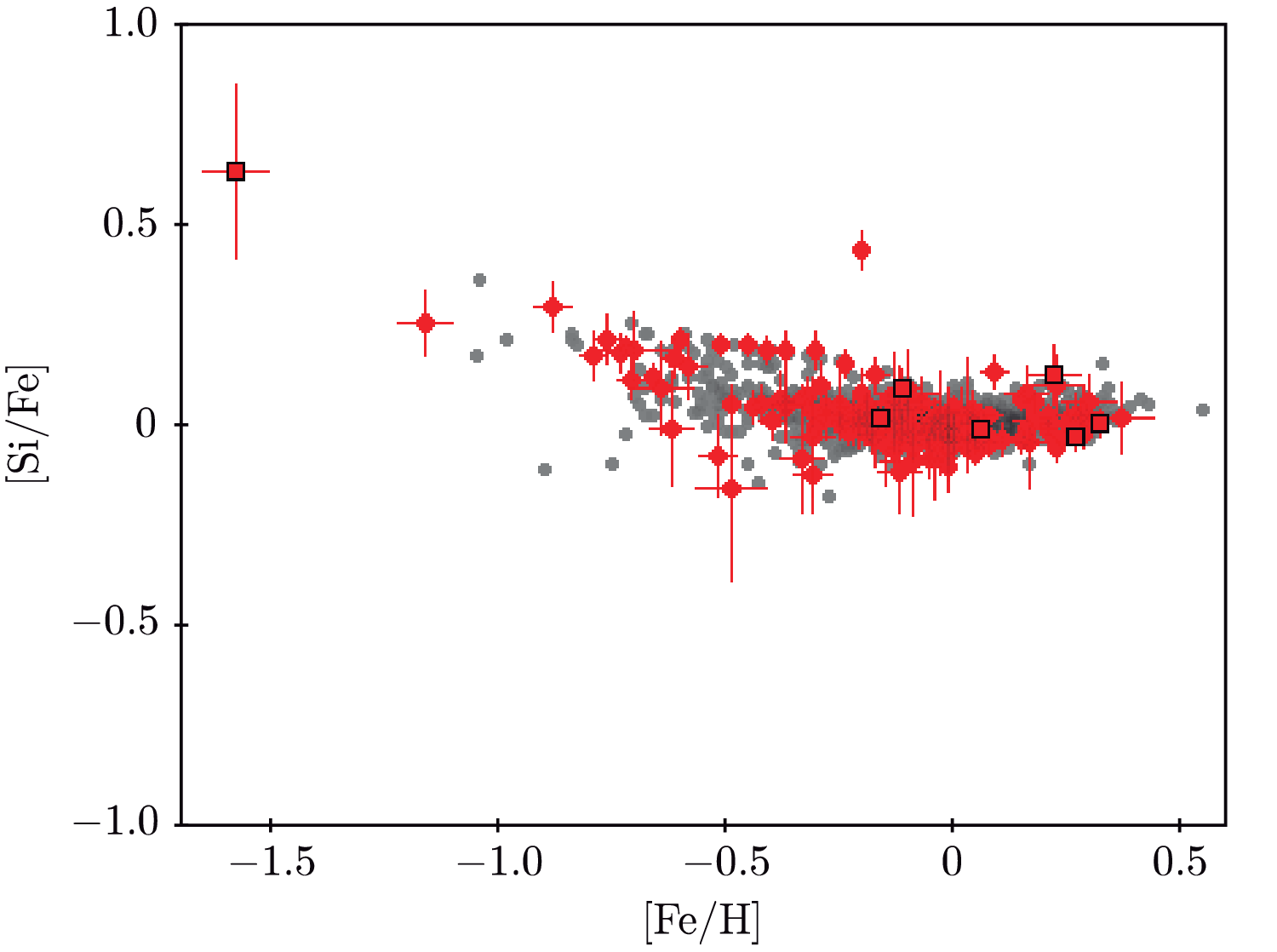}
	\includegraphics[width=\columnwidth]{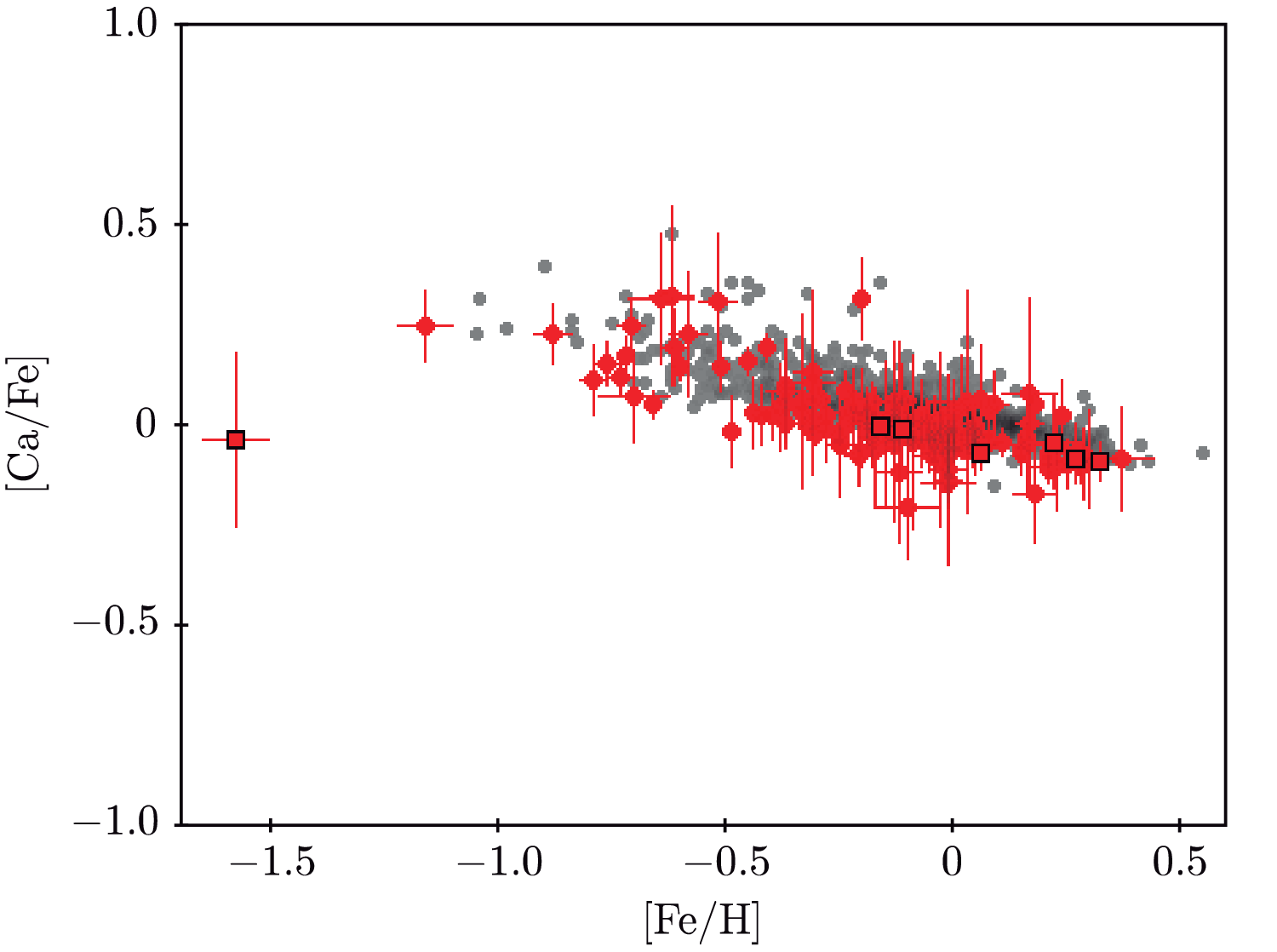}
	\includegraphics[width=\columnwidth]{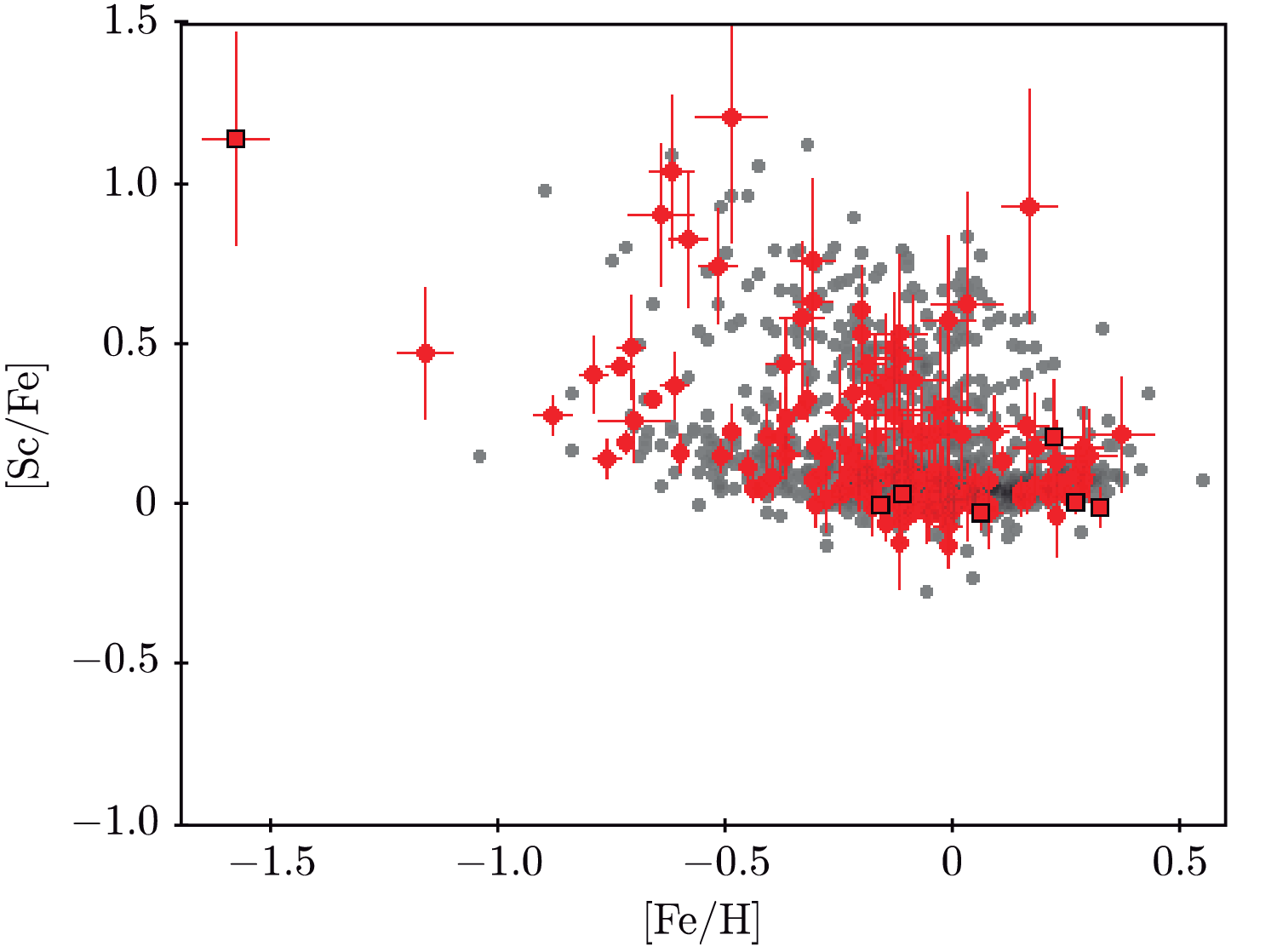}
    \caption{Abundance ratios of [X/Fe] versus [Fe/H], where X = Na, Mg, Al, Si, Ca, and Sc.
    Red circles: our stars; black-ensquared circles: low-gravity stars;
    small grey circles: stars from \citet{Adibekyan2012}.}
\label{fig:abunds-A1}
\end{figure*}

\begin{figure*}
	\includegraphics[width=\columnwidth]{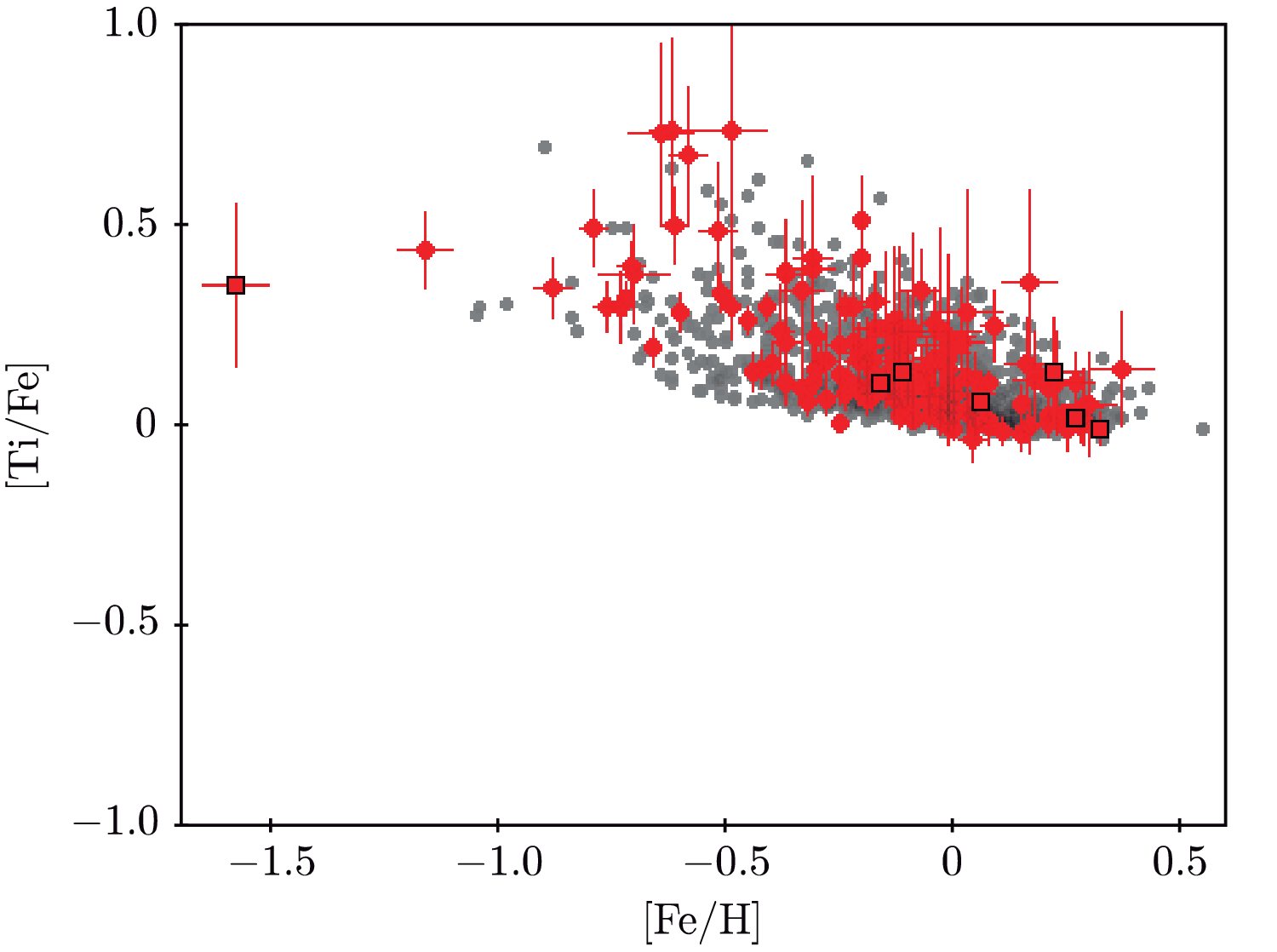}
	\includegraphics[width=\columnwidth]{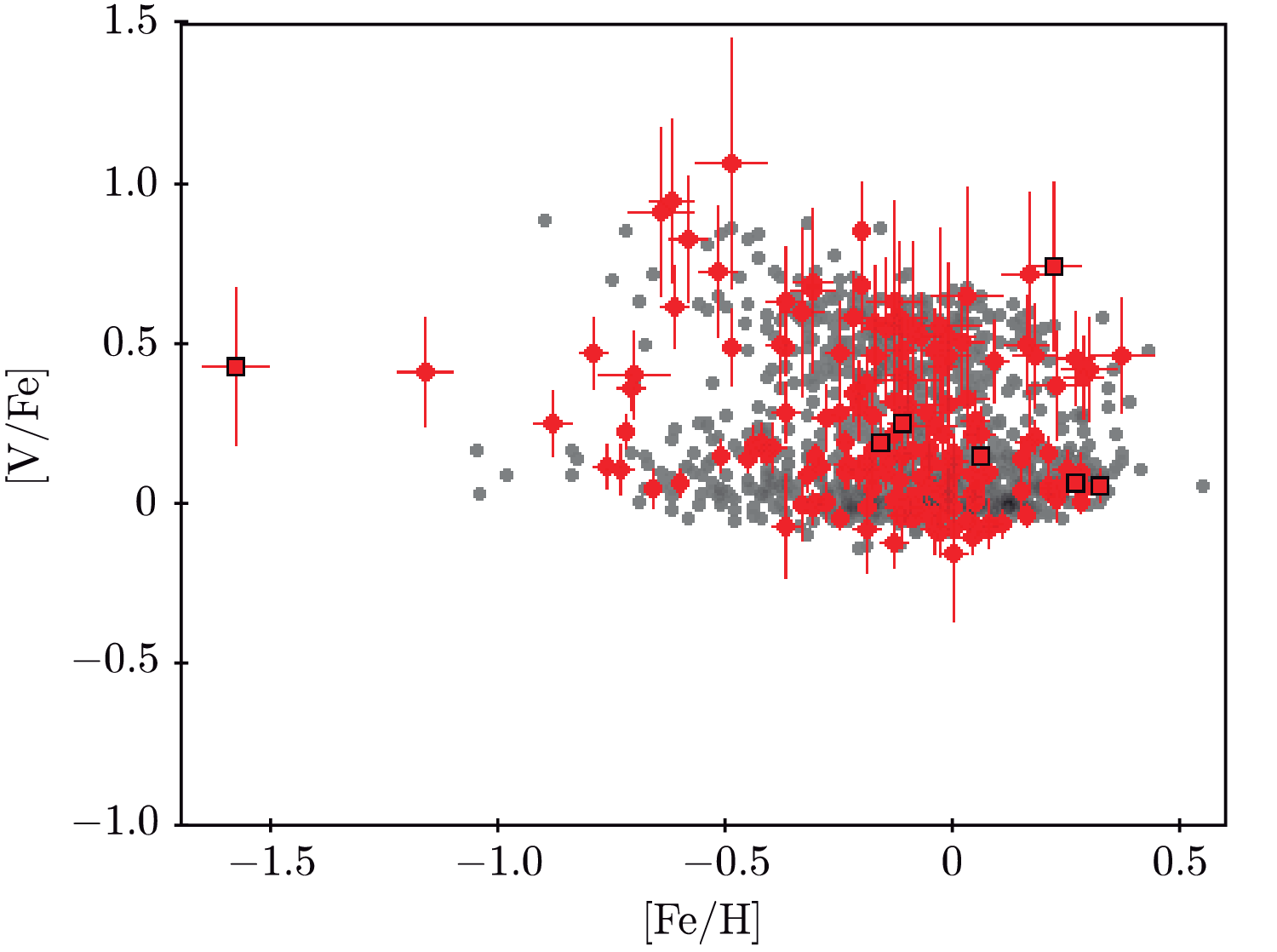}
	\includegraphics[width=\columnwidth]{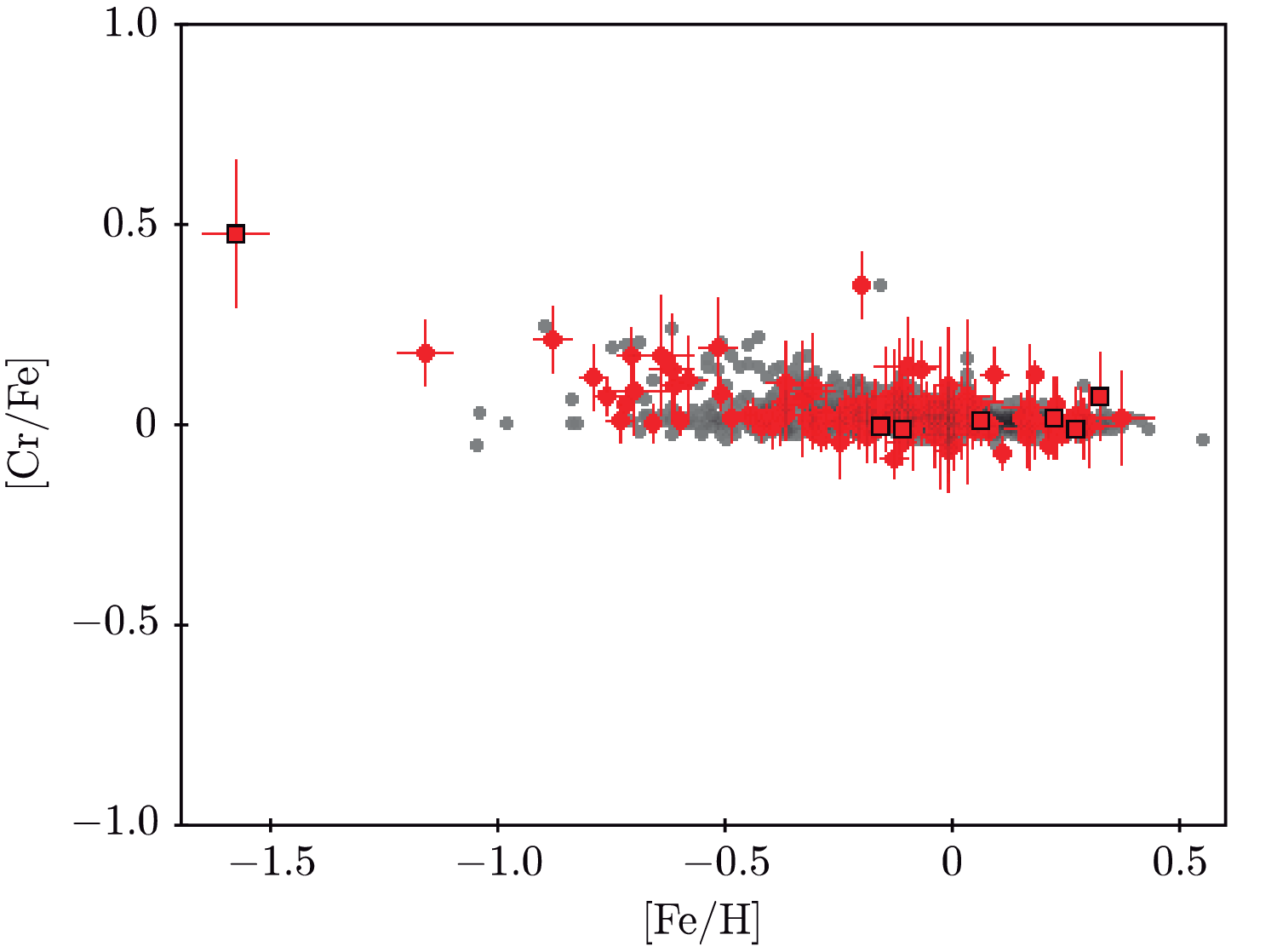}
	\includegraphics[width=\columnwidth]{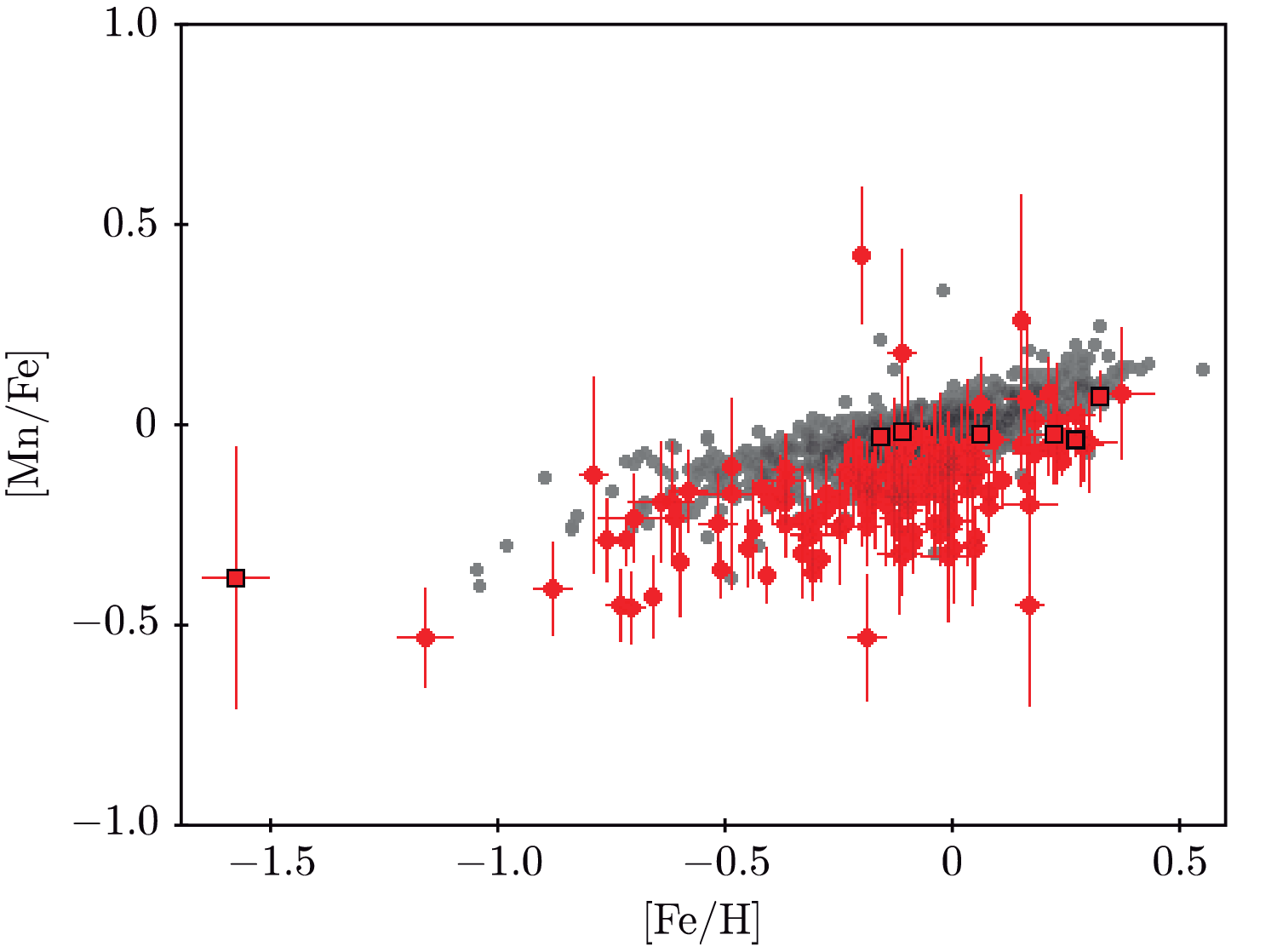}
	\includegraphics[width=\columnwidth]{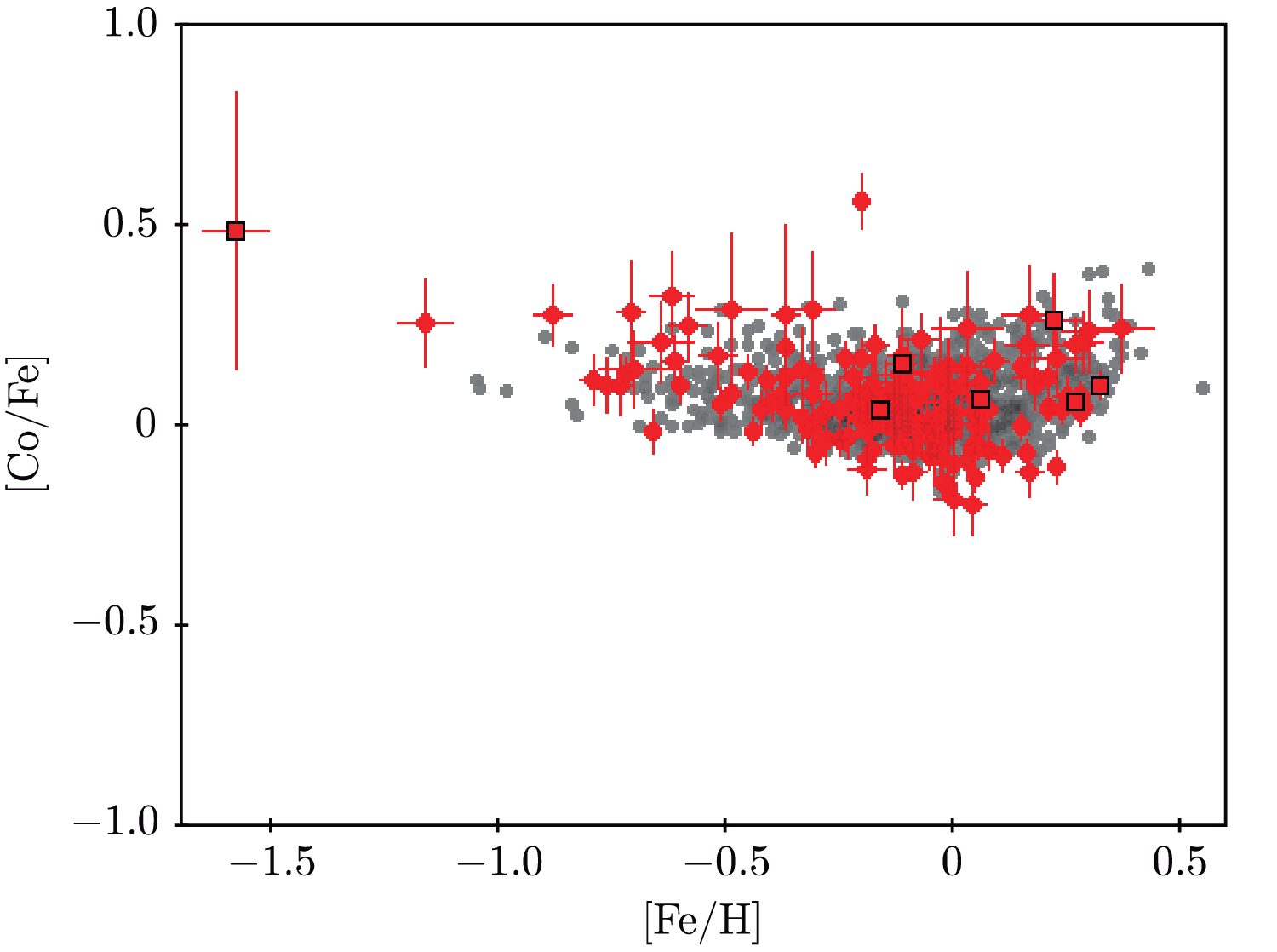}
	\includegraphics[width=\columnwidth]{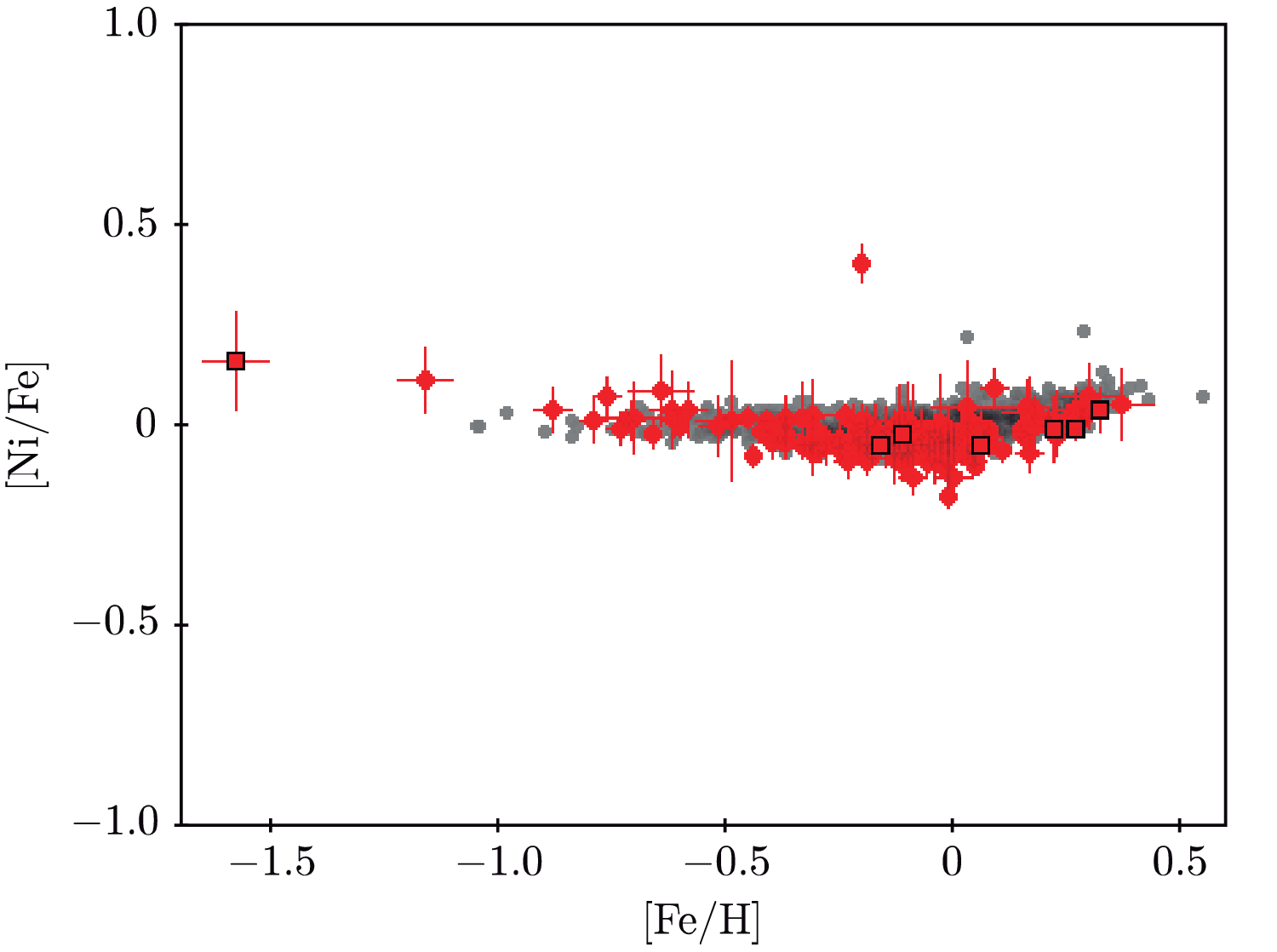}
    \caption{Same as Fig. A1, but for X = Ti, V, Cr, Mn, Co, and Ni.}
\label{fig:abunds-A2}
\end{figure*}

\section{Long tables}
\label{sec:longtables}

\begin{table*}
\caption{Basic properties of investigated systems and stars.}
\label{tab:investigated_stars}
\begin{tabular}{lllcclcccll} 
\hline
\hline
WDS & Comp. & Discoverer & $\rho$ & $\theta$ & Name$^a$  & $\alpha$ & $\delta$ & $J^b$ & Spectral & Obs.$^c$\\
 & & code & [arcsec] & [deg] &  & (J2000) & (J2000) & [mag] & type & \\
\hline
00153+5304 &A & &  &  & G 217-41 & 00:15:14.8 & +53:04:27 & 8.84$\pm$0.02 & K3\,V & H\\
 & B & GIC 5 & 18.90$\pm$0.08 & 354.43$\pm$0.23 & G 217-40 & 00:15:14.6 & +53:04:46 & 10.82$\pm$0.02 & M2.5\,V & C\\
 \noalign{\smallskip}
 00385+4300 &A & & & & BD+42 126  & 00:38:29.2  & +43:00:00 & 8.41$\pm$0.02 & G5\,V & H\\
 &B & LDS 5176 & 53.09$\pm$0.08 & 124.45$\pm$0.02 & LP 193-345 & 00:38:33.2 & +42:59:30 & 10.49$\pm$0.02 & M0.5\,V & C*\\
 \noalign{\smallskip}
00452+0015 &A &  &  &  & HD 4271 Aa,Ab & 00:45:11.0 & +00:15:12 & 5.98$\pm$0.02 & F8\,V & H\\
 &BC & LDS 836 & 55.35$\pm$0.17 & 44.90$\pm$0.24 & HD 4271 B & 00:45:13.6 & +00:15:51 & 10.11$\pm$0.02 & M4.0\,V+ & C\\
\hline
 \end{tabular}
\end{table*}

\begin{table*}
\contcaption{Basic properties of investigated systems and stars.}
\begin{tabular}{lllcclcccll} 
\hline
\hline
WDS & Comp. & Discoverer & $\rho$ & $\theta$ & Name$^a$  & $\alpha$ & $\delta$ & $J^b$ & Spectral & Obs.$^c$\\
 & & code & [arcsec] & [deg] &  & (J2000) & (J2000) & [mag] & type & \\
 \hline
03321+4340 & A &  &  &  & HD 21727 A & 03:32:05.1 & +43:40:12 & 7.34$\pm$0.02 & G5\,V & H\\
 & B& LDS 9155 & 14.50$\pm$0.08 & 136.91$\pm$0.02 & HD 21727 B & 03:32:05.9 & +43:40:01 & 9.24$\pm$0.02 & K7\,V & C\\
 \noalign{\smallskip}
03332+4615 &A & &  &  & V577 Per & 03:33:13.5 & +46:15:27 & 6.84$\pm$0.02 & G5\,V & H\\
 & B& ES 560 & 9.46$\pm$0.08 & 142.28$\pm$0.09 & HD 21845 B & 03:33:14.0 & +46:15:19 & 8.38$\pm$0.03 & M0.0\,V & C\\
 \noalign{\smallskip}
03356+4253 &A & &  &  & HD 22122 & 03:35:35.8 & +42:53:15 & 6.30$\pm$0.02 & F8\,V & H\\
 & B& BUP 45 & 46.02$\pm$0.08 & 212.36$\pm$0.15 & J03353356+4252364 & 03:35:33.6 & +42:52:36 & 12.38$\pm$0.02 & ... & ...\\
\hline
 \end{tabular}
\end{table*}

\begin{table*}
\contcaption{Basic properties of investigated systems and stars.}
\begin{tabular}{lllcclcccll} 
\hline
\hline
WDS & Comp. & Discoverer & $\rho$ & $\theta$ & Name$^a$  & $\alpha$ & $\delta$ & $J^b$ & Spectral & Obs.$^c$\\
 & & code & [arcsec] & [deg] &  & (J2000) & (J2000) & [mag] & type & \\
 \hline
05413+5329 &A &  &  &  & V538 Aur & 05:41:20.3 & +53:28:52 & 4.30$\pm$0.26 & K1\,V & H\\
 &B & ENG 22 & 98.04$\pm$0.13 & 71.24$\pm$0.09 & HD 233153 & 05:41:30.7 & +53:29:23 & 6.59$\pm$0.02 & M1.0\,V & C\\
 & C& BUP 82 & 115.14$\pm$0.13 & 322.68$\pm$0.01 & J05411251+5330239 & 05:41:12.5 & +53:30:24 & 13.25$\pm$0.02 & ... & ...\\
  &D & BUP 82 & 697.49$\pm$0.13 & 282.57$\pm$0.01 & HD 37229 & 05:40:04.0 & +53:31:24 & 8.40$\pm$0.03 & F5 & ...\\
  \noalign{\smallskip}
 05427+0241 & A & & &  & HD 38014 & 05:42:45.8 & +02:40:45 & 6.97$\pm$0.02 & K1\,V & H\\
 & B&LDS 6192 & 57.07$\pm$0.14 & 354.91$\pm$0.13 & G 99-27 & 05:42:45.5 & +02:41:42 & 9.45$\pm$0.03 & M3.0\,V & C\\
\hline
 \end{tabular}
\end{table*}

\begin{table*}
\contcaption{Basic properties of investigated systems and stars.}
\begin{tabular}{lllcclcccll} 
\hline
\hline
WDS & Comp. & Discoverer & $\rho$ & $\theta$ & Name$^a$  & $\alpha$ & $\delta$ & $J^b$ & Spectral & Obs.$^c$\\
 & & code & [arcsec] & [deg] &  & (J2000) & (J2000) & [mag] & type & \\
 \hline
08138+6306& A&  &  &  & HD 67850 & 08:13:45.8 & +63:06:14 & 7.17$\pm$0.02 & G0\,V & H\\
 & B&LDS 2564 & 244.54$\pm$0.10 & 112.74$\pm$0.01 & NLTT 19115 & 08:14:19.0 & +63:04:40 & 9.91$\pm$0.02 & M1.5\,V & C\\
  & C&RAO 60 & 249.36$\pm$0.10 & 109.32$\pm$0.01 & J08142041+6304518 & 08:14:20.4 & +63:04:52 & 14.38$\pm$0.03 & ... & ...\\
 \noalign{\smallskip}
08161+5706& A&  &  &  & HD 68638 & 08:16:06.3 & +57:05:39 & 6.07$\pm$0.02 & G8\,V & H\\
 & B& ENG 34 & 127.21$\pm$0.11 & 145.96$\pm$0.01 & HD 237688 & 08:16:15.1 & +57:03:54 & 8.34$\pm$0.02 & F8 & ...\\
 & C & BUP 113 & 226.89$\pm$0.11 & 150.14$\pm$0.01 & J08162022+5702224$^k$ & 08:16:20.2 & +57:02:23 & 10.34$\pm$0.03 & ... & ... \\
\hline
 \end{tabular}
\end{table*}

\begin{table*}
\contcaption{Basic properties of investigated systems and stars.}
\begin{tabular}{lllcclcccll} 
\hline
\hline
WDS & Comp. & Discoverer & $\rho$ & $\theta$ & Name$^a$  & $\alpha$ & $\delta$ & $J^b$ & Spectral & Obs.$^c$\\
 & & code & [arcsec] & [deg] &  & (J2000) & (J2000) & [mag] & type & \\
 \hline
11218+1811&A &  &  &  & HD 98736 & 11:21:49.3 & +18:11:24 & 6.49$\pm$0.02 & K0\,V & H\\
 & B&STF 1534 & 5.15$\pm$0.14 & 316.36$\pm$0.05 & BD+19 2443B & 11:21:49.1 & +18:11:28 & 7.65$\pm$0.08 & M0.0\,V & C\\
 \noalign{\smallskip}
11378+4150&A &  &  &  & BD+42 2230A & 11:37:50.8 & +41:49:32 & 8.38$\pm$0.03 & G6\,V & H\\
 & B&LDS 5735 & 28.49$\pm$0.18 & 340.37$\pm$0.22 & BD+42 2230B & 11:37:49.9 & +41:50:00 & 11.04$\pm$0.02 & M2.0\,V & C\\
  & C&DAL 24 & 5.68$\pm$0.21 & 172.32$\pm$1.86 & J11375084+4149269 & 11:37:50.8 & +41:49:27 & 10.57 & ... & ...\\
 \noalign{\smallskip}
11403+0931&A &  &  &  & BD+10 2321 & 11:40:16.6 & +09:30:44 & 8.05$\pm$0.02 & K0\,V & H\\
\hline
 \end{tabular}
\end{table*}

\begin{table*}
\contcaption{Basic properties of investigated systems and stars.}
\begin{tabular}{lllcclcccll} 
\hline
\hline
WDS & Comp. & Discoverer & $\rho$ & $\theta$ & Name$^a$  & $\alpha$ & $\delta$ & $J^b$ & Spectral & Obs.$^c$\\
 & & code & [arcsec] & [deg] &  & (J2000) & (J2000) & [mag] & type & \\
 \hline
13321-1115&Aa,Ab &  &  &  &  HD 117676 & 13:32:04.7 & --11:15:23 & 7.55$\pm$0.03 & G8\,V+ & H\\
 &B &TOK 291 & 83.98$\pm$0.13 & 157.60$\pm$0.05 & TYC 5548-829-1 & 13:32:06.9 & --11:16:41 & 9.45$\pm$0.03 & M0.0\,V & C\\
 \noalign{\smallskip}
13470+0621&A &  &  &  & HD 120066 & 13:46:57.1 & +06:21:01 & 5.21$\pm$0.02 & G0\,V & H\\
 &B &LDS 3101 & 488.44$\pm$0.15 & 104.82$\pm$0.22 & BD+07 2692 & 13:47:28.8 & +06:18:56 & 7.76$\pm$0.04 & K7\,V & C\\
 \noalign{\smallskip}
14050+0157&A &  &  &  & HD 122972 & 14:04:58.7 & +01:56:59 & 7.51$\pm$0.03 & G6\,V & H\\
 &B &LDS 5807 & 49.15$\pm$0.16 & 299.34$\pm$0.07 & Ross 799 & 14:04:55.8 & +01:57:23 & 10.13$\pm$0.02 & M2 & ...\\
\hline
 \end{tabular}
\end{table*}

\begin{table*}
\contcaption{Basic properties of investigated systems and stars.}
\begin{tabular}{lllcclcccll} 
\hline
\hline
WDS & Comp. & Discoverer & $\rho$ & $\theta$ & Name$^a$  & $\alpha$ & $\delta$ & $J^b$ & Spectral & Obs.$^c$\\
 & & code & [arcsec] & [deg] &  & (J2000) & (J2000) & [mag] & type & \\
 \hline
16024+0339&Aa,Ab &  &  &  & HD 143809 & 16:02:22.4 & +03:39:07 & 7.74$\pm$0.02 & G0\,V+ & H\\
 &B &CAB 4 & 86.40$\pm$0.11 & 252.57$\pm$0.09 & BD+04 3100s & 16:02:16.9 & +03:38:41 & 10.35$\pm$0.03 & M1.5\,V & C\\
 \noalign{\smallskip}
16048+3910&A &  &  &  & HD 144579 A & 16:04:56.8 & +39:09:23 & 5.18$\pm$0.02 & G8\,V & H\\
 &B &WNO 47 & 70.40$\pm$0.11 & 280.22$\pm$0.07 & HD 144579 B & 16:04:50.9 & +39:09:36 & 9.90$\pm$0.02 & M4.0\,V & C\\
  \noalign{\smallskip}
 16147+3352&A &  &  &  & $\sigma$ CrB Aa,Ab & 16:14:40.9 & +33:51:31 & 3.95$\pm$0.98 & F6\,V+ & ...\\
 &B &STF 2032 & 6.25$\pm$0.19 & 236.40$\pm$2.29 & $\sigma$ CrB B & 16:14:40.4 & +33:51:27 & 6.83 & G1\,V & H\\
\hline
 \end{tabular}
\end{table*}

\begin{table*}
\contcaption{Basic properties of investigated systems and stars.}
\begin{tabular}{lllcclcccll} 
\hline
\hline
WDS & Comp. & Discoverer & $\rho$ & $\theta$ & Name$^a$  & $\alpha$ & $\delta$ & $J^b$ & Spectral & Obs.$^c$\\
 & & code & [arcsec] & [deg] &  & (J2000) & (J2000) & [mag] & type & \\
 \hline
19510+1025$^{o}$&A &  &  & & $o$ Aql A & 19:51:01.6 & +10:24:57 & 4.23$\pm$0.32 & F8\,V & H\\
 &B &J 124 & 17.65$\pm$0.31 & 243.55$\pm$1.35 & ``$o$ Aql B'' & 19:51:00.6 & +10:24:48 & 10.31$\pm$0.04 & M0.0\,V & C*\\
&C &J 124 & 22.02$\pm$0.31 & 220.89$\pm$1.15 & $o$ Aql C & 19:51:00.7 & +10:24:40 & 8.89$\pm$0.03 & M3.5\,V & C*\\
& D & POP 1228 & 52.92$\pm$0.31 & 121.13$\pm$0.11 & J19510473+1024293 & 19:51:04.7 & +10:24:29 & 8.38$\pm$0.02 & ... & ... \\
& E & POP 1228 & 84.75$\pm$0.31 & 146.99$\pm$0.06 & J19510479+1023456 & 19:51:04.8 & +10:23:46 & 9.78$\pm$0.02 & ... & ... \\
\noalign{\smallskip}
19553+0624&A &  &  &  & $\beta$ Aql A & 19:55:18.8 & +06:24:24 & 2.29$\pm$0.25 & G9.5\,IV & H\\
\hline
  \end{tabular}
\end{table*}

 \begin{table*}
\contcaption{Basic properties of investigated systems and stars.}
\begin{tabular}{lllcclcccll} 
\hline
\hline
WDS & Comp. & Discoverer & $\rho$ & $\theta$ & Name$^a$  & $\alpha$ & $\delta$ & $J^b$ & Spectral & Obs.$^c$\\
 & & code & [arcsec] & [deg] &  & (J2000) & (J2000) & [mag] & type & \\
\hline
22311+4509&A &  &  &  & HD 213519 A & 22:31:05.7 & +45:08:42 & 6.47$\pm$0.02 & G5 & H\\
 &BC &LEP 108 & 62.24$\pm$0.08 & 7.62$\pm$0.08 & HD 213519 B & 22:31:06.5 & +45:09:44 & 10.34$\pm$0.02 & M3+ & ...\\
 & D & RAO 29 & 61.58$\pm$0.09 & 354.10$\pm$0.08 & J22310515+4509435 & 22:31:05.2 & +45:09:44 & 14.97$\pm$0.04 & ... & ... \\
 \noalign{\smallskip}
22467+1210&A &  &  &  & $\xi$ Peg A & 22:46:41.6 & +12:10:22 & 3.36$\pm$0.25 & F6\,V & H\\
 &B &HJ 301 & 11.19$\pm$0.30 & 97.08$\pm$1.38 & $\xi$ Peg B & 22:46:42.3 & +12:10:21 & 7.94$\pm$0.02 & M1.5 & C*\\
 &C &HJ 301 & 176.99$\pm$0.30 & 5.98$\pm$0.11 & BPS CS 30332-0037 & 22:46:42.8 & +12:13:19 & 11.73$\pm$0.02 &  & ...\\
\end{tabular}
\end{table*}

\begin{table*}
\caption{Distances, proper motions, and remarks.}
\label{tab:pms}
\begin{tabular}{llclccll} 
\hline
\hline
WDS & Name & $d$ & Ref. & $\mu_\alpha \cos{\delta}$ & $\mu_{\delta}$ & Ref. & Remarks\\
 &  & [pc] &  & [mas\,yr$^{-1}$] & [mas\,yr$^{-1}$] &  & \\
\hline

00153+5304 & G 217-41 & 75.7$\pm$1.7 & TGAS & +222.00$\pm$0.12 & +45.89$\pm$0.10 & TGAS & \\
& G 217-40 & 57.0$\pm$8.6 & This work & +213.5$\pm$2.4 & +55.5$\pm$2.4 & HSOY & \\
 \noalign{\smallskip}
00385+4300 & BD+42 126 & 52.91$\pm$0.62 & TGAS & +188.61$\pm$0.14 & --81.45$\pm$0.14 & TGAS & \\
& LP 193-345 & 77.3$\pm$6.2 & This work & +194.00$\pm$8.00 & --79.00$\pm$8.00 & UCAC4 & Wrong PPMXL $\mu$\\
\noalign{\smallskip}
00452+0015 & HD 4271 Aa,Ab & 41.37$\pm$0.94 & HIP2 & +266.80$\pm$0.17 & --51.19$\pm$0.17 & HSOY & SB1\\
& HD 4271 B & 24.1$\pm$4.9 & This work & +265.6$\pm$1.9 & --53.4$\pm$1.9 & HSOY & WDS close\\
\hline
\end{tabular}
\end{table*}

\begin{table*}
\contcaption{Distances, proper motions, and remarks.}
\begin{tabular}{llclccll} 
\hline
\hline
WDS & Name & $d$ & Ref. & $\mu_\alpha \cos{\delta}$ & $\mu_{\delta}$ & Ref. & Remarks\\
 &  & [pc] &  & [mas\,yr$^{-1}$] & [mas\,yr$^{-1}$] &  & \\
\hline
03321+4340 & HD 21727 A & 54.50$\pm$0.74 & TGAS & +298.57$\pm$0.12 & --119.53$\pm$0.09 & TGAS & \\
& HD 21727 B & 50.8$\pm$1.4 & This work & +299.00$\pm$8.00 & --119.00$\pm$8.00 & UCAC4 & \\
\noalign{\smallskip}
03332+4615 & V577 Per & 36.30$\pm$0.32 & TGAS & +68.59$\pm$0.09 & --175.38$\pm$0.06 & TGAS & \\
& HD 21845 B & 31.4$\pm$2.0 & This work & +64.7$\pm$4.0 & --172.4$\pm$3.6 & UCAC4 & \\
\noalign{\smallskip}
03356+4253 & HD 22122 & 68.2$\pm$1.4 & TGAS & +147.44$\pm$0.37 & --155.06$\pm$0.35 & TGAS & \\
& J03353356+4252364 & ... & ... & --1.7$\pm$2.2 & --3.3$\pm$2.2 & HSOY & No common $\mu$\\
\hline
\end{tabular}
\end{table*}

\begin{table*}
\contcaption{Distances, proper motions, and remarks.}
\begin{tabular}{llclccll} 
\hline
\hline
WDS & Name & $d$ & Ref. & $\mu_\alpha \cos{\delta}$ & $\mu_{\delta}$ & Ref. & Remarks\\
 &  & [pc] &  & [mas\,yr$^{-1}$] & [mas\,yr$^{-1}$] &  & \\
\hline
05413+5329 & V538 Aur & 12.28$\pm$0.08 & HIP2 & +1.82$\pm$0.48 & --523.99$\pm$0.31 & HIP2 & \\
& HD 233153 & 12.44$\pm$0.26 & HIP2 & +3.38$\pm$0.99 & --515.2$\pm$1.2 & HSOY & \\
& J05411251+5330239 & ... & ... & --6.5$\pm$2.2 & --2.0$\pm$2.2 & HSOY & No common $\mu$\\
& HD 37229 & ... & ... & +8.49$\pm$0.32 & --32.41$\pm$0.41 & HSOY & No common $\mu$\\
\noalign{\smallskip}
05427+0241 & HD 38014 & 32.39$\pm$0.33 & TGAS & +253.82$\pm$0.09 & --526.59$\pm$0.07 & TGAS & \\
& G 99-27 & 25.8$\pm$4.4 & This work & +243.8$\pm$2.3 & --530.6$\pm$2.4 & HSOY & \\
\hline
\end{tabular}
\end{table*}

\begin{table*}
\contcaption{Distances, proper motions, and remarks.}
\begin{tabular}{llclccll} 
\hline
\hline
WDS & Name & $d$ & Ref. & $\mu_\alpha \cos{\delta}$ & $\mu_{\delta}$ & Ref. & Remarks\\
 &  & [pc] &  & [mas\,yr$^{-1}$] & [mas\,yr$^{-1}$] &  & \\
\hline
08138+6306 & HD 67850 & 44.39$\pm$0.53 & TGAS & +143.69$\pm$0.06 & +112.23$\pm$0.05 & TGAS & \\
& NLTT 19115 & 48.8$\pm$5.6 & This work & +142.7$\pm$2.5 & +110.6$\pm$2.5 & HSOY & \\
& J08142041+6304518 & ... & ... & +3.8$\pm$2.3 & --4.2$\pm$2.3 & HSOY & No common $\mu$. No Simbad entry\\
\noalign{\smallskip}
08161+5706 & HD 68638 & 32.21$\pm$0.34 & TGAS & --314.98$\pm$0.04 & --222.58$\pm$0.03 & TGAS & \\
& HD 237688 & 177.6$\pm$7.9 & TGAS & +34.49$\pm$0.08 & +7.66$\pm$0.07 & TGAS & No common $\mu$, $d$ \\
& J08162022+5702224 & ... & ... & --4.4$\pm$2.1 & --4.7$\pm$2.1 & HSOY & No common $\mu$\\
\hline
\end{tabular}
\end{table*}

\begin{table*}
\contcaption{Distances, proper motions, and remarks.}
\begin{tabular}{llclccll} 
\hline
\hline
WDS & Name & $d$ & Ref. & $\mu_\alpha \cos{\delta}$ & $\mu_{\delta}$ & Ref. & Remarks\\
 &  & [pc] &  & [mas\,yr$^{-1}$] & [mas\,yr$^{-1}$] &  & \\
\hline
\noalign{\smallskip}
11218+1811 & HD 98736 & 32.26$\pm$0.25 & TGAS & --151.67$\pm$0.12 & --92.68$\pm$0.08 & TGAS & \\
& BD+19 2443B & 22.5$\pm$1.6 & This work & --151.1$\pm$2.5 & --95.7$\pm$2.5 & UCAC4 & \\
\noalign{\smallskip}
11378+4150 & BD+42 2230A & 71.1$\pm$3.1 & TGAS & +64.33$\pm$0.07 & --189.57$\pm$0.08 & TGAS & \\
& BD+42 2230B & 72.5$\pm$9.6 & This work & +58.7$\pm$2.2 & --190.2$\pm$2.2 & HSOY & \\
& J11375084+4149269 & ... & ... & ... & ... & ... & Not enough information. No Simbad entry\\
\noalign{\smallskip}
11403+0931 & BD+10 2321 & 43.4$\pm$3.5 & HIP2 & --148.2$\pm$1.8 & --94.3$\pm$1.4 & HIP2 & \\
\hline
\end{tabular}
\end{table*}

\begin{table*}
\contcaption{Distances, proper motions, and remarks.}
\begin{tabular}{llclccll} 
\hline
\hline
WDS & Name & $d$ & Ref. & $\mu_\alpha \cos{\delta}$ & $\mu_{\delta}$ & Ref. & Remarks\\
 &  & [pc] &  & [mas\,yr$^{-1}$] & [mas\,yr$^{-1}$] &  & \\
\hline
13321-1115 & HD 117676 & 53.22$\pm$0.68 & TGAS & --33.46$\pm$0.09 & --81.02$\pm$0.06 & TGAS & WDS close\\
& TYC 5548-829-1 & 52.2$\pm$1.7 & TGAS & --24.1$\pm$3.5 & --78.72$\pm$0.56 & TGAS & \\
\noalign{\smallskip}
13470+0621 & HD 120066 & 31.67$\pm$0.44 & HIP2 & --509.71$\pm$0.37 & --110.51$\pm$0.25 & HIP2 & \\
& BD+07 2692 & 30.8$\pm$1.5 & HIP2 & --508.72$\pm$0.82 & --110.91$\pm$0.78 & HSOY & \\
\noalign{\smallskip}
14050+0157 & HD 122972 & 51.76$\pm$0.72 & TGAS & --209.39$\pm$0.08 & +5.04$\pm$0.06 & TGAS & \\
& Ross 799 & 47.7$\pm$6.4 & This work & --210.1$\pm$2.4 & +5.8$\pm$2.4 & HSOY & \\
\hline
\end{tabular}
\end{table*}

\clearpage

\begin{table*}
\contcaption{Distances, proper motions, and remarks.}
\begin{tabular}{llclccll} 
\hline
\hline
WDS & Name & $d$ & Ref. & $\mu_\alpha \cos{\delta}$ & $\mu_{\delta}$ & Ref. & Remarks\\
 &  & [pc] &  & [mas\,yr$^{-1}$] & [mas\,yr$^{-1}$] &  & \\
\hline
16024+0339 & HD 143809 & 80.7$\pm$2.1 & TGAS & --48.09$\pm$0.09 & --41.74$\pm$0.07 & TGAS & WDS close\\
& BD+04 3100s & 59.7$\pm$6.9 & This work & --49.2$\pm$3.6 & --43.6$\pm$3.7 & HSOY & \\
\noalign{\smallskip}
16048+3910 & HD 144579 A & 14.38$\pm$0.05 & TGAS & --570.79$\pm$0.04 & +52.65$\pm$0.04 & TGAS & \\
& HD 144579 B & 13.79$\pm$0.65 & vAl95 & --547.0$\pm$8.0 & +55.0$\pm$8.0 & UCAC4 & \\
\noalign{\smallskip}
16147+3352 & $\sigma$ CrB Aa,Ab & 22.71$\pm$0.47 & TGAS & --267.53$\pm$0.05 & --86.68$\pm$0.07 & TGAS & SB2\\
& $\sigma$ CrB B & 21.08$\pm$0.54 & HIP2 & --292.9$\pm$1.2 & --80.7$\pm$1.1 & HSOY & \\
\hline
\end{tabular}
\end{table*}

\clearpage

\begin{table*}
\contcaption{Distances, proper motions, and remarks.}
\begin{tabular}{llclccll} 
\hline
\hline
WDS & Name & $d$ & Ref. & $\mu_\alpha \cos{\delta}$ & $\mu_{\delta}$ & Ref. & Remarks\\
 &  & [pc] &  & [mas\,yr$^{-1}$] & [mas\,yr$^{-1}$] &  & \\
\hline
19510+1025 & $o$ Aql A & 19.19$\pm$0.11 & HIP2 & +241.64$\pm$0.13 & --137.11$\pm$0.13 & HSOY & \\
& ``$o$ Aql B'' & 76.4$\pm$4.8 & This work & --6.4$\pm$6.1 & --15$\pm$14 & This work & No common $\mu, d$\\
& $o$ Aql C & 16.7$\pm$3.1 & This work & +261.0$\pm$2.9 & --159.9$\pm$2.8 & UCAC4 & \\
& J19510473+1024293 & ... & ... & +5.1$\pm$4.6 & --3.0$\pm$4.6 & HSOY & No common $\mu$. No Simbad entry \\
& J19510479+1023456 & ... & ... & +1.91$\pm$0.95 & --9.36$\pm$0.95 & HSOY & No common $\mu$. No Simbad entry \\
\noalign{\smallskip}
19553+0624 & $\beta$ Aql A & 13.70$\pm$0.04 & HIP2 & +45.27$\pm$0.18 & --481.90$\pm$0.20 & HIP2 & \\
\hline
\end{tabular}
\end{table*}

\clearpage

\begin{table*}
\contcaption{Distances, proper motions, and remarks.}
\begin{tabular}{llclccll} 
\hline
\hline
WDS & Name & $d$ & Ref. & $\mu_\alpha \cos{\delta}$ & $\mu_{\delta}$ & Ref. & Remarks\\
 &  & [pc] &  & [mas\,yr$^{-1}$] & [mas\,yr$^{-1}$] &  & \\
\hline
22311+4509 & HD 213519 A & 40.72$\pm$0.40 & TGAS & --175.51$\pm$0.04 & +34.94$\pm$0.05 & TGAS & \\
& HD 213519 B & 38.9$\pm$6.6 & This work & --179.0$\pm$1.9 & +45.5$\pm$2.0 & HSOY & \\
& J22310515+4509435 & ... & ... & --7.7$\pm$2.2 & --3.5$\pm$2.2 & HSOY & No common $\mu$. No Simbad entry \\
\noalign{\smallskip}
22467+1210 & $\xi$ Peg A & 16.30$\pm$0.05 & HIP2 & +234.18$\pm$0.21 & --493.29$\pm$0.17 & HIP2 & \\
& $\xi$ Peg B & 19.7$\pm$2.3 & This work & +233 & --453 & This work & \\
& BPS CS 30332-0037 & ... & ... & --3.55$\pm$0.64 & --20.03$\pm$0.64 & HSOY & No common $\mu$ \\
\end{tabular}
\end{table*}

\newpage

\begin{table*}
\caption{Stellar atmospheric parameters of late-F, G-, and early-K type stars.}
\label{tab:parameters}
\begin{tabular}{llccccccc} 
\hline
\hline
WDS & Name & SNR or & \textit{T$_{\rm eff}$} & $\log{g}$ & $\xi$ & [Fe/H] & [Fe/H]$_{\rm lit}$ & Ref.$^b$\\
 &  & type$^a$ & [K] & & [km\,s$^{-1}$] &  &  & \\
\hline

00153+5304 & G 217-41 & 66 & 5160$\pm$73 & 4.43$\pm$0.17 & 0.59$\pm$0.20 & 0.03$\pm$0.05 & 0.07$\pm$0.03 & Man13\\
00385+4300 & BD+42 126 & 73 & 5057$\pm$60 & 4.55$\pm$0.17 & 0.26$\pm$0.28 & --0.70$\pm$0.08 & .. & ...\\
00452+0015 & HD 4271 Aa,Ab & 152 & 6273$\pm$34 & 4.52$\pm$0.09 & 1.29$\pm$0.04 & 0.08$\pm$0.03 & 0.00$\pm$0.02 & Ram14\\
00467-0426 & HD 4449 & 69 & 5613$\pm$94 & 4.70$\pm$0.22 & 0.96$\pm$0.20 & --0.10$\pm$0.07 & ... & ...\\
00491+5749 & Achird Aa,Ab & 242 & 5907$\pm$15 & 4.41$\pm$0.05 & 1.03$\pm$0.03 & --0.25$\pm$0.01 & --0.25 & VF05\\
01055+1523 & HD 6440 A & 68 & 4704$\pm$68 & 4.41$\pm$0.19 & 0.63$\pm$0.26 & --0.25$\pm$0.03 & --0.09$\pm$0.03 & Man13\\
\hline
\end{tabular}
\end{table*}

\begin{table*}
\contcaption{Stellar atmospheric parameters of late-F, G-, and early-K type stars.}
\begin{tabular}{llccccccc} 
\hline
\hline
WDS & Name & SNR or & \textit{T$_{\rm eff}$} & $\log{g}$ & $\xi$ & [Fe/H] & [Fe/H]$_{\rm lit}$ & Ref.$^b$\\
 &  & type$^a$ & [K] & & [km\,s$^{-1}$] &  &  & \\
\hline
07191+6644 & HD 55745 A & 78 & 6238$\pm$27 & 4.56$\pm$0.07 & 1.31$\pm$0.04 & 0.23$\pm$0.02 & 0.21$\pm$0.03 & Man13 \\
07321-0853 & HD 59984 & 167 & 5985$\pm$36 & 4.22$\pm$0.08 & 1.34$\pm$0.06 & --0.66$\pm$0.02 & --0.69$\pm$0.02& Sou11 \\
07400-0336 & V869 Mon & 96 & 4918$\pm$63 & 4.36$\pm$0.06 & 0.80$\pm$0.17 & --0.11$\pm$0.03 & 0.01$\pm$0.08 & San05 \\ 
08082+2106 & BD+21 1764A & 38 & 4719$\pm$197 & 4.69$\pm$0.58 & 1.07$\pm$0.71 & --0.45$\pm$0.08 & ... & ... \\
08082+7155 & HD 66171 & 126 & 5735$\pm$28 & 4.30$\pm$0.07 & 0.87$\pm$0.04 & --0.28$\pm$0.02 & --0.29$\pm$0.03 & Ram13\\
08107-1348 & 18 Pup A & 197 & 6414$\pm$51 & 4.65$\pm$0.09 & 1.70$\pm$0.08 & --0.01$\pm$0.03 & --0.07$\pm$0.07 & Fuh08\\
\hline
\end{tabular}
\end{table*}

\begin{table*}
\contcaption{Stellar atmospheric parameters of late-F, G-, and early-K type stars.}
\begin{tabular}{llccccccc} 
\hline
\hline
WDS & Name & SNR or & \textit{T$_{\rm eff}$} & $\log{g}$ & $\xi$ & [Fe/H] & [Fe/H]$_{\rm lit}$ & Ref.$^b$\\
 &  & type$^a$ & [K] & & [km\,s$^{-1}$] &  &  & \\
\hline
14260+3422 & BD+35 2558 & 64 & 5018$\pm$55 & 4.59$\pm$0.18 & 0.38$\pm$0.25 & --0.79$\pm$0.03 & ... & ... \\
14336+0920 & HD 127871 A & 83 & 4942$\pm$51 & 4.30$\pm$0.14 & 0.45$\pm$0.18 & --0.11$\pm$0.03 & ... & ... \\
14415+1336 & HD 129290 A & 130 & 5848$\pm$20 & 4.32$\pm$0.05 & 0.96$\pm$0.03 & --0.12$\pm$0.02 & --0.15$\pm$0.03 & Ram13\\
14446-2215 & HD 129715 & 38 & 4831$\pm$123 & 4.40$\pm$0.32 & 0.83$\pm$0.29 & 0.17$\pm$0.06 & ... & ...\\
14493+4950 & HD 130986 A & 112 & 6043$\pm$26 & 4.41$\pm$0.06 & 1.05$\pm$0.03 & 0.05$\pm$0.02 & ... & ...\\
14575-2125 & HD 131977 & 70 & 4772$\pm$103 & 4.38$\pm$0.27 & 0.89$\pm$0.35 & --0.12$\pm$0.06 & 0.07$\pm$0.10 & San05\\
\hline
\end{tabular}
\end{table*}

\begin{table*}
\contcaption{Stellar atmospheric parameters of late-F, G-, and early-K type stars.}
\begin{tabular}{llccccccc} 
\hline
\hline
WDS & Name & SNR or & \textit{T$_{\rm eff}$} & $\log{g}$ & $\xi$ & [Fe/H] & [Fe/H]$_{\rm lit}$ & Ref.$^b$\\
 &  & type$^a$ & [K] & & [km\,s$^{-1}$] &  &  & \\
\hline
22589+6902 & BD+68 1345A & 106 & 5277$\pm$38 & 4.36$\pm$0.10 & 0.56$\pm$0.10 & --0.01$\pm$0.03 & ... & ... \\
23026+2948 & BD+29 4841Aa,Ab & SB2 & ... & ... & ... & ... & ... & ...\\
23104+4901 & HD 218790 & 113 & 6008$\pm$17 & 4.40$\pm$0.05 & 1.14$\pm$0.03 & 0.29$\pm$0.01 & ... & ...\\
23194+7900 & V368 Cep & Fast & ... & ... & ... & ... & -0.03$\pm$0.05 & Mis12\\
23235+4548 & HD 220445 & 77 & 5064$\pm$74 & 4.37$\pm$0.19 & 0.70$\pm$0.22 & 0.02$\pm$0.05 & ... & ...\\
23266+4520 & HD 220821 & 140 & 5721$\pm$22 & 4.27$\pm$0.06 & 0.94$\pm$0.03 & --0.18$\pm$0.02 & --0.22$\pm$0.04 & Ram12 \\
\end{tabular}
\end{table*}

\begin{landscape}
\begin{table}
\caption{Abundances [X/H] with respect to the Sun for the four $\alpha$-elements (Mg, Si, Ca and Ti), the Fe-peak elements (Cr, Mn, Co, and Ni), and the odd-Z elements (Na, Al, Sc, and V).}
\fontsize{8}{8}\selectfont
\label{tab:abundances}
\begin{tabular}{llcccccccccccc}
\hline
\hline
WDS & Name & [Na/H] &  [Mg/H] &  [Al/H] &  [Si/H] & [Ca/H] &  [Sc/H] & [Ti/H] & [V/H] & [Cr/H] & [Mn/H] & [Co/H] & [Ni/H] \\
\hline
00153+5304 & G 217-41 & 0.05$\pm$0.07 & --0.07$\pm$0.04 & 0.04$\pm$0.05 & 0.03$\pm$0.04 & --0.03$\pm$0.10 & 0.04$\pm$0.11 & 0.15$\pm$0.11 & 0.36$\pm$0.14 & 0.10$\pm$0.10 & --0.03$\pm$0.10 & 0.17$\pm$0.06 & 0.03$\pm$0.04 \\
00385+4300 & BD+42 126 & --0.70$\pm$0.10 & --0.48$\pm$0.16 & --0.47$\pm$0.04 & --0.52$\pm$0.04 & --0.63$\pm$0.08 & --0.43$\pm$0.10 & --0.33$\pm$0.09 & --0.30$\pm$0.11 & --0.61$\pm$0.07 & --0.93$\pm$0.08 & --0.56$\pm$0.05 & --0.69$\pm$0.04 \\
00452+0015 & HD 4271 Aa,Ab & 0.02$\pm$0.04 & --0.04$\pm$0.03 & --0.06$\pm$0.03 & 0.05$\pm$0.03 & 0.04$\pm$0.03 & 0.05$\pm$0.10 & 0.08$\pm$0.03 & 0.02$\pm$0.07 & 0.09$\pm$0.04 & --0.13$\pm$0.05 & 0.02$\pm$0.04 & 0.05$\pm$0.02 \\
00467-0426 & HD 4449 & --0.04$\pm$0.07 & --0.14$\pm$0.20 & 0.01$\pm$0.09 & --0.02$\pm$0.09 & --0.31$\pm$0.10 & 0.13$\pm$0.12 & 0.10$\pm$0.11 & 0.29$\pm$0.17 & 0.04$\pm$0.10 & --0.15$\pm$0.15 & 0.01$\pm$0.08 & --0.09$\pm$0.06 \\
00491+5749 & Achird Aa,Ab & --0.30$\pm$0.01 & --0.31$\pm$0.06 & --0.33$\pm$0.03 & --0.24$\pm$0.03 & --0.28$\pm$0.02 & --0.22$\pm$0.02 & --0.25$\pm$0.02 & --0.29$\pm$0.03 & --0.25$\pm$0.02 & --0.51$\pm$0.04 & --0.29$\pm$0.03 & --0.30$\pm$0.01 \\
01055+1523 & HD 6440 A & --0.22$\pm$0.09 & --0.27$\pm$0.15 & --0.18$\pm$0.05 & --0.20$\pm$0.08 & --0.30$\pm$0.13 & 0.04$\pm$0.17 & --0.05$\pm$0.13 & 0.23$\pm$0.16 & --0.29$\pm$0.09 & --0.51$\pm$0.14 & --0.21$\pm$0.07 & --0.29$\pm$0.05 \\
01076+2257 & HD 6660 A & 0.37$\pm$0.23 & --0.07$\pm$0.09 & 0.10$\pm$0.15 & 0.05$\pm$0.12 & 0.09$\pm$0.27 & 0.66$\pm$0.34 & 0.31$\pm$0.29 & 0.68$\pm$0.33 & 0.09$\pm$0.19 & --0.09$\pm$0.22 & 0.27$\pm$0.12 & 0.07$\pm$0.09 \\
\hline
\end{tabular}
\end{table}
\end{landscape}

\begin{landscape}
\begin{table}
\contcaption{Abundances [X/H] with respect to the Sun for the four $\alpha$-elements (Mg, Si, Ca and Ti), the Fe-peak elements (Cr, Mn, Co, and Ni), and the odd-Z elements (Na, Al, Sc, and V).}
\fontsize{8}{8}\selectfont
\begin{tabular}{llcccccccccccc}
\hline
\hline
WDS & Name & [Na/H] &  [Mg/H] &  [Al/H] &  [Si/H] & [Ca/H] &  [Sc/H] & [Ti/H] & [V/H] & [Cr/H] & [Mn/H] & [Co/H] & [Ni/H] \\
\hline
07191+6644 & HD 55745 A & 0.11$\pm$0.01 & --0.28$\pm$0.42 & 0.09$\pm$0.05 & 0.17$\pm$0.03 & 0.17$\pm$0.03 & 0.20$\pm$0.12 & 0.26$\pm$0.04 & 0.25$\pm$0.06 & 0.28$\pm$0.04 & 0.17$\pm$0.08 & 0.13$\pm$0.04 & 0.20$\pm$0.02 \\
07321-0853 & HD 59984 & --0.78$\pm$0.04 & --0.62$\pm$0.07 & ... & --0.54$\pm$0.03 & --0.61$\pm$0.03 & --0.33$\pm$0.10 & --0.47$\pm$0.04 & --0.61$\pm$0.06 & --0.66$\pm$0.04 & --1.09$\pm$0.10 & --0.68$\pm$0.05 & --0.68$\pm$0.03 \\
07400-0336 & V869 Mon & --0.13$\pm$0.05 & --0.19$\pm$0.05 & --0.13$\pm$0.11 & --0.04$\pm$0.05 & --0.04$\pm$0.08 & 0.01$\pm$0.13 & 0.05$\pm$0.10 & 0.29$\pm$0.14 & --0.06$\pm$0.07 & --0.29$\pm$0.09 & --0.07$\pm$0.04 & --0.15$\pm$0.03 \\
08082+2106 & BD+21 1764A & --0.42$\pm$0.24 & --0.57$\pm$0.28 & --0.02$\pm$0.18 &  --0.41$\pm$0.21 & --0.19$\pm$0.35 & 0.43$\pm$0.38 & 0.02$\pm$0.37 & 0.31$\pm$0.42 & --0.29$\pm$0.21 & --0.87$\pm$0.23 & --0.18$\pm$0.17 & --0.42$\pm$0.13 \\
08082+7155 & HD 66171 & --0.38$\pm$0.02 & --0.28$\pm$0.05 & --0.25$\pm$0.01 & --0.24$\pm$0.02 & --0.26$\pm$0.03 & --0.12$\pm$0.07 & --0.22$\pm$0.03 & --0.27$\pm$0.05 & --0.26$\pm$0.03 & --0.49$\pm$0.06 & --0.24$\pm$0.06 & --0.32$\pm$0.02 \\
08107-1348 & 18 Pup A & --0.07$\pm$0.03 & --0.23$\pm$0.08 & --0.01$\pm$0.05 & --0.01$\pm$0.04 & --0.07$\pm$0.04 & 0.09$\pm$0.10 & 0.08$\pm$0.07 & 0.02$\pm$0.10 & 0.09$\pm$0.14 & --0.27$\pm$0.17 & --0.03$\pm$0.07 & --0.10$\pm$0.05 \\
08110+7955 & BD+80 245 & --1.65$\pm$0.06 & --1.25$\pm$0.40 & --1.52$\pm$0.05 & --0.95$\pm$0.21 & --1.62$\pm$0.21 & --0.44$\pm$0.32 & --1.23$\pm$0.19 & --1.15$\pm$0.23 & --1.11$\pm$0.17 & --1.96$\pm$0.32 & --1.10$\pm$0.34 & --1.42$\pm$0.10 \\
\hline
\end{tabular}
\end{table}
\end{landscape}

\begin{landscape}
\begin{table}
\contcaption{Abundances [X/H] with respect to the Sun for the four $\alpha$-elements (Mg, Si, Ca and Ti), the Fe-peak elements (Cr, Mn, Co, and Ni), and the odd-Z elements (Na, Al, Sc, and V).}
\fontsize{8}{8}\selectfont
\begin{tabular}{llcccccccccccc}
\hline
\hline
WDS & Name & [Na/H] &  [Mg/H] &  [Al/H] &  [Si/H] & [Ca/H] &  [Sc/H] & [Ti/H] & [V/H] & [Cr/H] & [Mn/H] & [Co/H] & [Ni/H] \\
\hline
14260+3422 & BD+35 2558 & --0.83$\pm$0.08 & --0.75$\pm$0.04 & --0.54$\pm$0.06 & --0.62$\pm$0.05 & --0.68$\pm$0.08 & --0.38$\pm$0.11 & --0.30$\pm$0.09 & --0.32$\pm$0.10 & --0.67$\pm$0.07 & --0.92$\pm$0.24 & --0.68$\pm$0.06 & --0.78$\pm$0.04 \\ 
14336+0920 & HD 127871 A & --0.11$\pm$0.07 & --0.09$\pm$0.09 & --0.02$\pm$0.04 & --0.08$\pm$0.04 & --0.06$\pm$0.08 & 0.04$\pm$0.14 & 0.07$\pm$0.09 & 0.37$\pm$0.13 & --0.09$\pm$0.06 & --0.19$\pm$0.08 & --0.03$\pm$0.05 & --0.11$\pm$0.03\\
14415+1336 & HD 129290 A & --0.20$\pm$0.01 & --0.21$\pm$0.11 & --0.10$\pm$0.04 & --0.04$\pm$0.05 & --0.13$\pm$0.02 & --0.13$\pm$0.06 & --0.10$\pm$0.02 & --0.15$\pm$0.03 & --0.12$\pm$0.02 & --0.27$\pm$0.06 & --0.16$\pm$0.02 & --0.16$\pm$0.02\\
14446-2215 & HD 129715 & 0.56$\pm$0.18 & 0.05$\pm$0.16 & 0.52$\pm$0.13 & 0.13$\pm$0.10 & 0.25$\pm$0.23 & 1.10$\pm$0.36 & 0.53$\pm$0.22 & 0.89$\pm$0.25 & 0.21$\pm$0.15 & --0.03$\pm$0.16 & 0.45$\pm$0.11 & 0.20$\pm$0.07\\
14493+4950 & HD 130986 A & --0.07$\pm$0.02 & --0.08$\pm$0.06 & --0.05$\pm$0.06 & --0.02$\pm$0.02 & 0.05$\pm$0.04 & 0.12$\pm$0.05 & 0.06$\pm$0.03 & 0.07$\pm$0.05 & 0.08$\pm$0.03 & --0.26$\pm$0.10 & 0.05$\pm$0.08 & --0.03$\pm$0.02\\
14575-2125 & HD 131977 & 0.10$\pm$0.14 & --0.19$\pm$0.14 & 0.08$\pm$0.09 & --0.08$\pm$0.09 & --0.10$\pm$0.18 & 0.41$\pm$0.24 & 0.11$\pm$0.20 & 0.46$\pm$0.23 & --0.05$\pm$0.13 & --0.30$\pm$0.15 & 0.01$\pm$0.07 & --0.11$\pm$0.07\\
14595+4528 & HD 132830 & --0.17$\pm$0.05 & --0.23$\pm$0.03 & --0.10$\pm$0.06 & --0.08$\pm$0.03 & --0.07$\pm$0.06 & --0.05$\pm$0.08 & 0.08$\pm$0.07 & 0.23$\pm$0.10 & --0.04$\pm$0.05 & --0.20$\pm$0.12 & --0.09$\pm$0.03 & --0.15$\pm$0.03\\
\hline
\end{tabular}
\end{table}
\end{landscape}

\begin{landscape}
\begin{table}
\contcaption{Abundances [X/H] with respect to the Sun for the four $\alpha$-elements (Mg, Si, Ca and Ti), the Fe-peak elements (Cr, Mn, Co, and Ni), and the odd-Z elements (Na, Al, Sc, and V).}
\fontsize{8}{8}\selectfont
\begin{tabular}{llcccccccccccc}
\hline
\hline
WDS & Name & [Na/H] &  [Mg/H] &  [Al/H] &  [Si/H] & [Ca/H] &  [Sc/H] & [Ti/H] & [V/H] & [Cr/H] & [Mn/H] & [Co/H] & [Ni/H] \\
\hline
22589+6902 & BD+68 1345A & --0.19$\pm$0.04 & --0.09$\pm$0.04 & --0.02$\pm$0.02 & --0.07$\pm$0.03 & --0.15$\pm$0.06 & --0.08$\pm$0.07 & 0.03$\pm$0.05 & 0.11$\pm$0.08 & --0.08$\pm$0.04 & --0.20$\pm$0.05 & --0.04$\pm$0.03 & --0.13$\pm$0.03\\
23026+2948 & BD+29 4841Aa,Ab & ... &  ... &  ... &  ... & ... & ... & ... & ... & ... & ... & ... & ... \\
23104+4901 & HD 218790 & 0.37$\pm$0.02 & 0.29$\pm$0.03 & 0.29$\pm$0.01 & 0.27$\pm$0.03 & 0.21$\pm$0.02 & 0.37$\pm$0.04 & 0.30$\pm$0.02 & 0.34$\pm$0.03 & 0.29$\pm$0.02 & 0.26$\pm$0.03 & 0.33$\pm$0.02 & 0.30$\pm$0.02\\
23194+7900 & V368 Cep & ... &  ... &  ... &  ... & ... & ... & ... & ... & ... & ... & ... & ... \\
23235+4548 & HD 220445 & 0.03$\pm$0.10 & --0.08$\pm$0.04 & 0.15$\pm$0.06 & 0.04$\pm$0.04 & 0.06$\pm$0.12 & 0.24$\pm$0.15 & 0.23$\pm$0.13 & 0.52$\pm$0.16 & 0.04$\pm$0.09 & --0.05$\pm$0.11 & 0.12$\pm$0.06 & --0.01$\pm$0.05\\
23266+4520 & HD 220821 & --0.21$\pm$0.02 & --0.21$\pm$0.04 & --0.11$\pm$0.02 & --0.20$\pm$0.02 & --0.13$\pm$0.04 & --0.08$\pm$0.04 & --0.10$\pm$0.03 & --0.13$\pm$0.04 & --0.18$\pm$0.02 & --0.36$\pm$0.06 & --0.25$\pm$0.04 & --0.25$\pm$0.02\\
23355+3101 & HD 221830 A & --0.39$\pm$0.01 & --0.21$\pm$0.02 & --0.15$\pm$0.01 & --0.22$\pm$0.03 & --0.22$\pm$0.03 & --0.20$\pm$0.10 & --0.11$\pm$0.02 & --0.25$\pm$0.03 & --0.39$\pm$0.02 & --0.78$\pm$0.06 & --0.30$\pm$0.03 & --0.40$\pm$0.02\\
\hline
\end{tabular}
\end{table}
\end{landscape}

\begin{table*}
\caption{Kinematics of late-F-, G-, and late-K stars.}
\label{tab:kinematics}
\begin{tabular}{llccccllll}
\hline
\hline
WDS & Name & $V_{r}$ & $U$ & $V$ & $W$ & Pop.$^{b}$ & SKG$^{c}$ & SKG$_{lit}$ & Ref.$^{d}$\\
 &  & [km s$^{-1}$] & [km s$^{-1}$] & [km s$^{-1}$] & [km s$^{-1}$] & & & & \\

\hline

  00153+5304 & G 217-41 & --13.96$\pm$0.18 & --66.1$\pm$1.7 & --48.91$\pm$0.85 & 7.21$\pm$0.12 & D & $\times$ & ... & ...\\
  00385+4300 & BD+42 126 & --55.83$\pm$0.21 & --9.27$\pm$0.43 & --75.36$\pm$0.39 & --2.65$\pm$0.26 & D & $\times$ & ... & ...\\
  00452+0015 & HD 4271 Aa,Ab $^{a}$ & 12.45$\pm$0.04 & --42.24$\pm$0.90 & --30.38$\pm$0.81 & --16.94$\pm$0.14 & D & YD & Field & BAN$\Sigma$\\
  00467-0426 & HD 4449 & 6.26$\pm$0.19 & 13.7$\pm$1.2 & --27.3$\pm$2.3 & --19.6$\pm$1.1 & D & $\times$ & ... &... \\
  00491+5749 & Archid Aa,Ab $^{a}$ & 9.2 & --29.88$\pm$0.07 & --9.93$\pm$0.05 & --16.81$\pm$0.05 & D & YD & YD & Mon01\\
  01055+1523 & HD 6440 A & --5.68$\pm$0.32 & 11.45$\pm$0.15 & --19.73$\pm$0.21 & --13.1$\pm$0.26 & D & $\times$ & ... & ...\\
\hline
  \end{tabular}
 \end{table*}

\begin{table*}
\contcaption{Kinematics of late-F-, G-, and late-K stars.}
\begin{tabular}{llccccllll}
\hline
\hline
WDS & Name & $V_{r}$ & $U$ & $V$ & $W$ & Pop.$^{b}$ & SKG$^{c}$ & SKG$_{lit}$ & Ref.$^{d}$\\
 &  & [km s$^{-1}$] & [km s$^{-1}$] & [km s$^{-1}$] & [km s$^{-1}$] & & & & \\

\hline
  07191+6644 & HD 55745 A & 5.13$\pm$0.04 & --31.56$\pm$0.39 & --22.49$\pm$0.35 & --18.82$\pm$0.30 & D & YD & Field & BAN$\Sigma$\\
  07321-0853 & HD 59984 & 54.85$\pm$0.11 & --29.77$\pm$0.15 & --49.89$\pm$0.18 & --16.57$\pm$0.33 & D & $\times$ & ... &... \\
  07400-0336 & V869 Mon & --18.20$\pm$0.52 & 25.22$\pm$0.10 & --2.37$\pm$0.10 & --7.47$\pm$0.03 & D & UMa & UMa & Tab17\\
  08082+2106 & BD+21 1764A & 79.35$\pm$0.37 & --73.41$\pm$0.31 & --49.18$\pm$0.16 & 4.01$\pm$0.21 & D & $\times$ & ... &... \\
  08082+7155 & HD 66171 & 36.39$\pm$0.03 & --104.31$\pm$0.94 & --55.53$\pm$0.87 & --11.8$\pm$0.36 & TD--D & $\times$ &...  &... \\
  08107-1348 & 18 Pup A & 33.06$\pm$0.17 & --36.66$\pm$0.38 & --18.21$\pm$0.22 & --13.14$\pm$0.40 & D & HS & Field  & BAN$\Sigma$\\
\hline
  \end{tabular}
 \end{table*}

\begin{table*}
\contcaption{Kinematics of late-F-, G-, and late-K stars.}
\begin{tabular}{llccccllll}
\hline
\hline
WDS & Name & $V_{r}$ & $U$ & $V$ & $W$ & Pop.$^{b}$ & SKG$^{c}$ & SKG$_{lit}$ & Ref.$^{d}$\\
 &  & [km s$^{-1}$] & [km s$^{-1}$] & [km s$^{-1}$] & [km s$^{-1}$] & & & & \\

\hline
  14260+3422 & BD+35 2558 & 40.98$\pm$0.26 & --11.34$\pm$0.27 & --63.9$\pm$1.1 & 67.84$\pm$0.48 & TD & $\times$ & ... &... \\
  14336+0920 & HD 127871 A & 30.71$\pm$0.23 & 76.41$\pm$0.47 & --43.32$\pm$0.33 & --8.00$\pm$0.33 & D & $\times$ & ... &... \\
  14415+1336 & HD 129290 A & --18.41$\pm$0.08 & --74.4$\pm$4.5 & --104.7$\pm$7.0 & 30.3$\pm$3.2 & TD & $\times$ & ... &... \\
  14446-2215 & HD 129715 & 9.05$\pm$0.30 & 7.76$\pm$0.23 & --45.56$\pm$0.44 & --23.57$\pm$0.33 & D & $\times$ & ... &... \\
  14493+4950 & HD 130986 A & --7.98$\pm$0.04 & 0.47$\pm$0.02 & 6.33$\pm$0.17 & --13.37$\pm$0.11 & D & $\times$ & ... &... \\
  14575-2125 & HD 131977 & 27.00$\pm$0.32 & 48.11$\pm$0.29 & --21.87$\pm$0.13 & --32.3$\pm$0.31 & D & $\times$ & ... &... \\
\hline
  \end{tabular}
 \end{table*}

\begin{table*}
\contcaption{Kinematics of late-F-, G-, and late-K stars.}
\begin{tabular}{llccccllll}
\hline
\hline
WDS & Name & $V_{r}$ & $U$ & $V$ & $W$ & Pop.$^{b}$ & SKG$^{c}$ & SKG$_{lit}$ & Ref.$^{d}$\\
 &  & [km s$^{-1}$] & [km s$^{-1}$] & [km s$^{-1}$] & [km s$^{-1}$] & & & & \\

\hline
  22589+6902 & BD+68 1345A & --33.08$\pm$0.14 & --105.8$\pm$3.1 & --81.3$\pm$1.4 & --1.10$\pm$0.14 & TD & $\times$ &...  & ...\\
  23026+2948 & BD+29 4841Aa,Ab $^{a}$  & --7.09$\pm$0.25 & 23.74$\pm$0.75 & 2.34$\pm$0.35 & 14.8$\pm$0.39 & D & $\times$ & ... & ...\\
  23104+4901 & HD 218790 & --4.85$\pm$0.06 & --47.32$\pm$0.51 & --20.69$\pm$0.18 & --8.38$\pm$0.10 & D & HS & Field & BAN$\Sigma$ \\
  23194+7900 & V368 Cep & --16.8$\pm$2.0 & --9.48$\pm$0.91 & --23.2$\pm$1.7 & --5.46$\pm$0.58 & D & LA & LA & Mon01\\
  23235+4548 & HD 220445 & --49.43$\pm$0.22 & --4.67$\pm$0.47 & --53.83$\pm$0.28 & 3.98$\pm$0.21 & D & $\times$ & ... &... \\
  23266+4520 & HD 220821 & --2.50$\pm$0.07 & --70.0$\pm$2.0 & --26.85$\pm$0.71 & --6.50$\pm$0.21 & D & $\times$ & ... &... \\
\end{tabular}
\end{table*}




\bsp	
\label{lastpage}
\end{document}